\documentclass{article}

\usepackage{microtype}
\usepackage{graphicx}
\usepackage{subfigure}
\usepackage{booktabs} 

\usepackage{hyperref}

\usepackage[accepted]{icml2023}

\usepackage{amsmath}
\usepackage{amssymb}
\usepackage{mathtools}
\usepackage{amsthm}

\usepackage[capitalize,noabbrev]{cleveref}

\theoremstyle{plain}

\theoremstyle{definition}

\theoremstyle{remark}

\usepackage[textsize=tiny]{todonotes}

\icmltitlerunning{Von Mises Mixture Distributions for Molecular Conformation Generation}

\begin{document}

\twocolumn[
\icmltitle{Von Mises Mixture Distributions for Molecular Conformation Generation}



\icmlsetsymbol{equal}{*}

\begin{icmlauthorlist}
\icmlauthor{Kirk Swanson}{yyy}
\icmlauthor{Jake Williams}{yyy}
\icmlauthor{Eric Jonas}{yyy}
\end{icmlauthorlist}

\icmlaffiliation{yyy}{Department of Computer Science, University of Chicago, Chicago, USA}

\icmlcorrespondingauthor{Kirk Swanson}{swansonk1@uchicago.edu}

\icmlkeywords{Machine Learning, ICML}

\vskip 0.3in
]



\printAffiliationsAndNotice{} 

\begin{abstract}
Molecules are frequently represented as graphs, but the underlying 3D molecular geometry (the locations of the atoms) ultimately determines most molecular properties. However, most molecules are not static and at room temperature adopt a wide variety of geometries or \textit{conformations}. The resulting distribution on geometries $p(x)$ is known as the Boltzmann distribution, and many molecular properties are expectations computed under this distribution. Generating accurate samples from the Boltzmann distribution is therefore essential for computing these expectations accurately. Traditional sampling-based methods are computationally expensive, and most recent machine learning-based methods have focused on identifying \textit{modes} in this distribution rather than generating true \textit{samples}. Generating such samples requires capturing conformational variability, and it has been widely recognized that the majority of conformational variability in molecules arises from rotatable bonds. In this work, we present VonMisesNet, a new graph neural network that captures conformational variability via a variational approximation of rotatable bond torsion angles as a mixture of von Mises distributions. We demonstrate that VonMisesNet can generate conformations for arbitrary molecules in a way that is both physically accurate with respect to the Boltzmann distribution and orders of magnitude faster than existing sampling methods.
\end{abstract}

\section{Introduction}
\label{sec:sec1}
Accurate prediction of molecular properties is an important task in computational chemistry. Many physical and chemical properties are dependent on a molecule's 3D conformations, and
some properties, such as spectroscopic properties like nuclear magnetic resonance (NMR) chemical shifts, are measured on timescales considerably longer than the timescale of molecular motion. At room temperature, most molecules are not static and adopt a wide variety of conformations. These properties are therefore computed as expectations with respect to the probability density that governs molecular conformations, known as the Boltzmann distribution. This density is given by
\begin{align}
    \begin{split}
        p(x) = \frac{e^{-E(x)/{k_BT}}}{Z},
    \end{split}
\end{align}
where $x \in \mathbb{R}^{n \times 3}$ is the set of atomic coordinates that define a conformation over the $n$ atoms, $E(x)$ is the potential energy, $T$ is the temperature, $Z$ is a normalizing constant, and $k_B$ is the Boltzmann constant \cite{Tuckerman}. $E(x)$ has been modeled extensively using fundamental chemistry knowledge. For example, there are classical mechanics-based force fields, such as the widely-used Merck Molecular Force Field (MMFF) \cite{MMFF}, and quantum mechanical-based approaches, such as Density Functional theory (DFT).

Generating accurate samples from this distribution is essential for computing accurate molecular property expectations. However, sampling from $p(x)$ is very challenging because it is a high-dimensional and multi-modal distribution; generating samples from high-dimensional distributions is a classic problem in computational statistics \cite{JunSLiu}.

The complexity in the Boltzmann distribution is primarily due to the large number of degrees of freedom that a given molecule possesses. These degrees of freedom are typically defined in terms of three types of so-called internal coordinates: bond lengths, which are distances between pairs of bonded atoms; bond angles, which are angles between bonds connected to a common atom; and torsions, which are dihedral angles formed by four consecutively bonded atoms. The number of degrees of freedom therefore increases substantially as a function of molecule size.

There are a variety of classical approaches to conformation generation, including stochastic methods and systematic methods \cite{ConfGenReview}. Some stochastic methods, such as Markov Chain Monte Carlo and Molecular Dynamics, attempt to generate samples from the Boltzmann distribution (which we refer to below as ``Boltzmann samples") but can suffer from long runtimes on the order of minutes or even hours for large molecules \cite{Tuckerman}. 

Other stochastic methods, such as Monte Carlo Multiple Minimum and Low Mode Search, generate low energy (i.e., high probability) conformations and have shorter runtimes \cite{InternalMC,MCMethods,SimulatedAnnealing,LMS,MCMMLowMode,GA,MacroModel}. However, finding \textit{modes} is different than generating true \textit{samples} from the underlying distribution. Generating modes can be useful, and it is possible to utilize strategies such as importance sampling to compute expectations with them. However, this is not ideal for computing expectations with respect to the Boltzmann distribution, because there are lots of inherent symmetries and flexibility that can make mere mode identification insufficient.

Distance geometry methods, such as RDKit's ETKDG algorithm \cite{ETKDG}, use a stochastic approach in conjunction with knowledge-based rules to produce a diverse set of geometries, but these conformations are not Boltzmann samples. Systematic methods, such as Confab \cite{Confab}, aim to generate conformations via brute force search or using knowledge-based libraries, but this becomes intractable for large molecules and these methods also do not produce Boltzmann samples \cite{BCLConf,CSD}.

A number of machine learning approaches have been introduced which aim to improve the speed and accuracy of conformation generation. Most of these approaches, including DL4Chem \cite{DL4Chem}, CGCF \cite{CGCF}, GeoMol \cite{GeoMol}, and RMCF \cite{wang2022regularized}, have focused on generating low energy conformations \cite{LearningGradient,luo2021predicting,GeoDiff}. Similar to some of the classical methods, these methods do not generate Boltzmann samples. GraphDG was designed to generate such samples when combined with an importance sampling scheme, but it was only trained and evaluated on a limited set of 197 distinct molecular graphs from a single molecular formula \cite{GraphDG}. No\'{e} \textit{et al.} devised so-called Boltzmann generators, which utilize normalizing flows to generate Boltzmann samples \cite{BoltzmannGenerator}. These have shown promise on proteins as well as small organic molecules \cite{kohler2021smooth}, but a separate model needs to be trained for every molecule, because the normalizing flows operate on intrinsic coordinates specific to a given molecule. This limits their utility for high-throughput tasks \cite{jing2022torsional}.

A Torsional Diffusion method was recently introduced by Jing \textit{et al.} which exclusively models rotatable bond torsion angles \cite{jing2022torsional}. It is well-understood that the primary source of conformational variability arises from torsional rotations about so-called rotatable bonds \cite{Solved,Geom}. Rotatable bonds are canonically defined as single, non-ring bonds which are attached to atoms that are non-terminal and not triply bonded.\footnote{This definition is based on the canonical rotatable bond SMARTS string from RDKit:  \texttt{[!\$(\*\#\*)\&!D1]-!\@[!\$(\*\#\*)\&!D1]}.} Jing \textit{et al.} use a slightly more expansive definition of rotatable bond as any bond which, if severed, would produce two connected components in the molecular graph. The authors trained a Torsional Diffusion model to produce low energy conformations, and it showed state-of-the-art performance on standard datasets. Because the model provides exact likelihoods, the authors were also able to train a Torsional Diffusion model to generate independent samples from the marginal Boltzmann distribution of rotatable bond torsion angles for a variety of molecules. However, it was only trained and evaluated on molecules with between 3 and 7 rotatable bonds (using the authors' definition), which excludes large swaths of chemical space; ibuprofen, for example, has 8 rotatable bonds. Torsional Diffusion is also more computationally expensive than other machine learning models. 

In this work, we present VonMisesNet, a new graph neural network that captures conformational variability via a variational approximation of rotatable bond torsion angles as a mixture of von Mises distributions.\footnote{Code is available at \href{https://github.com/thejonaslab/vonmises-icml-2023}{https://github.com/thejonaslab/vonmises-icml-2023}.} Similar to Torsional Diffusion, we focus on modeling the rotatable bonds, although we use the canonical rotatable bond definition because it excludes some bonds, such as double or triple bonds, that may not be freely rotatable. VonMisesNet places no restriction on the number of rotatable bonds that it can process, and therefore it can be used to generate conformations for arbitrary molecular graphs. It is also the first machine learning method that specifically accounts for chirality inversion, a phenomenon that can strongly influence the local geometry about atoms with three neighbors and one lone pair of electrons, which is often the case with nitrogen atoms. We show that our method is not only more accurate than other methods in terms of its ability to generate samples from the marginal Boltzmann distribution of rotatable bond torsion angles, but that it is also orders of magnitude faster than diffusion-based methods.

\section{Methods}
\label{sec:sec2}

\subsection{Ground Truth Conformation Generation}\label{sec:sec21}
To train a machine learning model to generate Boltzmann samples, we require datasets that contain Boltzmann samples for a variety of molecules. However, most widely-used benchmark datasets consist of low energy conformations. For example, the popular datasets GEOM-QM9 \cite{GEOMQM9} and GEOM-DRUGS \cite{Geom} consist of low energy conformations generated from DFT calculations. Simm and Hernandez-Lobato introduced the CONF17 benchmark \yrcite{GraphDG}, which consists of conformations generated via \textit{ab initio} Molecular Dynamics simulations. Although these conformations are Boltzmann samples, the dataset only consists of 197 unique molecular graphs with the formula $\text{C}_7\text{H}_{10}\text{O}_2$.

Therefore, to train and evaluate on Boltzmann samples that represent a much larger and more diverse set of molecules, we construct our own ground truth dataset. We combine Parallel Tempering and Hamiltonian Monte Carlo, techniques that are both well-known to efficiently sample complex distributions \cite{JunSLiu,Tuckerman}. We refer to this method as PT-HMC (see Appendix \S \ref{app:appA} for details).  

In our PT-HMC simulations, we use MMFF to give the molecular potential energy, $E(x)$. The Torsional Diffusion model that was trained to generate Boltzmann samples for rotatable bonds also utilized MMFF. We use the MMFF implementation in RDKit to compute potential energies and gradients \cite{MMFFRdkit}. Although quantum mechanical approaches would yield more physically accurate geometries, classical force fields allow for more computationally efficient simulations while still providing good approximations to the true Boltzmann distribution. This lets us show proof of concept for our method; future work will investigate the use of quantum mechanical approaches. 

\subsection{Datasets}\label{sec:sec22}
We used PT-HMC to generate conformations for two datasets of molecules: NMRShiftDB and GDB-17 (see details in Appendix \S \ref{app:appB}). We use 32,171 molecules from NMRShiftDB and 134,228 molecules from GDB-17 that have at most 64 atoms and elements in the set \{H, C, O, N, F, S, P, Cl\}. We split each of these datasets into training (NMRShiftDB-train and GDB-17-train) and test (NMRShiftDB-test and GDB-17-test) datasets by computing a hash of the Morgan fingerprint with a radius of 4 and 2,048 bits for each molecule. If the last digit of this hash is a 0 or a 1, the molecule is in the test set, otherwise it is in the train set. This produces an 80/20 train/test split, and it allows us to consistently determine whether a given molecule is in the train set or the test set. Each molecule has approximately 560 conformations. 

\subsection{VonMisesNet}\label{sec:sec23}
In this section, we describe VonMisesNet and how it is used to generate conformations. In subsections \ref{sec:sec231}, \ref{sec:sec232}, and \ref{sec:sec233}, we describe the geometric components that VonMisesNet predicts and explain our modeling choices. In subsection \ref{sec:sec234}, we describe the VonMisesNet architecture and training procedure. In subsection \ref{sec:sec235}, we explain how the predictions from VonMisesNet can be used to rapidly generate accurate molecular conformations.

\subsubsection{Rotatable Bond Modeling}\label{sec:sec231}
Our primary goal is to model the marginal Boltzmann distribution of rotatable bond torsion angles in a given molecule. To do so, we need a canonical way of defining these angles, which are dihedral angles formed by four consecutively bonded atoms. The central two atoms are fixed as the begin and end atoms of the bond itself, but there can be multiple choices for the other two atoms. We use a breadth-first search approach based on the Cahn-Ingold-Prelog (CIP) \cite{CIP} rules to determine which four atoms, including the two central atoms, should be chosen (see Appendix \S \ref{app:appC} for details). 

\begin{figure}[!ht]
\vspace{-1mm}
\begin{center}
\subfigure[]{
    \includegraphics[scale = 0.1]{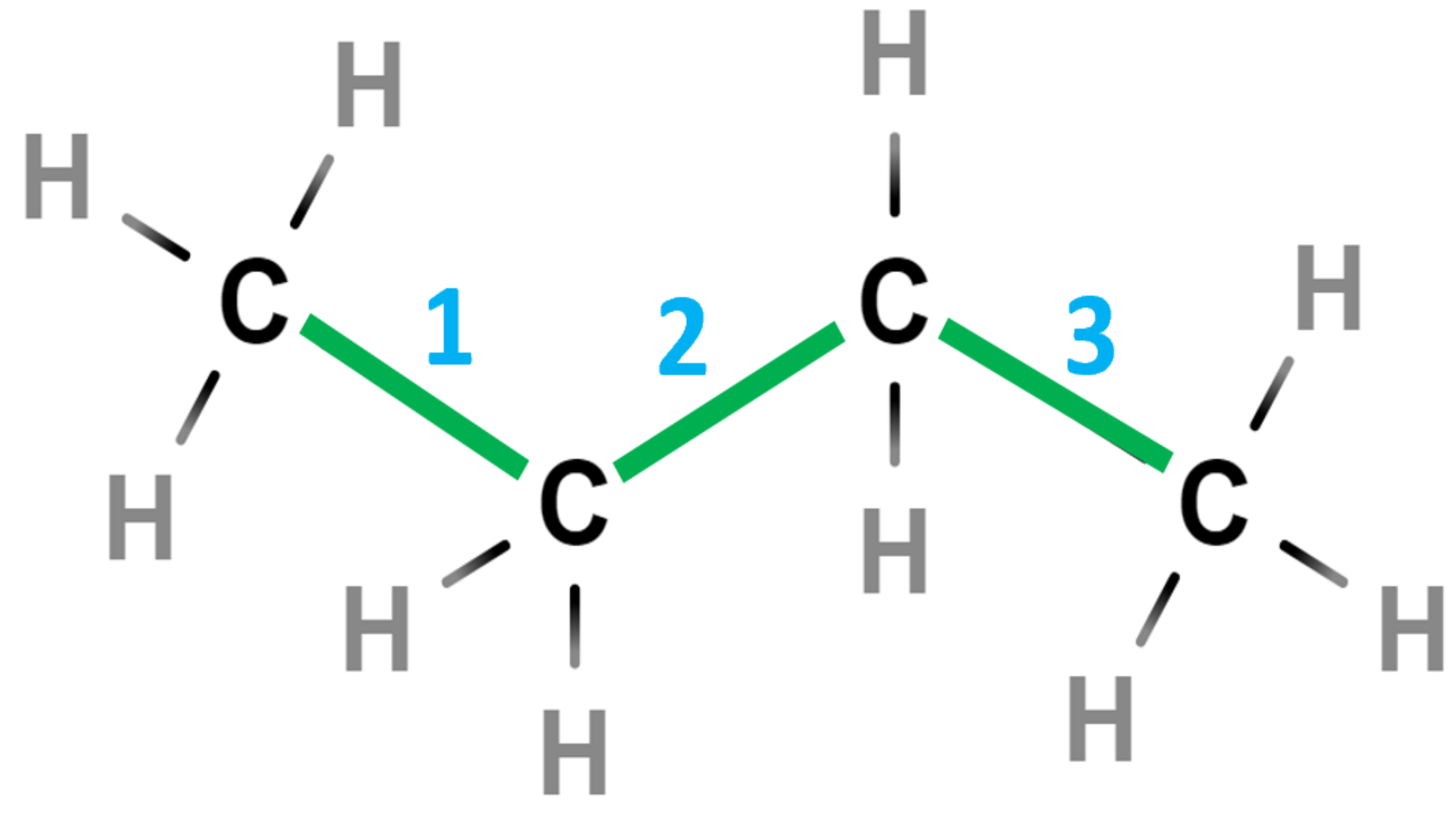}}%
\subfigure[]{
  \includegraphics[scale = 0.4]{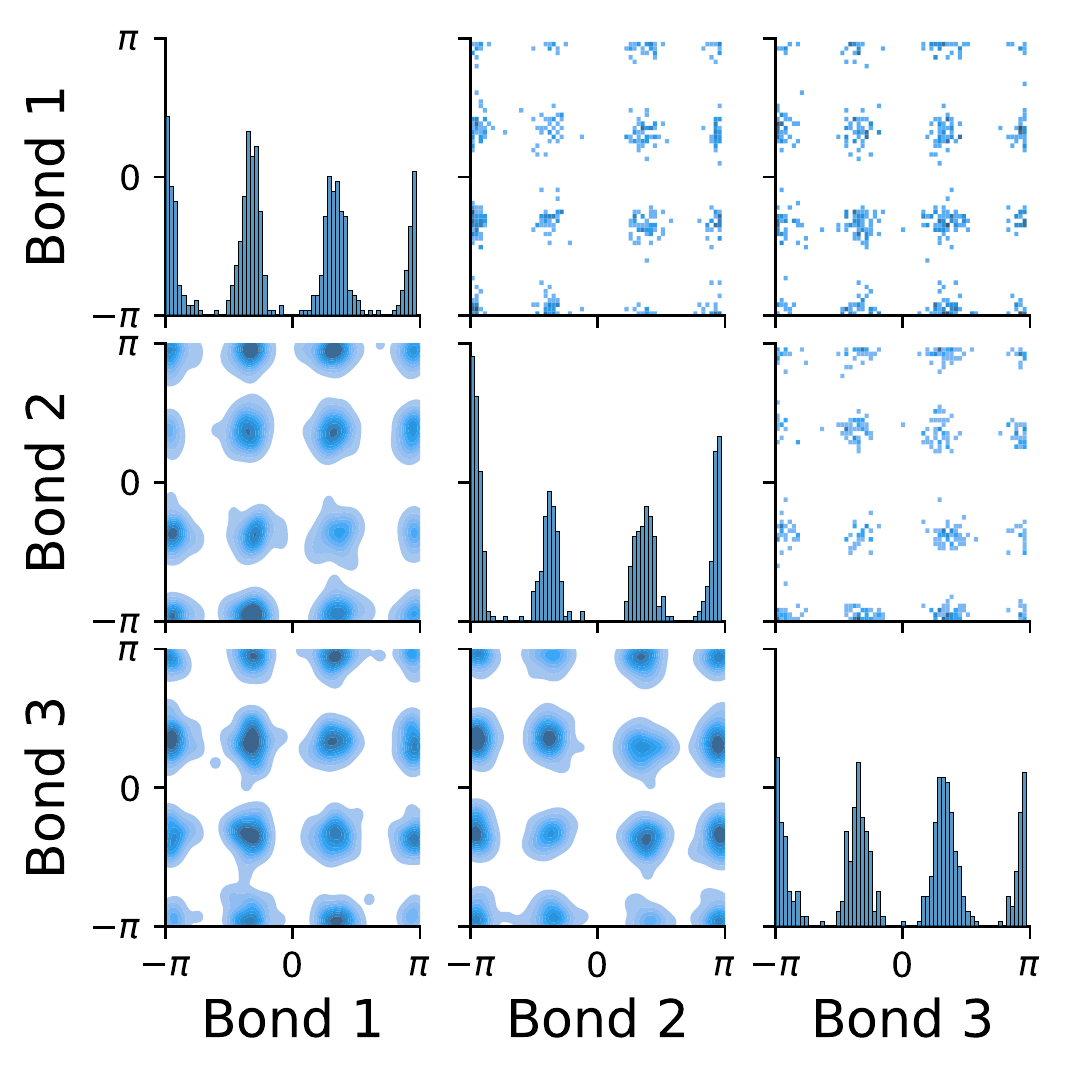}}%
\caption{\label{fig:figure1}\textit{Rotatable bonds example}. \textbf{(a)} Molecular graph of butane, with the three rotatable bonds highlighted. \textbf{(b)} Marginal and pairwise joint distributions of the three rotatable bond torsion angles in butane from 560 conformations generated via PT-HMC. The diagonal plots show the marginal distributions. Each of these distributions has three distinct modes (note that the axes represent a periodic angle variable ranging from $-\pi$ to $\pi$). The plots above and below the diagonal show scatter and kernel density estimates of the pairwise joint distributions, respectively. The pairwise joint distributions present clearly as product distributions, which indicates that the distributions are approximately independent.}
\end{center}
\end{figure}

If $\theta_i$ is the angle of the $i^{th}$ rotatable bond and there are $N$ rotatable bonds, then our target distribution is given by $p(\theta_1, ...,\theta_N)$. We make two primary assumptions about this distribution: independence and multi-modality. 

\textbf{Independence.} The first assumption is that the individual rotatable bond torsion angle distributions are approximately independent. Under this assumption, we express the target distribution as a product:
\vspace{-2mm}
\begin{align}
    \begin{split}
        p(\theta_1, ..., \theta_N) \approx \prod_{i=1}^N p(\theta_i)
    \end{split}
\end{align}
We find that this formulation facilitates straightforward modeling and rapid generation of conformations, but in future work we hope to relax this assumption and to explicitly model long-range interactions.

Butane is an example of a molecule that supports this assumption. In Figure \ref{fig:figure1}, the pairwise joint distributions of the rotatable bond torsion angles clearly present as product distributions. In Appendix \S \ref{app:appD}, we show that these distributions are approximately rank-1, which indicates independence. We also show that on average, for a larger set of molecules, pairs of rotatable bonds have a joint distribution that is approximately rank-1, with the approximation improving when the bonds are farther apart within a molecule. 

\begin{figure}[!ht]
\begin{center}
\subfigure[]{
\includegraphics[scale = 0.45]{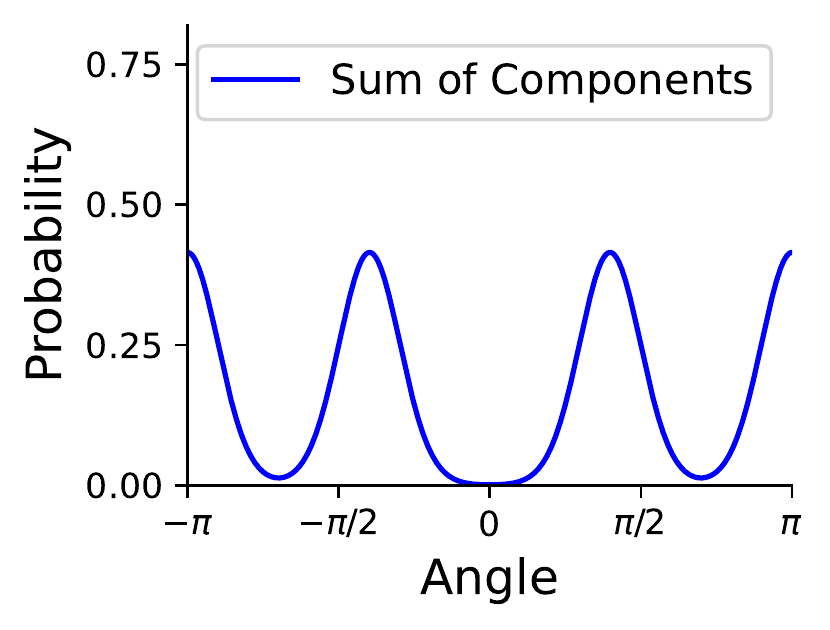}
}
\subfigure[]{
\includegraphics[scale = 0.45]{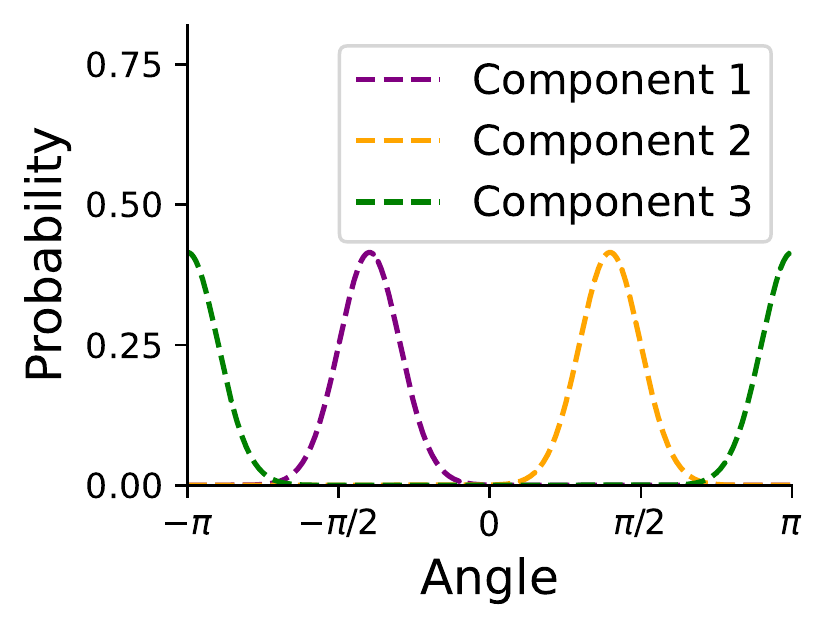}
}
\caption{\label{fig:figure2}\textit{von Mises distribution example.} \textbf{(a)} The probability density of a mixture of three von Mises distributions with means $\mu_1=-1.25$, $\mu_2=1.25$ and $\mu_3=\pi$. The concentrations $\kappa_i$ are all equal to 10, and the weights $w_i$ are all equal to 1/3. \textbf{(b)} The probability densities of the individual component distributions.}
\end{center}
\vskip -0.1in
\end{figure}

\textbf{Multi-modality.} The second assumption is that most individual rotatable bond torsion angle distributions are multi-modal, typically with up to four distinct modes. As shown in Figure \ref{fig:figure1}, butane also exemplifies this. In Appendix \S \ref{app:appE}, we show that in a larger set of molecules, the majority of rotatable bond torsion angle distributions have fewer than four distinct modes.  

With these assumptions, we model a given rotatable bond torsion angle distribution as a mixture of von Mises distributions. We use a von Mises distribution because it is a continuous probability distribution on the circle with support in $[-\pi, \pi]$, which is the periodic range of rotatable bond torsion angles. We use a mixture of von Mises distributions to capture the multi-modal nature of rotatable bond torsion angles. The weighted sum of $N$ von Mises distributions is given by 
\vspace{-3mm}
\begin{align}
    \begin{split}
        p(\theta) = \sum_{i=1}^Nw_i\frac{e^{\kappa_i\text{cos}(\theta - \mu_i)}}{2\pi \text{I}_0(\kappa_i)},
    \end{split}
\end{align}
where $\theta$ is the angle, $I_0$ is the modified Bessel function of order 0, $w_i$ is the weight, $\mu_i$ is the mean, and $\kappa_i$ is the concentration of the $i^{th}$ distribution.\footnote{The concentration is analagous to the inverse variance.} Figure \ref{fig:figure2} shows an example. In our experiments we use $N=4$ as most of the distributions have up to four distinct modes. VonMisesNet predicts the parameters of this weighted sum for each rotatable bond using the molecular graph as the input. Hence, we capture conformational variability via a variational approximation of rotatable bond torsion angles as a mixture of von Mises distributions.

\subsubsection{Chirality Inversion Modeling}\label{sec:sec232}
\begin{figure}[!hb]
\begin{center}
\includegraphics[scale = 0.15]{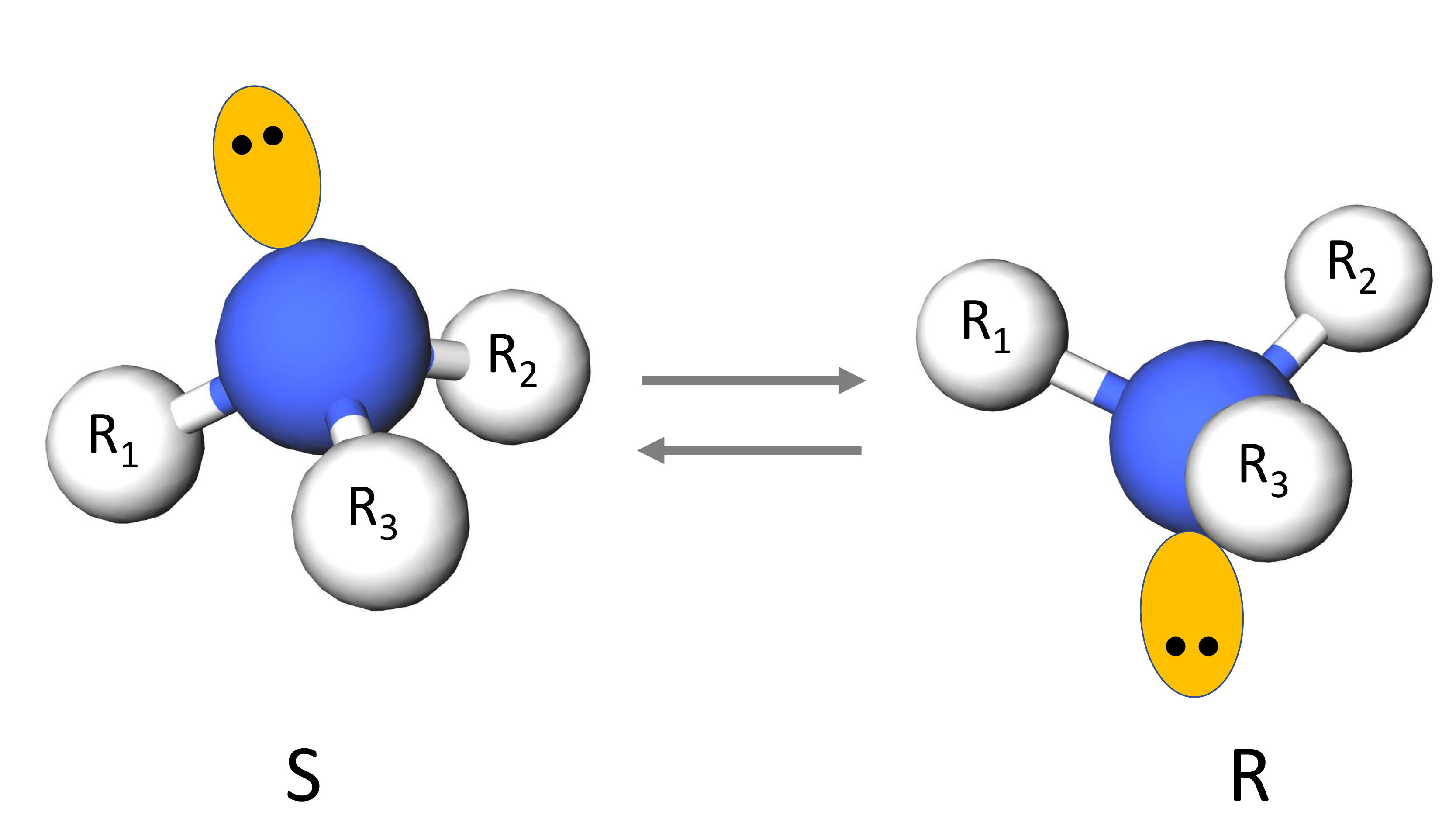}
\caption{\label{fig:figure3} \textit{Chirality inversion example.} A molecular fragment with a chirality inversion atom (blue) connected to three bonded neighbors (white) along with a lone pair (yellow). The chirality inversion atom moves through the plane formed by the three neighbors. Assuming that the neighbors are numbered in order of decreasing CIP priority and that the lone pair has the lowest priority, the configuration on the left has S chirality and the configuration on the right has R chirality; one cannot simply rotate the one on the left to achieve the one on the right.}
\end{center}
\vskip -0.1in
\end{figure}

Some atoms can undergo a specific type of transformation that influences the local geometry about a rotatable bond. Atoms that have three bonded neighbors and one lone pair of electrons can exhibit chirality inversion. (From here on we refer to such atoms as ``chirality inversion atoms.") This means that the atom moves through the plane formed by its three neighbors, which causes an oscillation between R and S chirality (see Appendix \S \ref{app:appG} for details). This can often occur with nitrogen atoms, where the thermodynamic barrier for this inversion, $\sim$ 25 kJ/mol, is low enough to allow rapid inversion at room temperature, leading to a racemic mixture of R and S configurations \cite{NitrogenInversion}. An example is shown in Figure \ref{fig:figure3}. 

\vspace{-1mm}

If a chirality inversion atom is an endpoint of a rotatable bond, then the inversion can influence the rotatable bond torsion angle distribution. 31\% of molecules in our NMRShiftDB data and 60\% of molecules in our GDB-17 data have at least one rotatable bond connected to a chirality inversion atom.

We model chirality inversion as follows. Let $\theta$ be a rotatable bond torsion angle where one of the endpoint atoms, A, is a chirality inversion atom.\footnote{There are some cases where both of the atom endpoints are chirality inversion atoms. There are only 203 molecules with such rotatable bonds in our NMRShiftDB data (0.6\%)  and none in our GDB-17 data, so we do not explicitly handle this special case. As described below, when making predictions for such cases, we condition on the chirality of just one of the endpoint atoms.} Let $p(\theta | R)$ be the angle distribution when A has R chirality,  $p(\theta | S)$ be the angle distribution when A has S chirality, and $p(R)$ be the probability that A has R chirality. Then, we can express the rotatable bond torsion angle distribution as
\vspace{-0.2mm}
\begin{align}
    \begin{split}
        p(\theta) = p(R)\cdot p(\theta | R) + (1 - p(R))\cdot p(\theta | S).
    \end{split}
\end{align}
We model $p(\theta | R)$ and $p(\theta | S)$, each, as mixtures of von Mises distributions, as described above. For each rotatable bond that has a chirality inversion endpoint atom, VonMisesNet predicts the von Mises mixture parameters for both $p(\theta | R)$ and $p(\theta | S)$. It additionally predicts $p(R)$ for each chirality inversion atom in the molecule.

\subsubsection{Bond Length and Bond Angle Modeling}\label{sec:sec233}
Although rotatable bonds are responsible for most of the conformational variability in a molecule, its geometry is also determined by bond lengths and bond angles. These are typically unimodal distributions with minor variance, which obviates the need to predict full distributions for them. Therefore, VonMisesNet predicts averages for bond lengths and bond angles. 

\subsubsection{Architecture and Training}\label{sec:sec234}
\begin{figure*}[!ht]
\begin{center}
\subfigure[]{
\includegraphics[scale = 0.14]{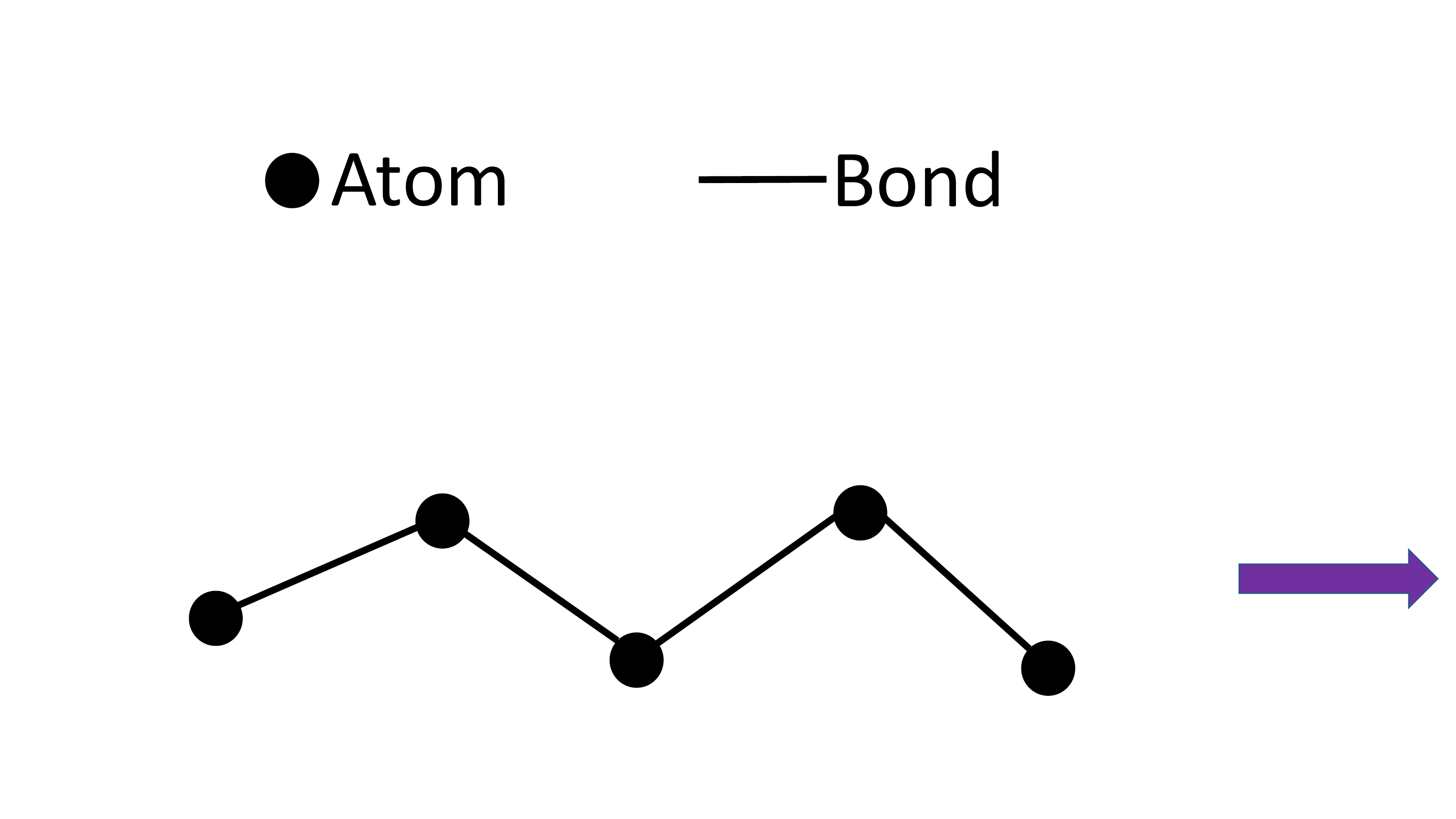}
}
\subfigure[]{
\includegraphics[scale = 0.14]{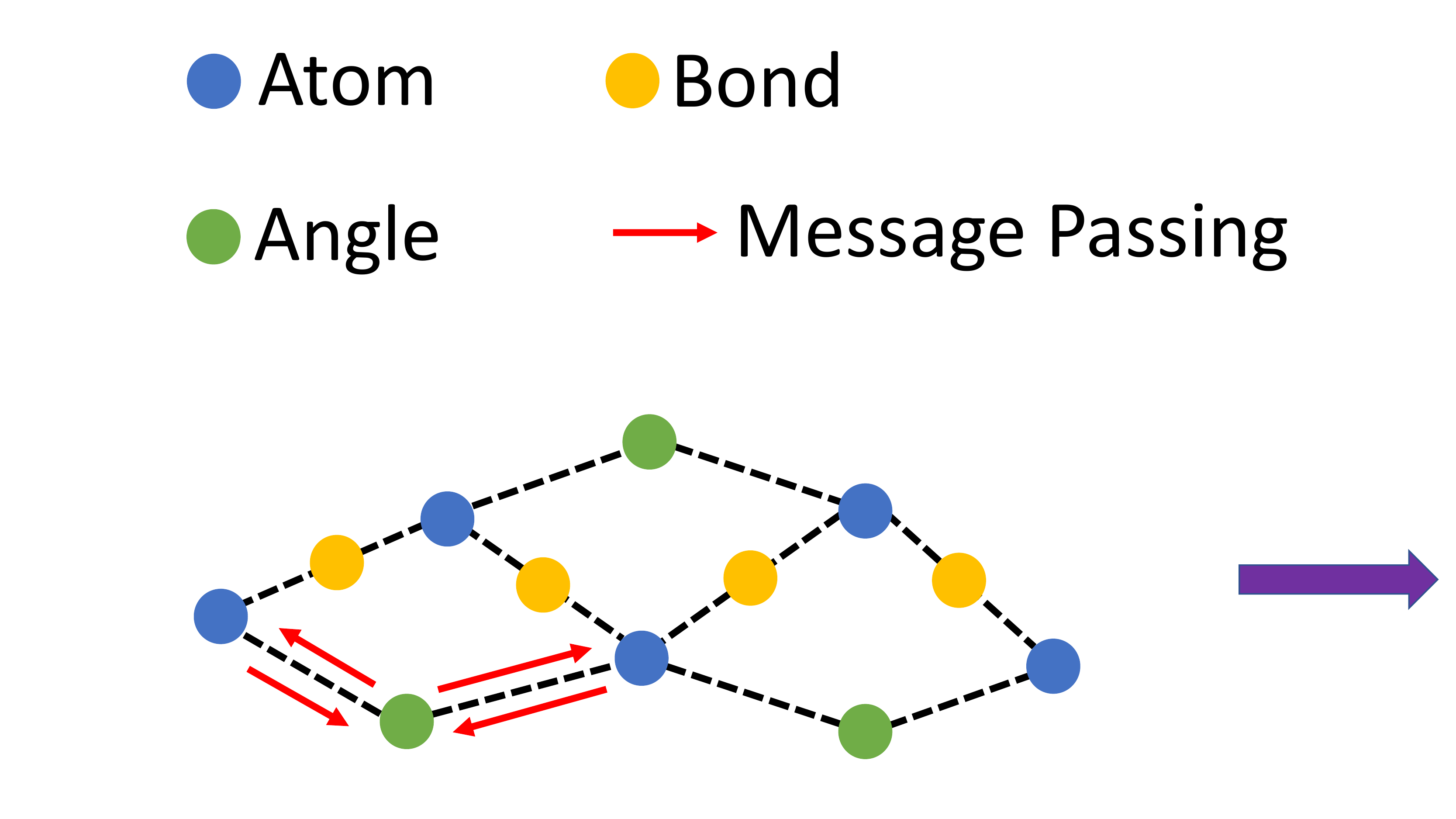}
}
\subfigure[]{
\includegraphics[scale = 0.14]{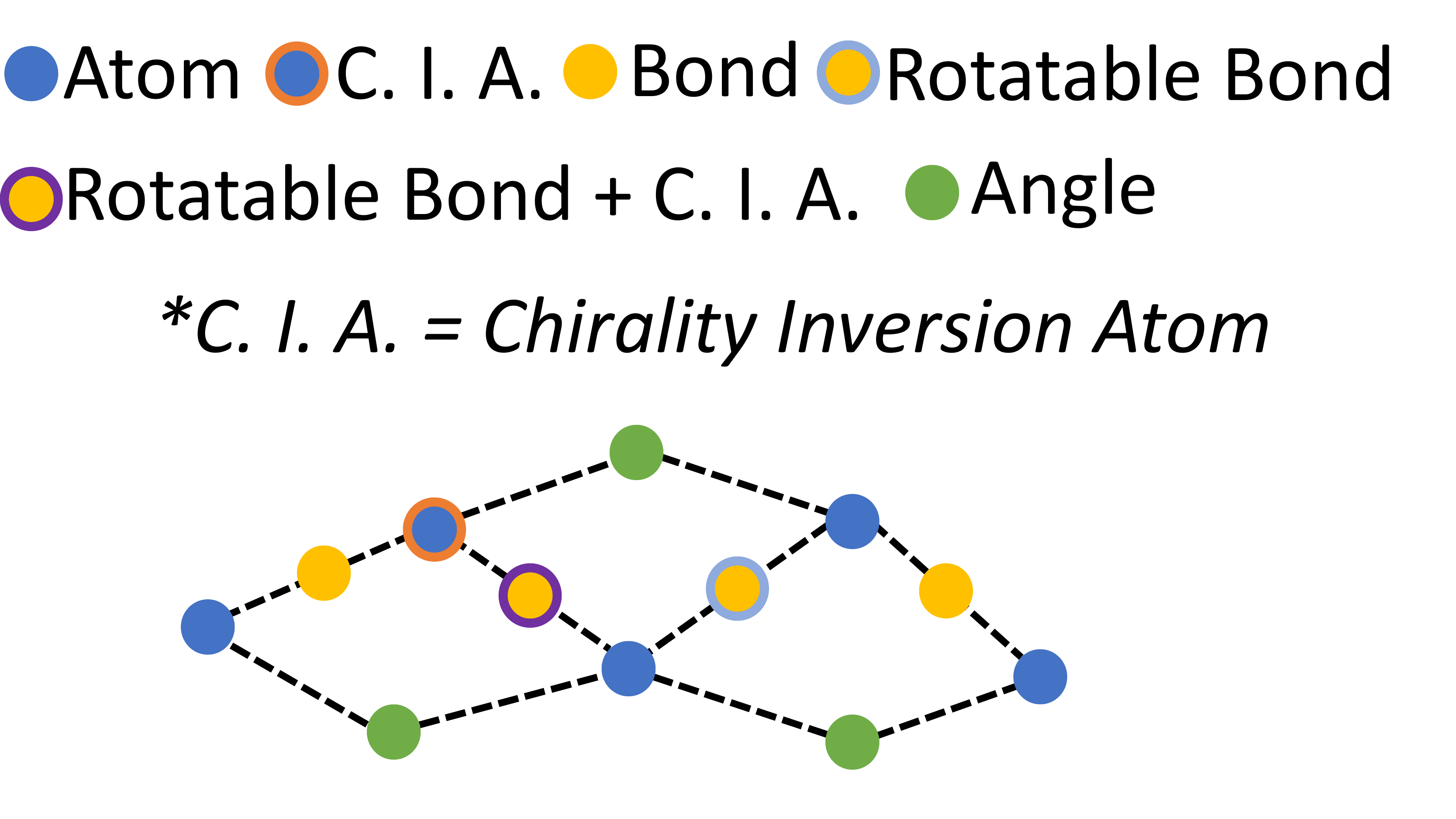}
}
\subfigure[]{
\includegraphics[scale = 0.1]{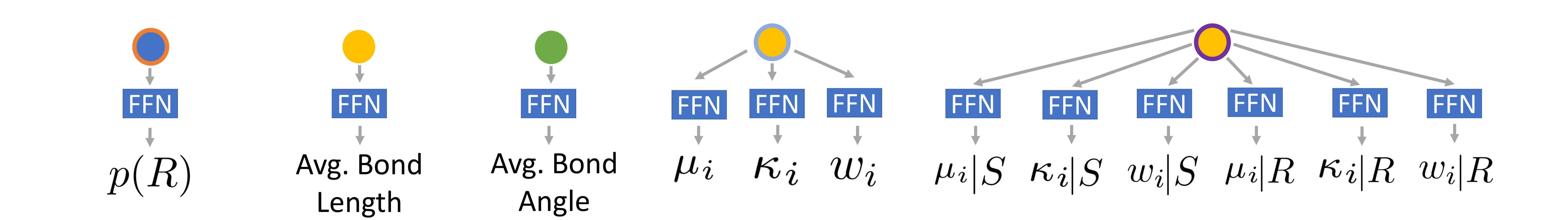}
}
\caption{\label{fig:figure4} \textit{VonMisesNet architecture.} \textbf{(a)} A fragment of a molecule represented in typical fashion with atoms corresponding to nodes and bonds corresponding to edges. \textbf{(b)} We transform this graph into a multi-partite representation that has atom nodes, bond nodes, and angle nodes. The atoms remain fixed as nodes, shown in blue, but the bond edges become two edges connected to a bond node, shown
in yellow. In addition, we add angle nodes, shown in green, which are connected by edges to the endpoints of a set of three consecutive atom nodes that define a bond angle. We perform message passing on this input graph, which results in node-level vectors. For each node type, shown in \textbf{(c)}, we generate the relevant predictions using Feed Forward Networks (FFN), shown in \textbf{(d)}. $p(R)$ is the probability of R chirality, and $\mu_i$, $k_i$, and $w_i$ are the von Mises parameters.} 
\end{center}
\vskip -0.2in
\end{figure*}

VonMisesNet is a graph neural network that takes a molecular graph with an initial 3D structure as input and predicts the following: the probability of R chirality for each chirality inversion atom, average length for all bonds, average angle for all bond angles, and parameters for a mixture of four von Mises distributions for each rotatable bond (including two sets of parameters for any rotatable bond that is connected to a chirality inversion atom, conditioned on R and S). These predictions can be used to rapidly generate conformations. We use a multi-partite graph representation for molecules, which removes the need to update edge or global features, allowing a simple message passing scheme to be used for nodel-level predictions of all quantities of interest. In Figure \ref{fig:figure4}, we illustrate the multi-partite graph representation and summarize the neural network architecture that we use to produce node-level predictions (see Appendix \S \ref{app:appF} for details on our node-level featurization and Appendix \S \ref{app:appJ} for further details on the neural network architecture). During training, we minimize the negative log likelihood of the ground truth angle samples for a given rotatable bond under a mixture of four von Mises distributions defined by the predicted parameters. For all of the other predictions, we minimize the mean squared error. 



\subsubsection{Conformation Generation}\label{sec:sec235}
To generate conformations, we create an initial geometry from ETKDG and run inference with VonMisesNet on the molecular graph. Using RDKit, we set all of the non-ring bonds in the initial geometry, as well as the bond angles containing at least one rotatable bond, to the respective predicted averages from VonMisesNet. Then, we generate a specified number of conformations using the following steps. For each chirality inversion atom, we randomly set the atom to have R or S chirality based on the predicted probability (see Appendix \S \ref{app:appG} for details). For each rotatable bond, we sample a value from its predicted von Mises mixture distribution and set the rotatable bond torsion angle in the initial geometry to that value. For rotatable bonds that are connected to a chirality inversion atom, we use the von Mises mixture distribution that is conditioned on the chirality of that atom. When both endpoints of the bond are chirality inversion atoms, we condition on the chirality of the ``begin atom" in RDKit. At this point, a full conformation has been generated. This process is repeated to generate the specified number of conformations.

There is an optional filtering step in this process. After a conformation is generated, we check if there are any two atoms separated by more than five bonds that are closer than the average of their van der Waals radii. If so, we discard the conformation and re-generate, up to a maximum number of total retries. This is inspired by the ETKDG algorithm, which uses the sum of the van der Waals radii of two atoms more than five bond lengths apart as a constraint in the distance geometry process.\footnote{We use a looser constraint as we find that it increases the ratio of preserved conformations while also improving evaluation metrics.} These constraints are useful because it is unphysical for two atoms to overlap spatially. We find that this filtering approach is a crude but effective way of mitigating the fact that VonMisesNet does not explicitly model long-range interactions.

\section{Results}\label{sec:sec3}
In this section, we evaluate the speed and accuracy of VonMisesNet. In subsection \ref{sec:sec31} we discuss baselines for comparison, and in subsection \ref{sec:sec32} we compare speed via runtime per conformer. For the problem of generating low energy conformations, accuracy evaluations typically focus on a set of reference conformations and use RMSD to measure the fraction that are recovered by a given method. Because our goal is to sample from a distribution rather than to generate low energy conformations, we use accuracy metrics that compare distributions rather than individual conformations. Specifically, we compare the generated distributions of rotatable bond torsion angles (subsection \ref{sec:sec33}) and the distributions of pairwise atomic distances (subsection \ref{sec:sec34}) to the respective ground truth Boltzmann distributions. 

\subsection{Baselines}\label{sec:sec31}
We use three baselines: ETKDG, GeoMol, and Torsional Diffusion. When ETKDG generates conformations, it occasionally swaps the indices of otherwise indistinguishable atoms, which can affect the comparison metrics we use. We made a small modification to enforce consistent atom indexing, which we call ETKDG-Clean (see Appendix \S \ref{app:appI} for details). For Torsional Diffusion, we use the Boltzmann generator model that was trained at 300 Kelvin, using the default 20 de-noising steps, a target sample temperature of 293 Kelvin, and the GEOM-DRUGS featurization configuration. Although GeoMol is designed to generate low energy conformations, we use it as a baseline to examine whether such a model can be easily repurposed to generate Boltzmann samples. Because GeoMol was originally trained on DFT geometries, we re-trained it on our ground truth data using the default hyperparameters and GEOM-DRUGS featurization configuration so that it generates geometries that are closer to our MMFF-based ground truth. When evaluating on molecules from NMRShiftDB-test, we use the VonMisesNet and GeoMol models that were trained on NMRShiftDB-train, and when evaluating on GDB-17-test, we use the corresponding models that were trained on GDB-17-train. Below, VonMisesNet-Filtered means that we use the optional filtering step when generating conformations. We generate 560 conformations per molecule for each method. 

\subsection{Runtime}\label{sec:sec32}
Runtime is an important metric because conformation generation is often an intermediate step in high-throughput molecular property prediction tasks. In Table \ref{tab:table1}, we compare the average time it takes to generate a single conformation. VonMisesNet is the fastest method and Torsional Diffusion is the slowest. Compared to Torsional Diffusion, VonMisesNet is about 47 times faster on a GPU and about 220 times faster on a CPU. 

\begin{table}[t]
\caption{Runtime per conformer in milliseconds averaged across 58 molecules from NMRShiftDB-test, along with standard error measurements. We selected 100 random molecules, 58 of which satisfed the Torsional Diffusion rotatable bond constraints. 100 conformers were generated for each molecule on a 64-core machine that has a single NVIDIA GeForce RTX 2080 Ti GPU. We do not include the start up times for any of the methods (e.g., loading neural network weights).}
\label{tab:table1}
\vskip 0.15in
\begin{center}
\begin{small}
\begin{sc}
\begin{tabular}{lrrr}
\toprule
Method & CPU (ms) & GPU (ms) \\
\midrule
ETKDG-Clean & 12.3 $\pm$ \hspace{1.7mm}1.5 & NA\\
GeoMol & 3.6 $\pm$ \hspace{1.7mm}0.2 & NA\\
Torsional Diffusion & 682.7 $\pm$ 46.9 & 140.0 $\pm$ 5.3\\
VonMisesNet & \textbf{3.1 $\pm$ \hspace{1.7mm}0.2} & \textbf{3.0 $\pm$ 0.2}\\
VonMisesNet-Filtered & 4.7 $\pm$ \hspace{1.7mm}0.7 & 4.9 $\pm$ 0.8\\
\bottomrule
\end{tabular}
\end{sc}
\end{small}
\end{center}
\vskip -0.1in
\end{table}

\subsection{Rotatable Bond Torsion Angle Distributions}\label{sec:sec33}
\begin{figure}[!ht]
\begin{center}
\subfigure[]{
  \includegraphics[scale = 0.2]{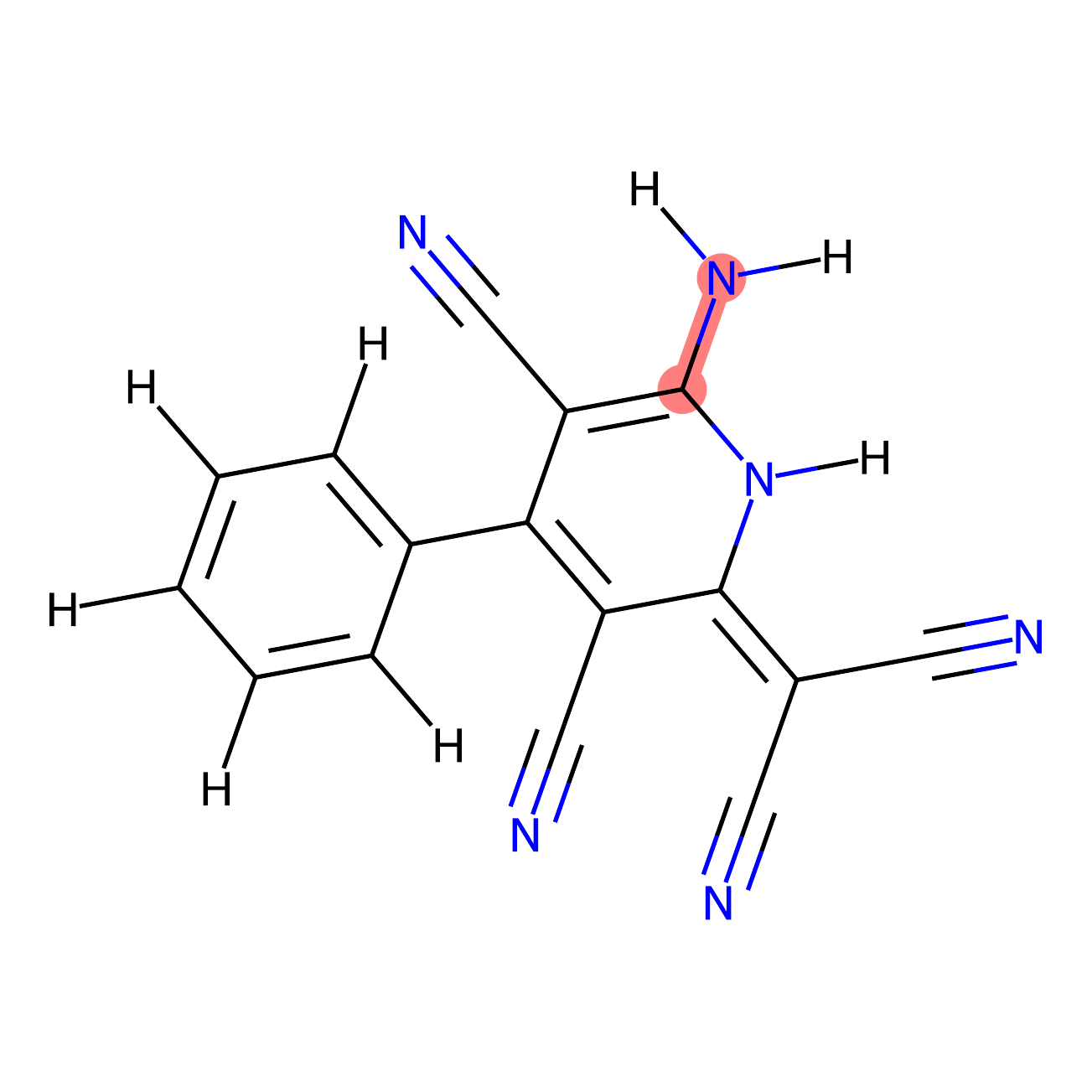}}%
  \hspace{4mm}
\subfigure[]{
  \includegraphics[scale = 0.45]{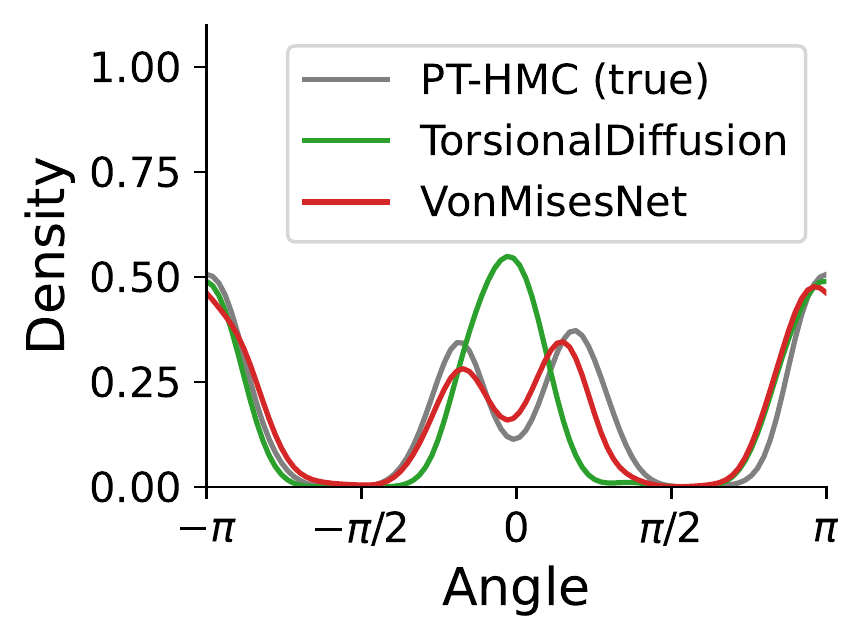}}\\%
\subfigure[]{
  \includegraphics[scale = 0.45]{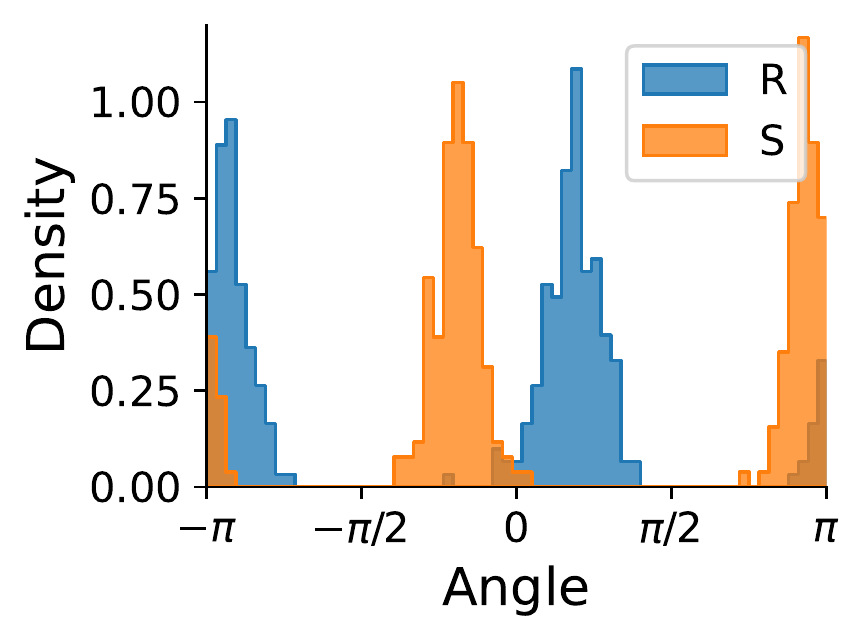}}%
\subfigure[]{
  \includegraphics[scale = 0.45]{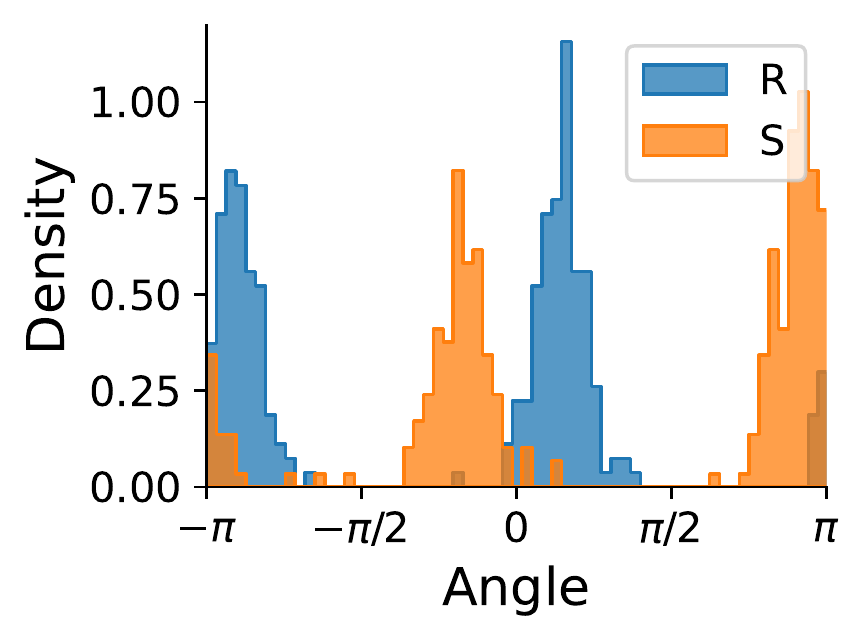}}%
  
\caption{\label{fig:figure6} \textit{Predicting chirality inversion example.} \textbf{(a)} A molecule from NMRShiftDB-test, where the highlighted rotatable bond is between a carbon atom and a chirality inversion nitrogen atom. The nitrogen atom has R chirality in 54.2\% of the PT-HMC conformations, 47.8\% of the VonMisesNet conformations, and 100\% of the Torsional Diffusion conformations. \textbf{(b)} Kernel density estimates of the rotatable bond torsion angle distribution from PT-HMC, VonMisesNet, and Torsional Diffusion. \textbf{(c)} and \textbf{(d)} show the distributions of this angle conditioned on the chirality of the nitrogen atom for PT-HMC and VonMisesNet, respectively. See Figure \ref{fig:figure22} in the Appendix for a plot that includes ETKDG-Clean and GeoMol as well.}
\end{center}
\vskip -0.2in
\end{figure}

\begin{figure*}[!ht]
\begin{center}
\includegraphics[width=0.88\textwidth]{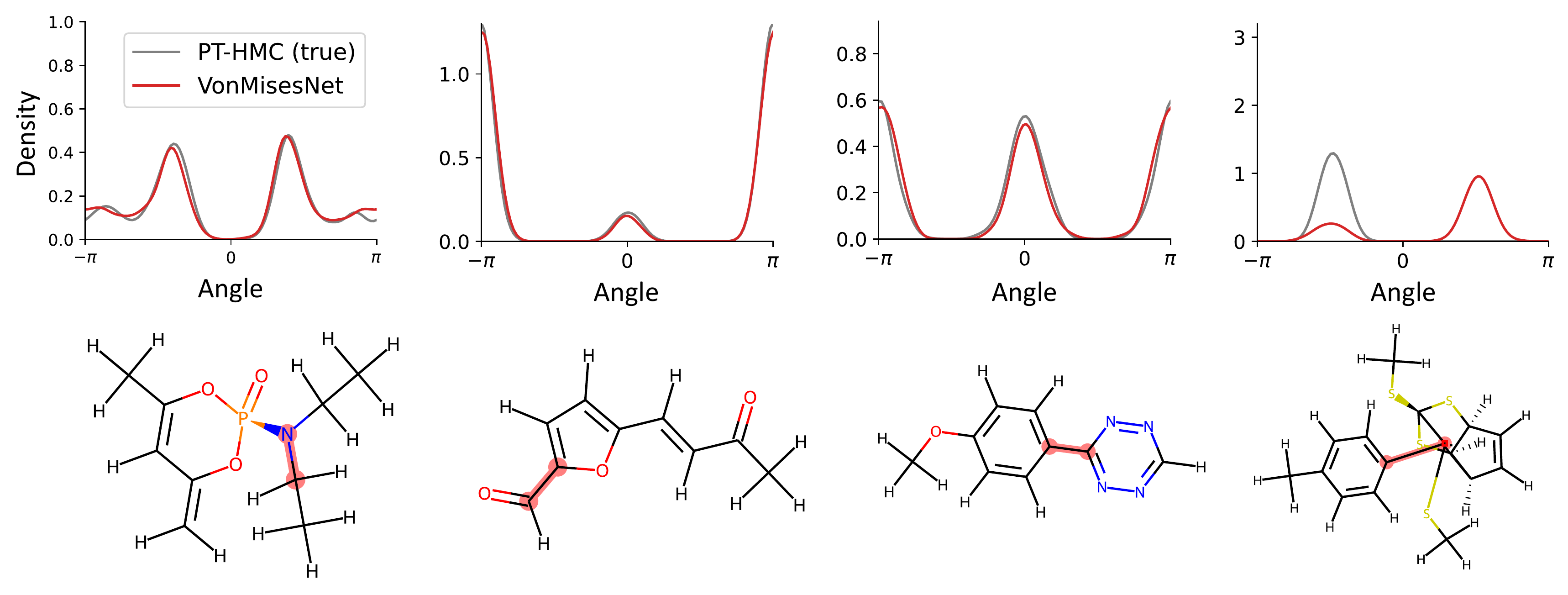}
\vspace{-0.1in}

\caption{\label{fig:figure7} \textit{VonMisesNet prediction examples.} We compare kernel density estimates of the torsion angle distributions of several rotatable bonds from PT-HMC and VonMisesNet. These molecules are taken from NMRShiftDB-test. The KL divergence of the VonMisesNet distributions relative to PT-HMC are, from left to right, 0.11, 0.024, 0.25, and 3.19. VonMisesNet performs well in the first three and poorly in the fourth. In the fourth example, the PT-HMC distribution is an odd function despite the presence of a methylbenzene group, which is symmetric under flipping of the aromatic ring. We find that the energy barrier is approximately 800 kcal/mol based on the MMFF implementation in RDKit, suggesting that such 180 flips may never occur in nature.}
\end{center}
\vskip -0.1in
\end{figure*}

To evaluate the accuracy of generated rotatable bond torsion angle distributions, we measure the KL divergence and Earth Mover's Distance (EMD) relative to the PT-HMC ground truth. These metrics show how well a model is capturing the distributions for individual rotatable bonds. Among 1000 random molecules from NMRShiftDB-test, we select 538 that satisfy the Torsional Diffusion rotatable bond constraints, and among 1000 random molecules from GDB-17-test, we select 610 that satisfy the constraints.\footnote{We do not use the entire test sets due to the large amount of time it would take to run inference for all methods.} As shown in Table \ref{fig:figure5}, VonMisesNet and VonMisesNet-Filtered outperform the other methods and have the lowest average KL and EMD values. In Figure \ref{fig:figure6}, we examine a specific example where chirality inversion has a nontrivial effect on the distribution, which is captured by VonMisesNet. In Figure \ref{fig:figure7}, we compare VonMisesNet and PT-HMC across several example rotatable bonds. Additional evaluations on the full sets of 1000 random molecules, without the Torsional Diffusion constraints, are shown in Appendix \S \ref{app:appH}.

\begin{table}[!ht]
\scriptsize
\centering
\caption{\label{fig:figure5}\textit{Rotatable bond torsion angle distributions evaluation.} Average KL divergence and EMD of rotatable bond torsion angle distributions, per molecule, relative to PT-HMC ground truth in 538 molecules from NMRSfhiftDB-test and 610 molecules from GDB-17-test, respectively. Standard error is in parentheses. The KL is measured with 32 bins. EC, GM, TD, VMN, and VMN-F stand for ETKDG-Clean, GeoMol, Torsional Diffusion, VonMisesNet, and VonMisesNet-Filtered, respectively. NMR and GDB stand for NMRShiftDB-test, and GDB-17-test, respectively.}
\begin{center}
\begin{tabular}{lccccc}
\toprule
& \multicolumn{5}{c}{KL Divergence}\\
& EC & GM & TD & VMN & VMN-F\\
\cmidrule(lr){2-6}
NMR & 2.80 (0.09) &  2.62 (0.08) &  1.14 (0.05) &  0.82 (0.04) &  \textbf{0.76 (0.04)} \\
GDB & 3.97 (0.09) &  2.70 (0.07) &  1.71 (0.04) &  1.36 (0.04) &  \textbf{1.31 (0.04)} \\
\midrule
& \multicolumn{5}{c}{Earth Mover's Distance (EMD)}\\
& EC & GM & TD & VMN & VMN-F\\
\cmidrule(lr){2-6}
NMR & 0.75 (0.03) &  0.92 (0.04) &  0.76 (0.03) &  0.47 (0.03) &  \textbf{0.46 (0.02)} \\
GDB & 1.13 (0.03) &  1.69 (0.07) &  1.02 (0.03) &  \textbf{0.93 (0.03)} &  \textbf{0.93 (0.03)} \\
\end{tabular}
\end{center}
\end{table}


\subsection{Pairwise Distance Distributions}\label{sec:sec34}
\begin{figure*}[!ht]
\begin{center}
\hspace{5mm}
\subfigure{
  \includegraphics[scale = 0.09]{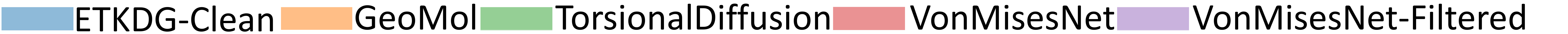}}%

\subfigure{
  \includegraphics[scale = 0.08]{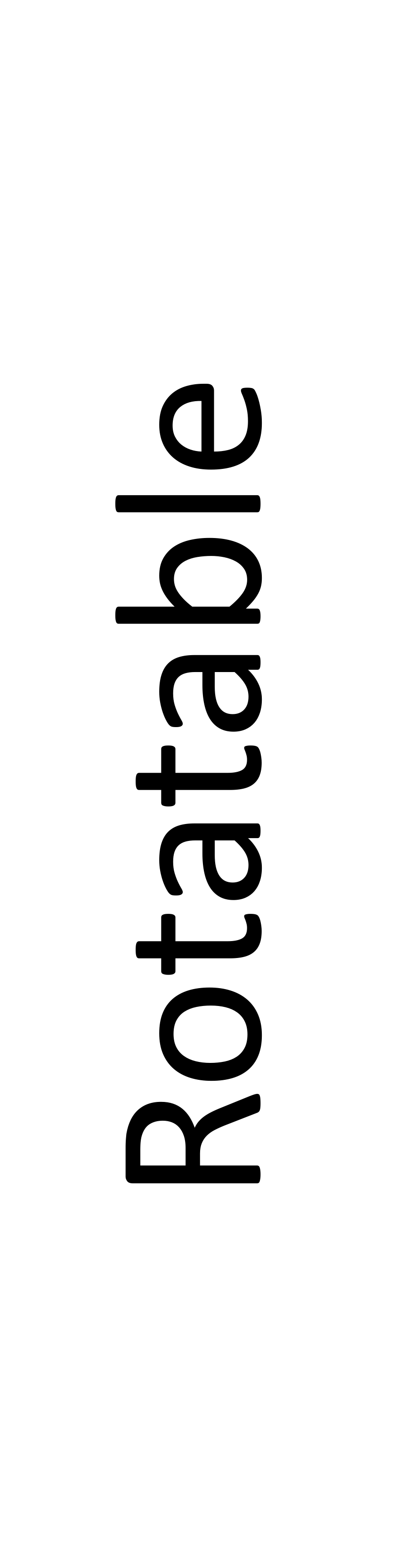}}%
\subfigure{
  \includegraphics[scale = 0.47]{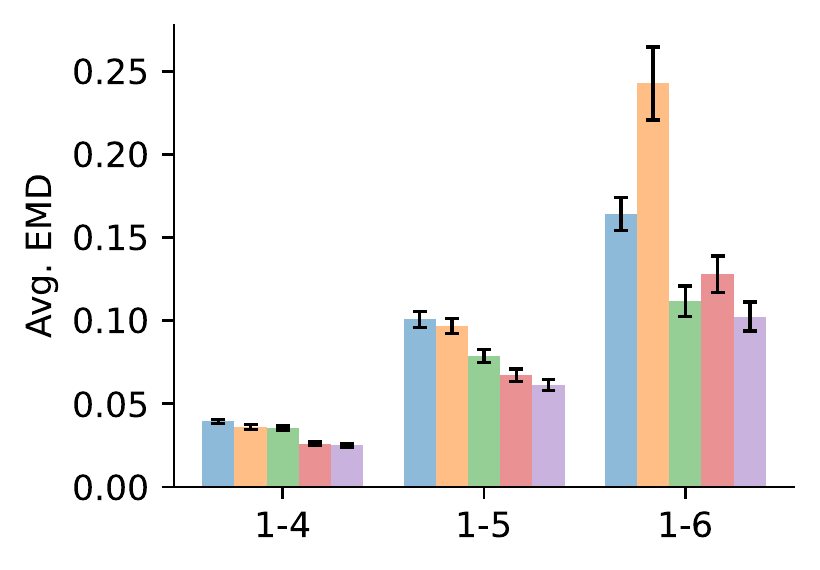}}%
\subfigure{
  \includegraphics[scale = 0.47]{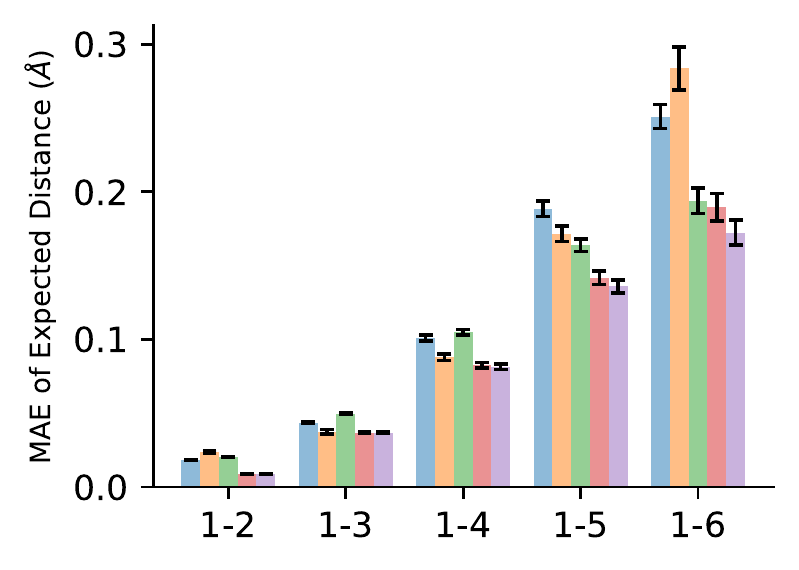}}%
\subfigure{
  \includegraphics[scale = 0.47]{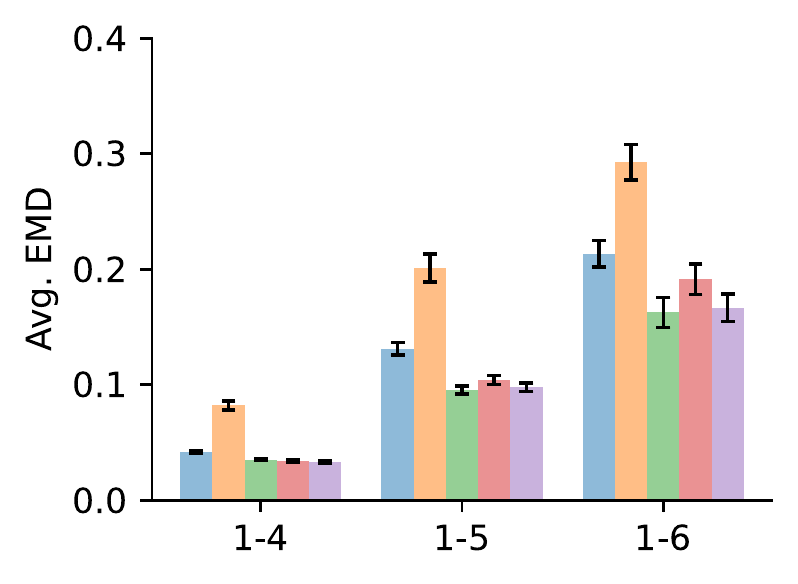}}
\subfigure{
  \includegraphics[scale = 0.47]{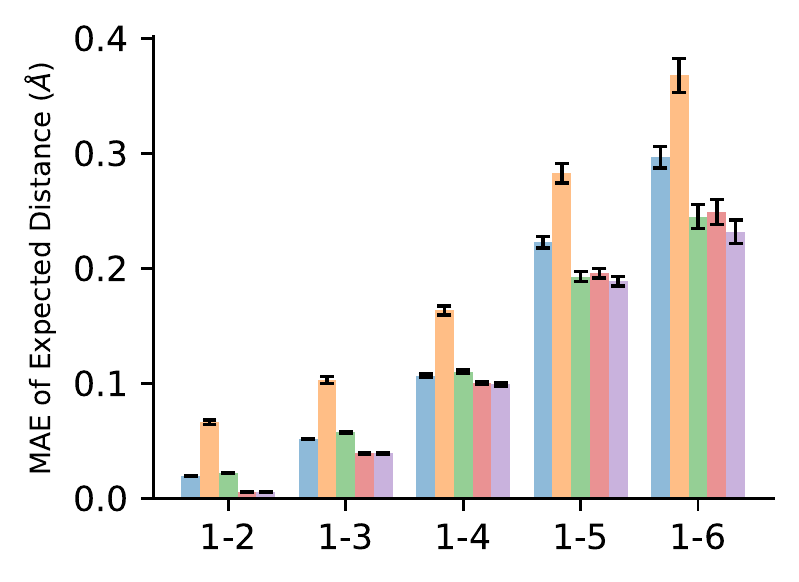}}%

\subfigure{
  \includegraphics[scale = 0.08]{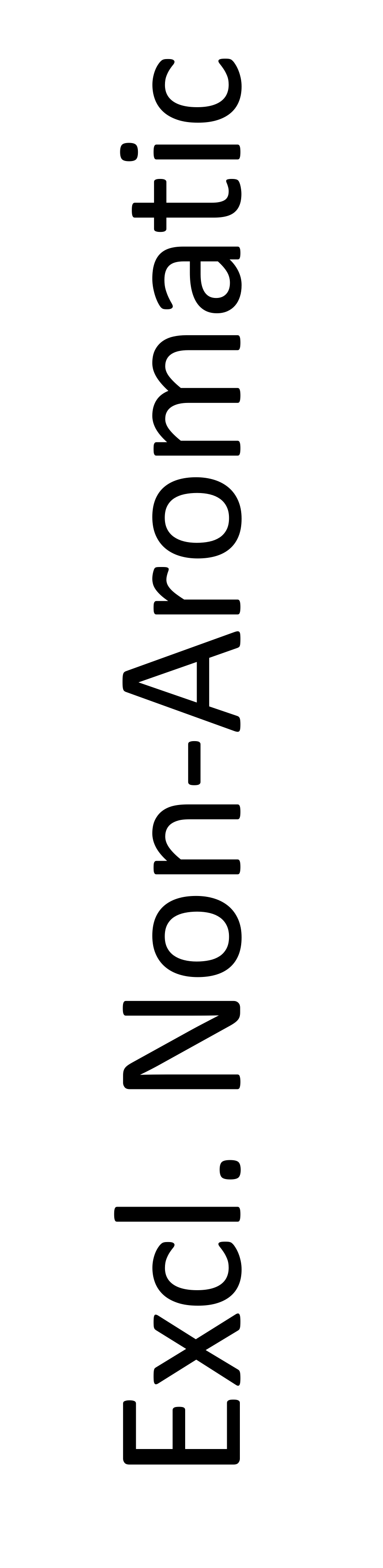}}%
  \subfigure{
  \includegraphics[scale = 0.47]{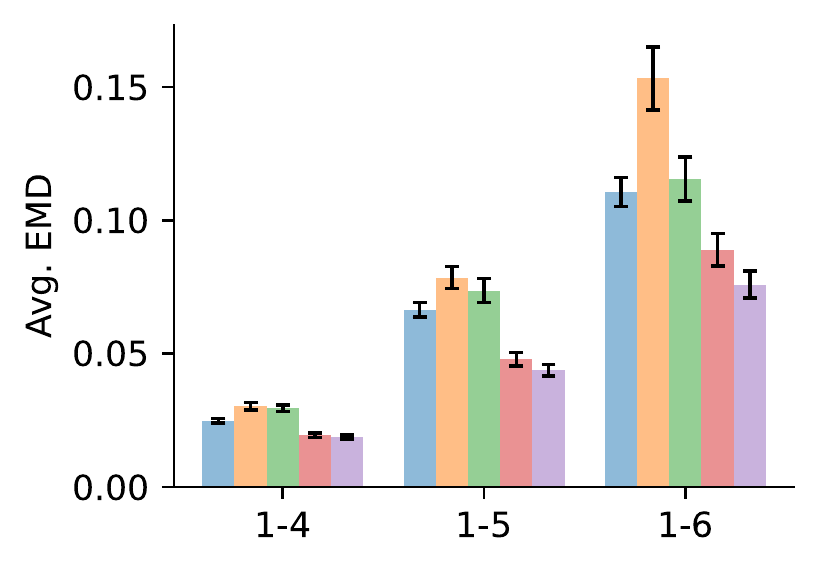}}%
\subfigure{
  \includegraphics[scale = 0.47]{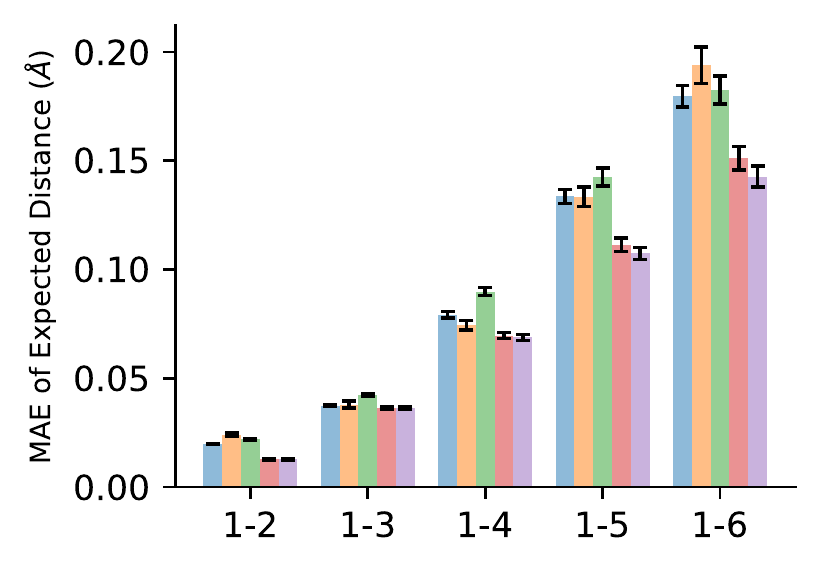}}%
\subfigure{
  \includegraphics[scale = 0.47]{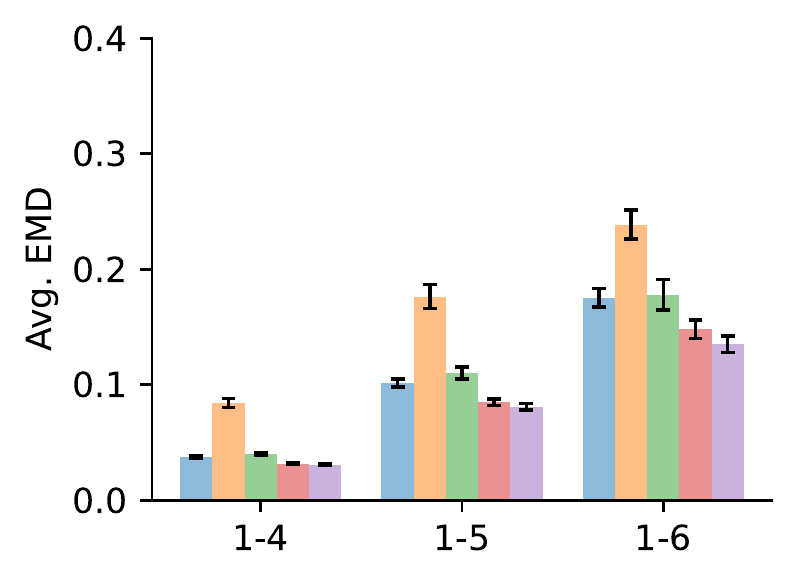}}
\subfigure{
  \includegraphics[scale = 0.47]{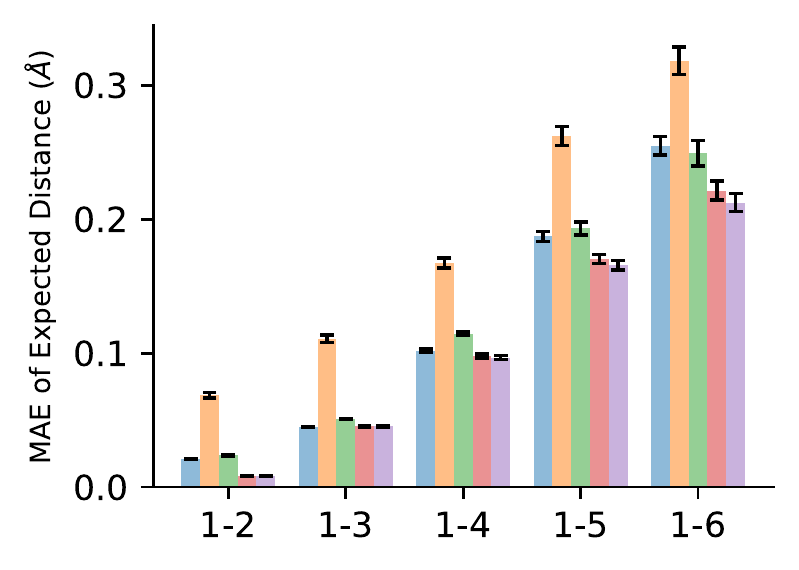}}%

\subfigure{
  \includegraphics[scale = 0.08]{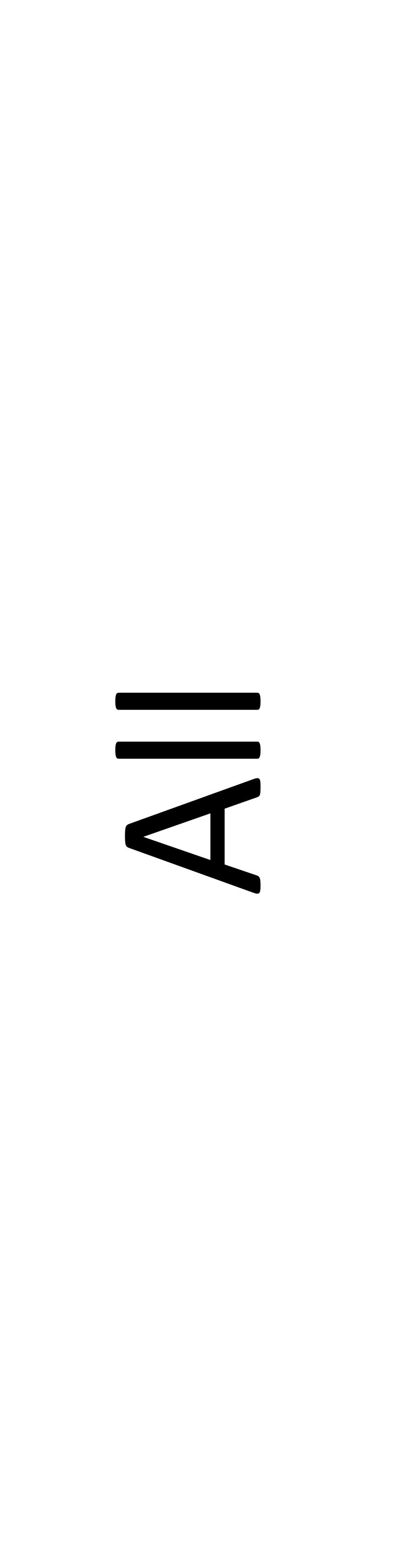}}%
  \subfigure{
  \includegraphics[scale = 0.47]{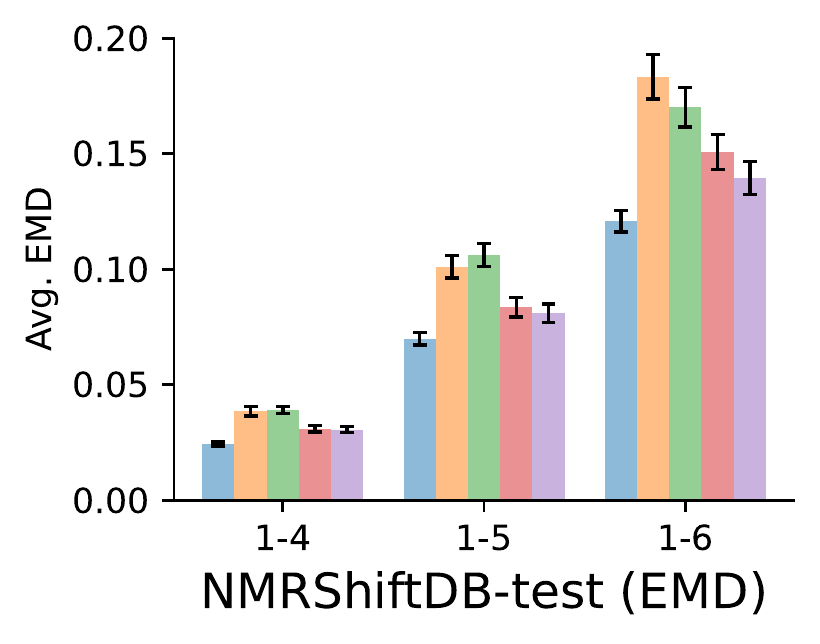}}%
\subfigure{
  \includegraphics[scale = 0.47]{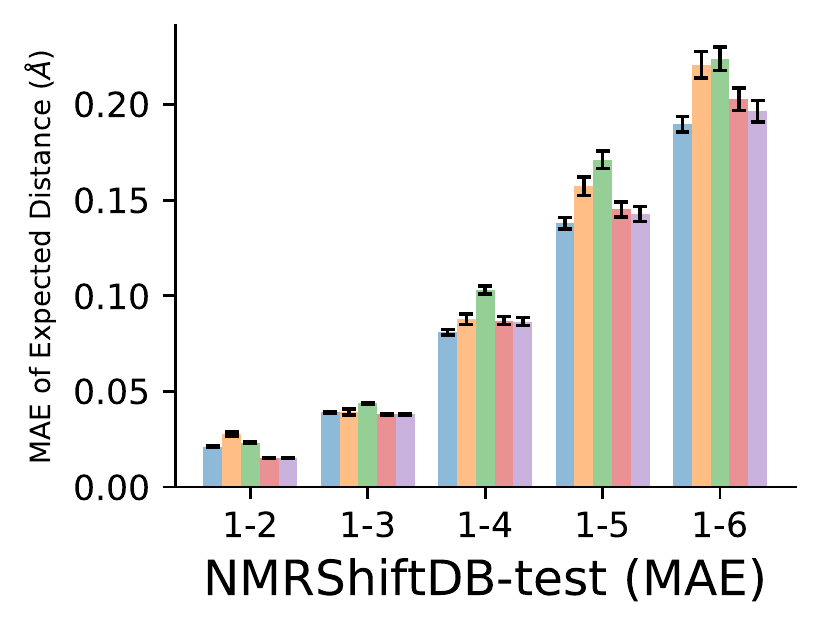}}%
\subfigure{
  \includegraphics[scale = 0.47]{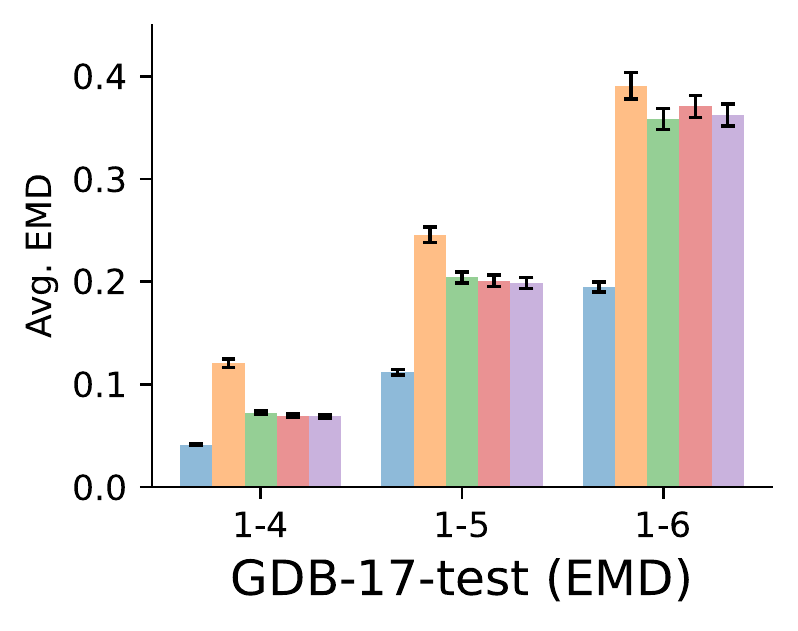}}
\subfigure{
  \includegraphics[scale = 0.47]{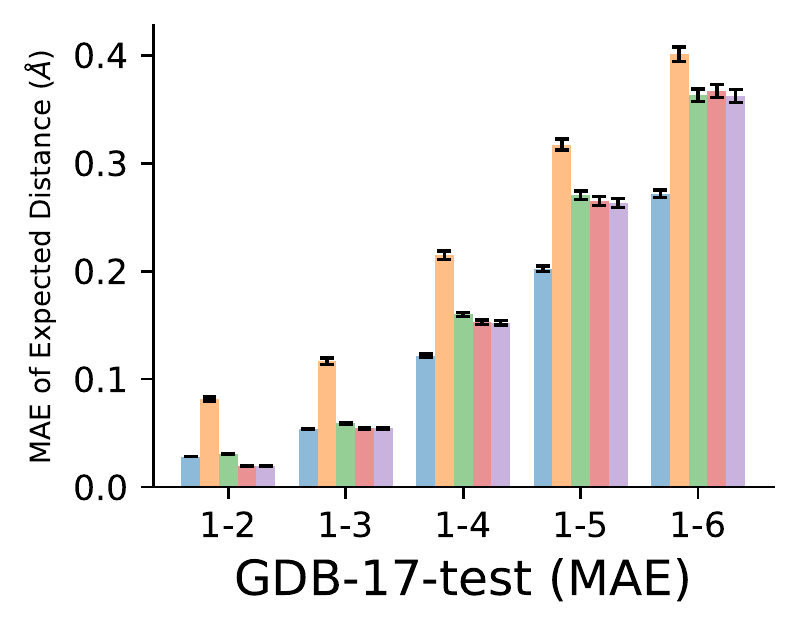}}%
\caption{\label{fig:figure8}\textit{Pairwise distance distributions evaluation.} We evaluate pairwise distance distributions relative to PT-HMC ground truth for 538 molecules from NMRShiftDB-test and 610 molecules from GDB-17-test. We evaluate 1-4, 1-5, and 1-6 distances for which every intermediate bond along the shortest path is rotatable in row 1, for which every intermediate bond along the shortest path is not part of a non-aromatic ring in row 2, and with no restrictions in row 3.  In columns 1 and 2 we compare the average EMD, per molecule, and the MAE of expected distance, per molecule, respectively, for the molecules from NMRShiftDB-test. Columns 3 and 4 show the same for molecules from GDB-17-test. For the expected distance evaluations, in row 1 we additionally include 1-2 distances that are not part of a ring and 1-3 distances for which at least one of the bonds is rotatable. In row 2, we include 1-2 distances that are not part of a non-aromatic ring and 1-3 distances for which neither of the bonds is in a non-aromatic ring. In row 3, there are no 1-2 or 1-3 restrictions. We exclude 1-2 and 1-3 distances from the EMD evaluations as these distributions are typically unimodal with small variance.}
\end{center}
\vskip -0.2in
\end{figure*}

To evaluate the accuracy of pairwise distance distributions, we measure the EMD of these distributions as well as the mean absolute error (MAE) of the expected distance relative to the PT-HMC ground truth.\footnote{We do not use KL because there are no consistent lower and upper bounds on the distances, and therefore we cannot easily assign bins for KL.}  When the shortest path between two atoms contains $n$ atoms, we say that the distance is a 1-$n$ distance. Comparing these 1-$n$ distance distributions gives us a way to evaluate the accuracy of generated geometries as a function of the graph-distance between pairs of atoms. We use the same sets of molecules as in \S \ref{sec:sec33}. In the first row in Figure \ref{fig:figure8}, we evaluate 1-$n$ distances for which every intermediate bond along the shortest path is rotatable, which lets us focus on how well models are capturing the marginal Boltzmann distribution of rotatable bond torsion angles.\footnote{By intermediate bonds, we mean all bonds along the shortest path except for the first and the last.} VonMisesNet, VonMisesNet-Filtered, and Torsional Diffusion outperform the other methods. VonMisesNet shows relatively stronger performance for $n < 6$ and weaker performance in some cases for $n = 6$, and  VonMisesNet-Filtered performs best overall. In the second row, we allow for any intermediate bond along the shortest path except for those that belong to a non-aromatic ring. There is a regression in the relative performance of GeoMol and Torsional Diffusion, and VonMisesNet and VonMisesNet-Filtered outperform all other methods on nearly all metrics. In the third row, we remove the non-aromatic restriction and consider all bonds along the shortest path, and ETKDG-Clean performs better overall. Future work will handle non-aromatic rings, which are an especially complex case for which, to the best of our knowledge, no full machine learning solution exists. Evaluations with $n > 6$ and without the Torsional Diffusion constraints are shown in Appendix \S \ref{app:appH}.


\section{Discussion}
In this work, we presented VonMisesNet, a graph neural network that models conformational variability with a variational approximation of rotatable bond torsion angles as a mixture of von Mises distributions. Conformations generated with VonMisesNet have more accurate rotatable bond torsion angle distributions with respect to the Boltzmann distribution than the other methods, and they also tend to have the most accurate 1-$n$ pairwise distance distributions, especially for $n < 7$ and when excluding non-aromatic rings. The performance of Torsional Diffusion tends to improve relative to VonMisesNet as a function of $n$, but VonMisesNet is orders of magnitude faster. To the best of our knowledge, it is also the only machine learning model that takes chirality inversion into account. 

There are several avenues for future work. First, we focused exclusively on rotatable bonds and have not yet modeled non-aromatic rings, which are often flexible and can contribute to conformational variability. Second, we showed proof of concept with MMFF-based energies, but training on ground truth data from quantum mechanical calculations would allow for generating conformations with more accurate geometries. Finally, explicitly modeling joint rotatable bond torsion angle distributions would likely yield more accurate long-range interactions. Torsional Diffusion achieves this with a diffusion model that operates on the hypertorus defined by the torsion angles. Some of the low energy conformation generation methods also model joint probabilities. For example, GeoMol jointly predicts all torsion angles with graph neural networks, and RMCF models the joint distribution of molecular fragments and dihedral angles via a Markov random field. However, these methods are not suitable for Boltzmann sampling. 

We believe that the variational inference framework introduced here could be extended to joint distributions, and therefore long-range interactions, while maintaining computational efficiency. For example, future ideas include modeling the joint distribution for all pairs of rotatable bond torsion angles, yielding a Markov random field which is still fast to sample from, in contrast to diffusion approaches that require multiple slow diffusion steps (multiple neural network forward passes) for each conformation.

\section*{Acknowledgements}
The authors thank all of the members of the Jonas Lab (University of Chicago), as well as Kyle Swanson and Melody Huang, for discussions and advice. This material is based upon work supported by the National Science Foundation under Award 2231634. Computing research and services were partly provided by the OSG Consortium which is supported by the National Science Foundation awards 2030508 and 1836650 and by Texas Advanced Computing Center (TACC) at the University of Texas, Austin via award CHE21008.

\bibliography{icml_2023_manuscript}
\bibliographystyle{icml2023}

\newpage
\appendix
\onecolumn

\section{Simulation Details for Ground Truth Conformation Generation}\label{app:appA}
To generate ground truth molecular conformations, we combine Parallel Tempering and Hamiltonian Monte Carlo (PT-HMC). In the context of molecular systems, Hamiltonian Monte Carlo is a Monte Carlo simulation where at each step, a short Molecular Dynamics simulation is used to propose a molecular geometry. Parallel Tempering refers to the technique of running multiple Monte Carlo simulations simultaneously at different temperatures and periodically proposing swaps between the states of adjacent temperatures. For the PT-HMC simulations, we used 70,000 total steps, sampling every 100 steps. We used eight different temperatures for parallel tempering: 293, 400, 500, 600, 700, 800, 900, and 1000 Kelvin. 293 Kelvin served as the target base temperature. In other words, the generated conformations are sampled from the Boltzmann distribution with $T = 293$ Kelvin. We used a Parallel Tempering temperature-swap probability of 0.2, and for the Hamiltonian Monte Carlo Molecular Dynamics trajectories, we used $L=100$ steps with a step size of $\epsilon=1.5$ femtoseconds. Empirically, we found that these parameters were sufficient to explore the potential energy landscape for a variety of molecules. Each molecule has approximately 560 conformations. There are minor differences in the total number of generated conformations for each molecule, because conformations are only proposed and then accepted with a certain probability in PT-HMC simulations. 

In the fourth example of Figure \ref{fig:figure7}, the left hand side of the molecule has a methylbenzene group, which is symmetric under flipping of the aromatic ring, but the ground truth distribution is an odd function. We carefully inspected our data and found that for nearly all molecules in our training data that contain relevant symmetry configurations, we properly recover even functions. We identified several with odd functions (12 in NMRShiftDB) and systematically investigated rotational barriers. We find that the energy barriers are so high (hundreds of kcal/mol) that it is possible that such 180 degree flips would never occur in nature. For this example, the energy barrier is approximately 800 kcal/mol based on the MMFF implementation in RDKit.

\section{Dataset Details}\label{app:appB}
For NMRShiftDB we took 32,171 molecules from NMRShiftDB \cite{kuhnjonas2019}, a popular database of nuclear magnetic resonance spectra, limiting it to molecules with up to 64 atoms of elements {H, C, O, N, F, S, P, Cl} and no formal charges or radicals. We ensured that every molecule had an assigned stereo chemistry.

For GDB-17 we took a random 134,228 molecule subset of the publically-available 50M ``lead-like" molecules made available by the GDB-17 enumeration of chemical space \cite{GDB17}. As GDB provides non-isomeric SMILES strings, we leveraged RDKit's stereo enumeration code and identified at least one stereoisomer with a reasonable force-field energy (that is, no pathological steric hindrance). 

\begin{figure}[!ht]
\vskip 0.2in
\begin{center}
\subfigure[][]{
    \includegraphics[scale=0.5]{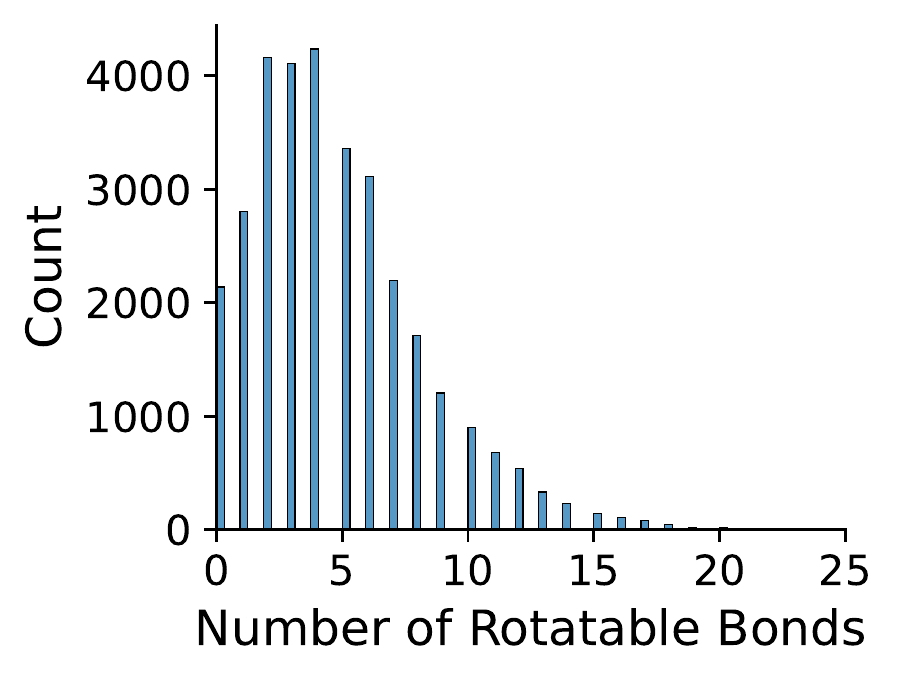}
}
\subfigure[][]{
    \includegraphics[scale=0.5]{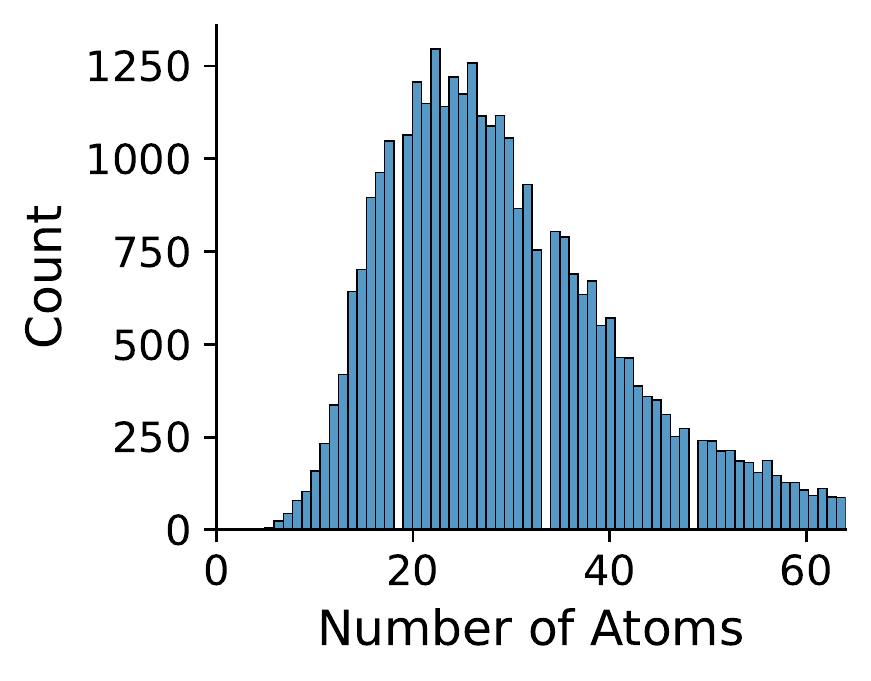}
}
\caption{\label{fig:figure9}\textit{NMRShiftDB dataset statistics.} \textbf{(a)} Distribution of number of rotatable bonds per molecule in our NMRShiftDB data. \textbf{(b)} Distribution of number of atoms per molecule in our NMRShiftDB data.}
\end{center}
\vskip -0.2in
\end{figure}

\begin{figure}[!ht]
\vskip 0.2in
\begin{center}
\subfigure[][]{
    \includegraphics[scale=0.5]{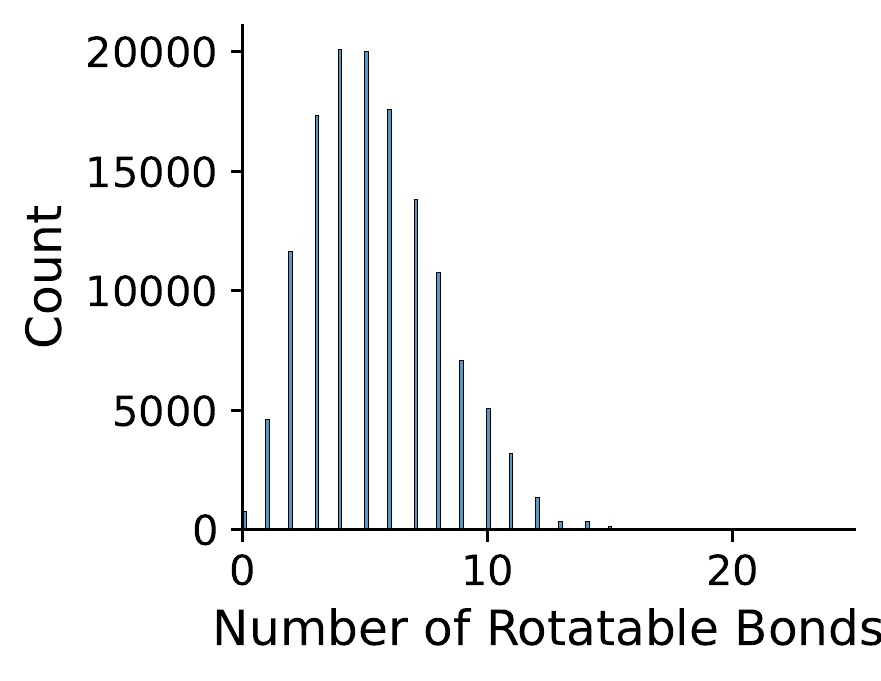}
}
\subfigure[][]{
    \includegraphics[scale=0.5]{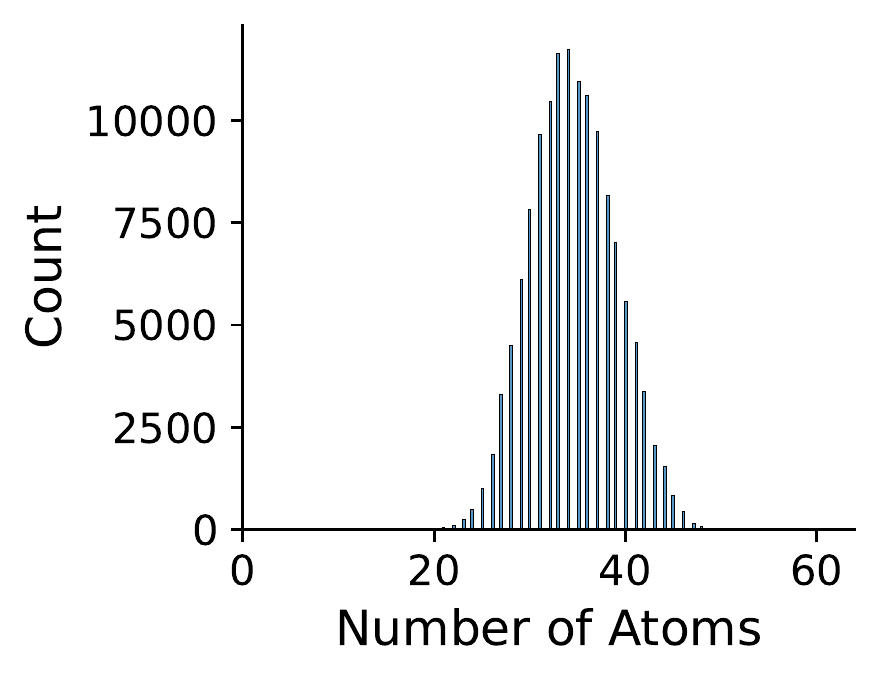}
}
\caption{\label{fig:figure10}\textit{GDB-17 dataset statistics.} \textbf{(a)} Distribution of number of rotatable bonds per molecule in our GDB-17 data. \textbf{(b)} Distribution of number of atoms per molecule in our GDB-17 data.}
\end{center}
\vskip -0.2in
\end{figure}

In Figures \ref{fig:figure9} and \ref{fig:figure10}, we examine the distribution of number of rotatable bonds and number of atoms in our NMRShiftDB and GDB-17 datasets.

\section{Dihedral Angle Encoding for Rotatable Bonds}\label{app:appC}
We use a breadth-first search approach based on the Cahn-Ingold-Prelog (CIP) \cite{CIP} rules to determine which four atoms, including the two central atoms, should be chosen to define a given rotatable bond torsion angle. For each of the two central atoms, we use the following procedure. We construct a list of neighbors excluding the other central atom. If, among this list, there is one with a unique largest atomic number, then we select that atom. If there is a tie, atoms at a distance of two bonds from the central atom are examined. We replace each atom in the original list with a list of the other atoms bonded to it that have not yet been examined. These lists are compared atom by atom, and at the earliest difference, the group with the highest atomic number is given higher priority. This process is repeated recursively until any ties are broken. In the case where there is a tie at distance $n$ but there are no bonded atoms at distance $n+1$, the first atom or list of atoms is given priority.

\section{Approximate Independence of Rotatable Bond Torsion Angle Distributions}\label{app:appD}
\begin{figure}[!ht]
\vskip 0.2in
\begin{center}
\subfigure[][]{
  \includegraphics[scale = 0.4]{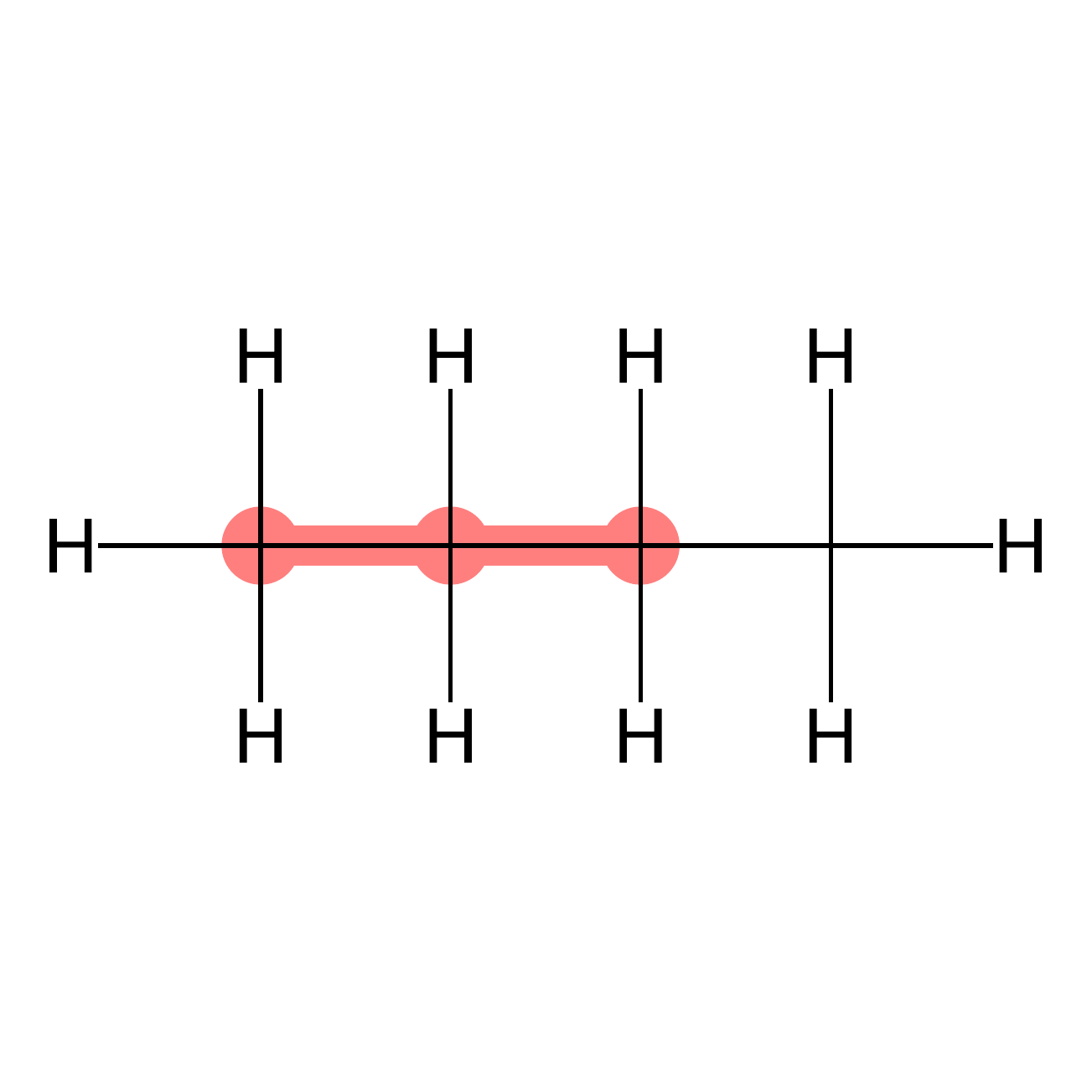}}\\
\subfigure[][]{
  \includegraphics[scale = 0.5]{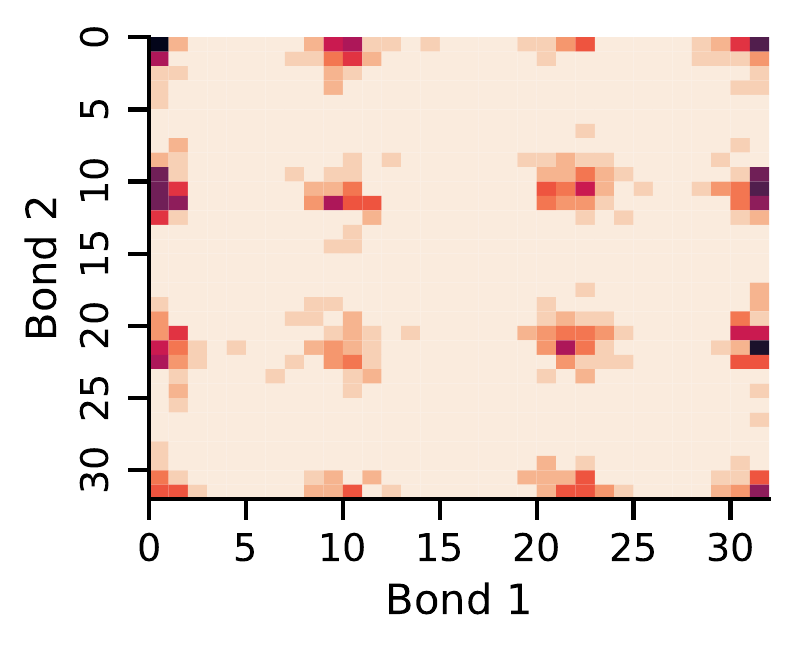}}
\subfigure[][]{
  \includegraphics[scale = 0.5]{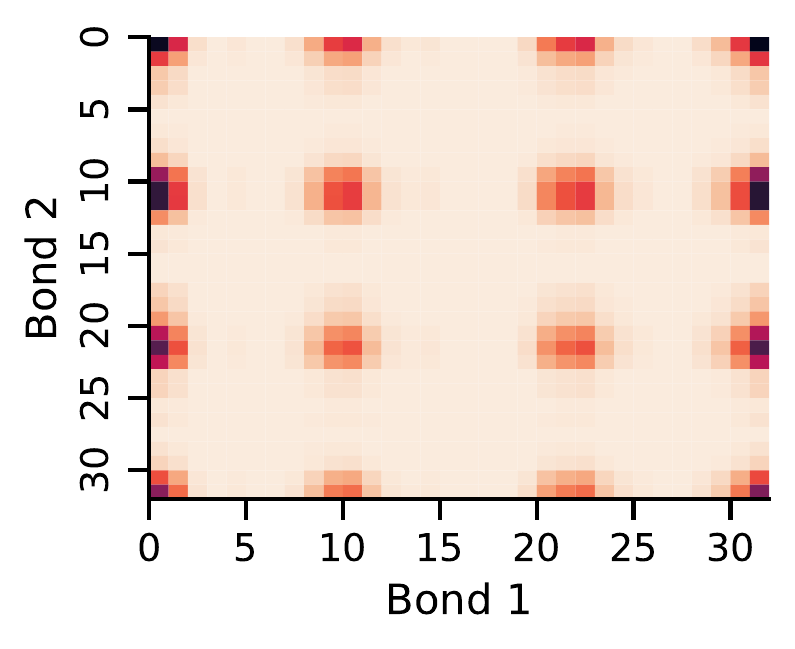}}\\
\subfigure[][]{
  \includegraphics[scale = 0.5]{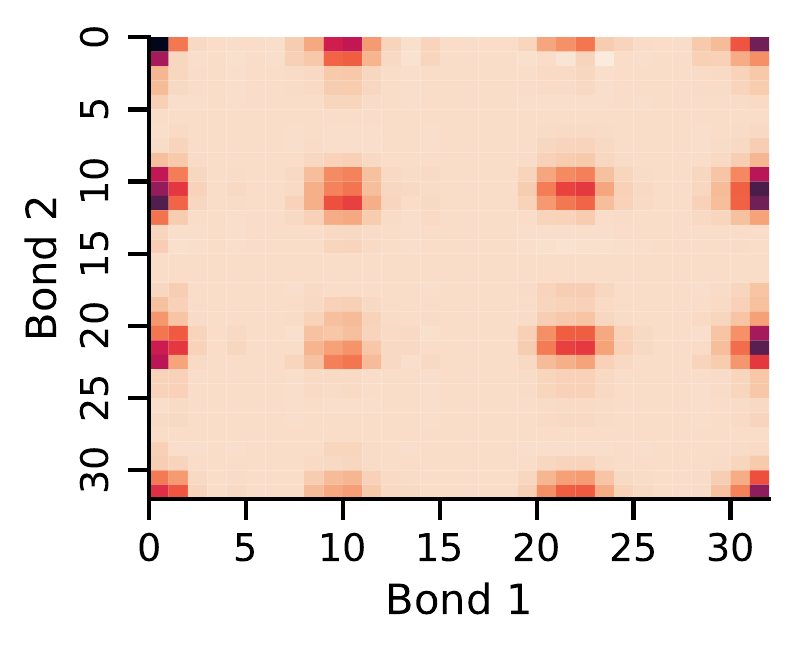}}
\subfigure[][]{
  \includegraphics[scale = 0.5]{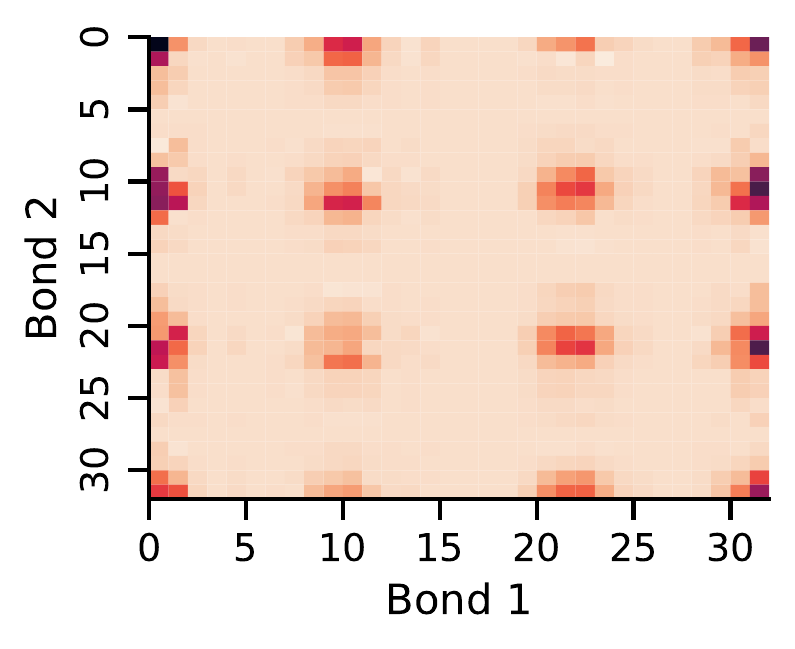}}
\caption{\label{fig:figure11}\textit{Rank-1 example.} \textbf{(a)} Molecular graph of butane along with two adjacent rotatable bonds highlighted. \textbf{(b)} A histogram of the torsion angle joint distribution of the highlighted rotatable bonds, using 32 bins. \textbf{(c)} shows a rank-1 approximation to this histogram with reconstruction mean squared error (MSE) 0.47, \textbf{(d)} shows a rank-2 approximation with MSE 0.31, and \textbf{(e)} shows a rank-3 approximation with MSE 0.21. Although the rank-1 MSE is higher, it’s clear from the visualization that the ground truth distribution is approximately rank-1, indicating independence.}
\end{center}
\vskip -0.2in
\end{figure}

\begin{figure}[!ht]
\vskip 0.2in
\begin{center}
\subfigure[][]{
  \includegraphics[scale = 0.4]{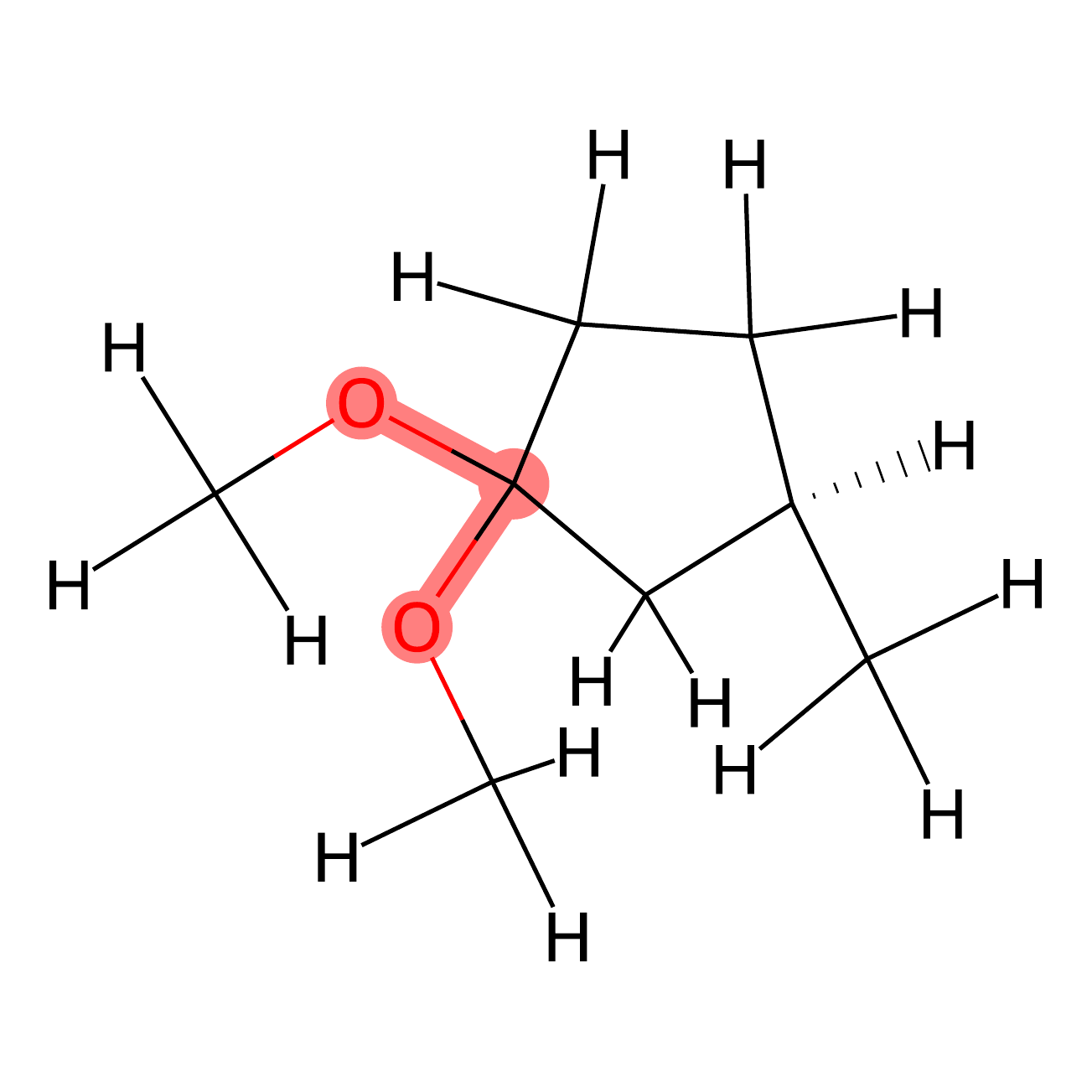}}\\
\subfigure[][]{
  \includegraphics[scale = 0.5]{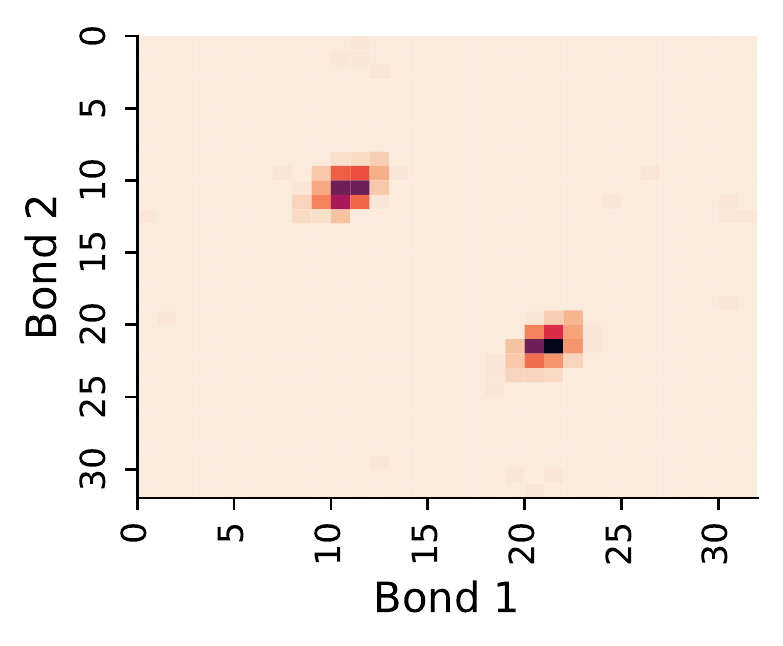}}
\subfigure[][]{
  \includegraphics[scale = 0.5]{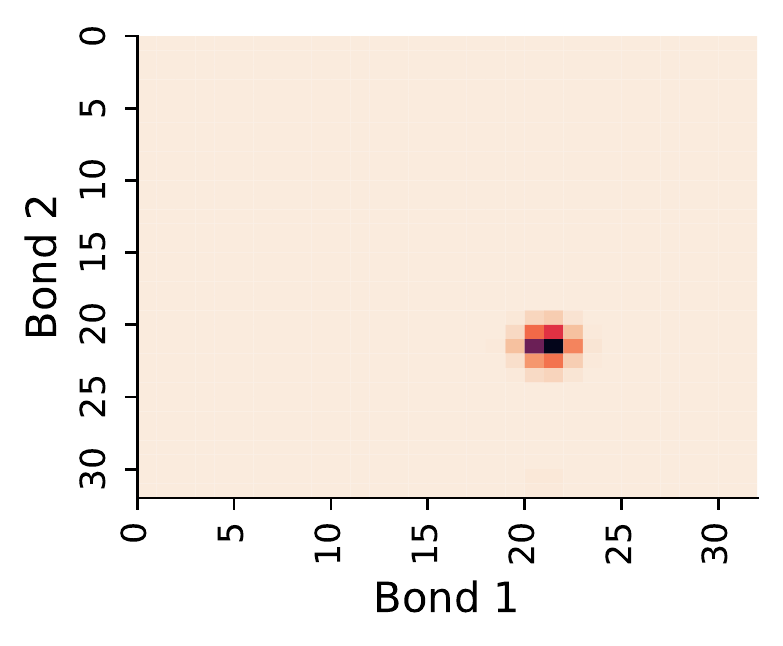}}\\
\subfigure[][]{
  \includegraphics[scale = 0.5]{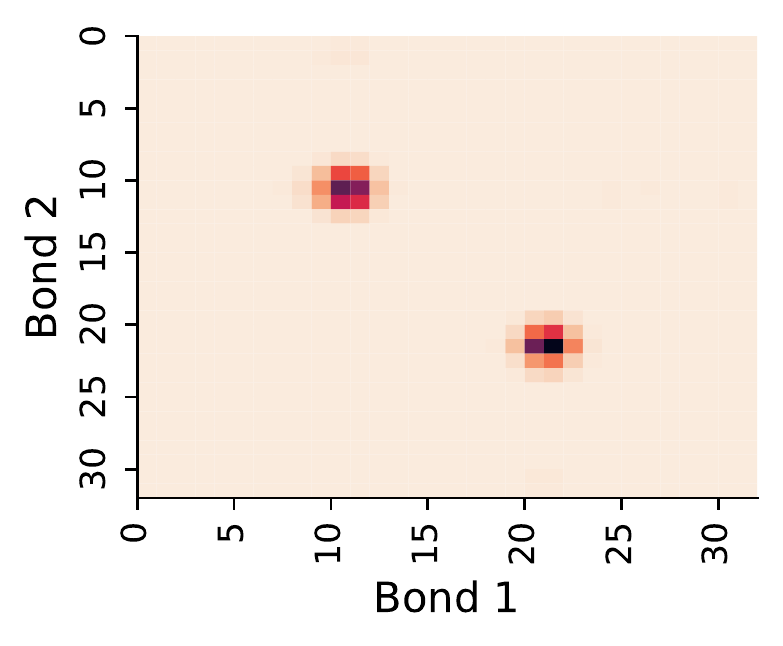}}
\subfigure[][]{
  \includegraphics[scale = 0.5]{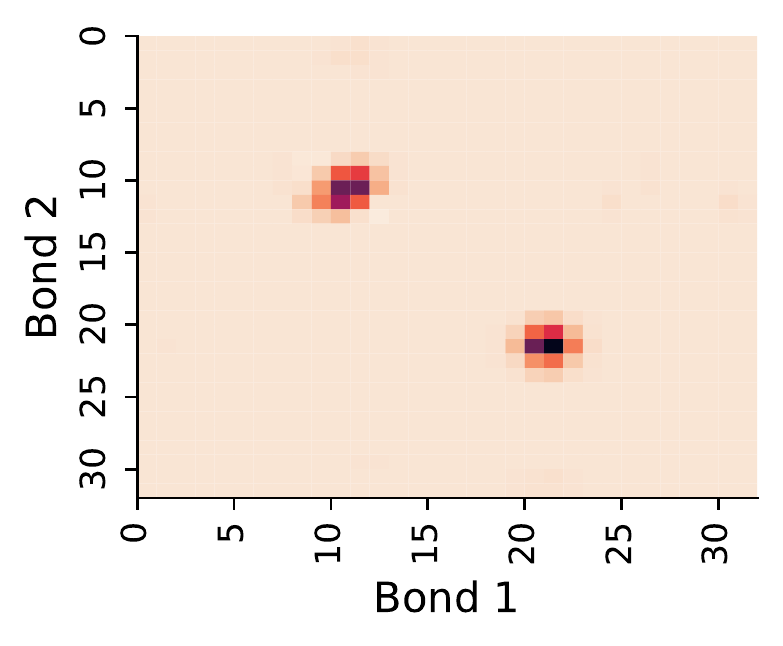}}
\caption{\textit{Non-rank-1 example.}\label{fig:figure12} \textbf{(a)} Molecular graph of a molecule from NMRShiftDB along with two adjacent rotatable bonds highlighted. \textbf{(b)} A histogram of the torsion angle joint distribution of the highlighted rotatable bonds, using 32 bins. \textbf{(c)} shows a rank-1 approximation to this histogram with reconstruction mean squared error (MSE) 8.01, \textbf{(d)} shows a rank-2 approximation with MSE 0.53, and \textbf{(e)} shows a rank-3 approximation with MSE 0.28. This is a clear example where the joint distribution is not approximately rank-1.}
\end{center}
\vskip -0.2in
\end{figure}

\begin{figure}[!ht]
\vskip 0.2in
\begin{center}
\subfigure[][]{
  \includegraphics[scale = 0.8]{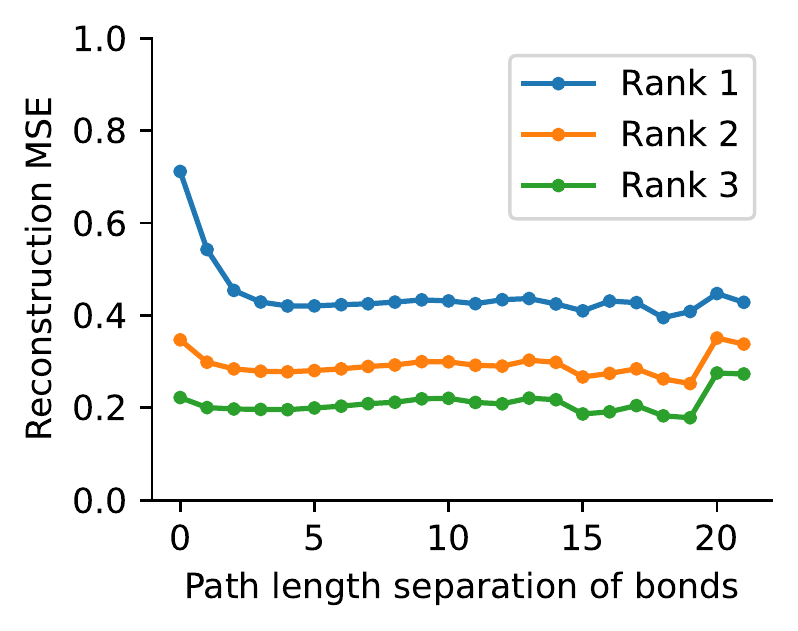}}
\subfigure[][]{
\includegraphics[scale = 0.8]{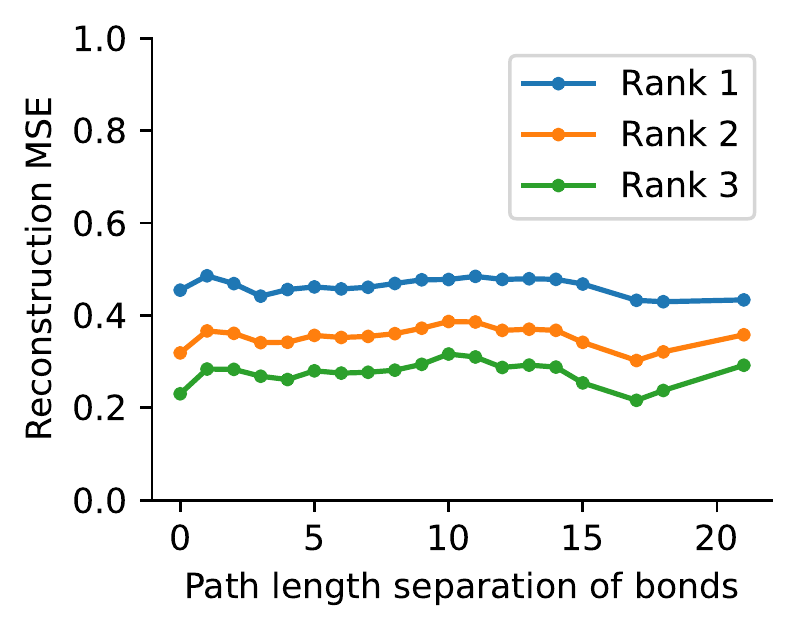}}
\caption{\textit{Large-scale rank-1 analysis.}\label{fig:figure13} For each pair of rotatable bonds in each molecule of a random subset of 4,225 molecules from our NMRShiftDB data, we computed a 2D histogram of the torsion angle joint distribution using 32 bins. We then computed the reconstruction mean squared error (MSE) of the rank-1, rank-2, and rank-3 approximations to this histogram. We show these averages as a function of the graph distance between rotatable bond pairs. \textbf{(a)} shows that the rank-1 value rapidly converges to approximately 0.4, which is below the rank-1 value from butane in Figure \ref{fig:figure11}. The average MSE value across all pairs is 0.51. This suggests that pairs of rotatable bonds have a joint distribution that is approximately rank-1, with the approximation improving when the bonds are farther apart within a molecule. \textbf{(b)} shows a similar plot that is restricted to rotatable bonds attached to methyl groups, which we would expect to be pairwise independent. The rank-1 value indicates approximate independence across all path lengths.}
\end{center}
\vskip -0.2in
\end{figure}

In this work, we make the assumption that individual rotatable bond torsion angle distributions are approximately independent. Butane is an example of a molecule that supports this assumption. Figure \ref{fig:figure11} shows an example pair of rotatable bonds in butane whose joint torsion angle distribution is approximately rank-1, which indicates independence. There are, of course, cases where pairs of rotatable bonds are not approximately rank-1, as exhibited in Figure \ref{fig:figure12}. However, we show in Figure \ref{fig:figure13} that on average, across a large set of molecules, pairs of rotatable bonds have a joint distribution that is approximately rank-1, with the approximation improving when the bonds are farther apart within a molecule. 

\section{Mode Analysis for Rotatable Bond Torsion Angle Distributions}\label{app:appE}
\begin{figure}[!ht]
\vskip 0.2in
\begin{center}
\includegraphics[scale = 0.6]{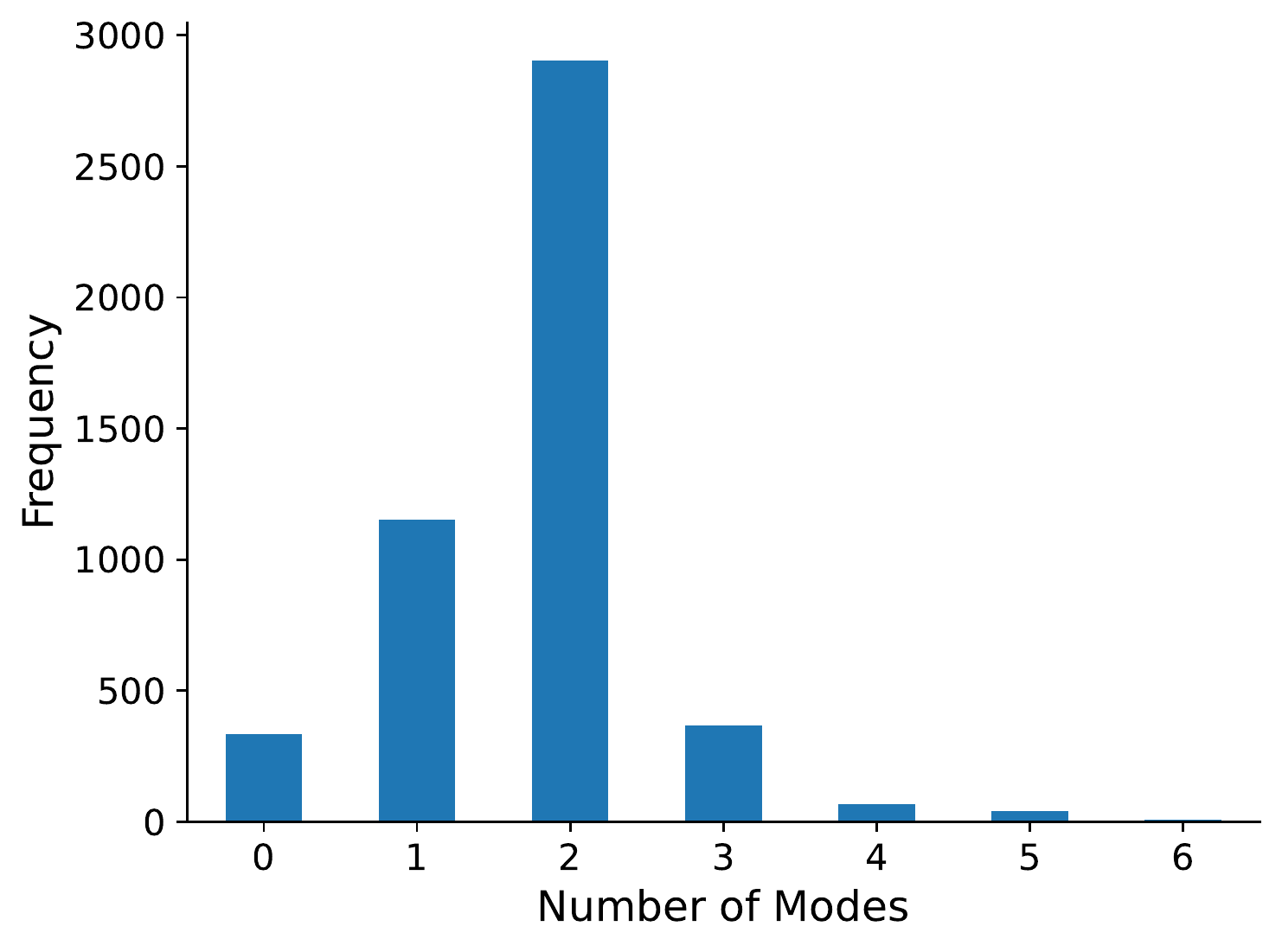}
\caption{\label{fig:figure14} For each rotatable bond in a random subset of 4,870 molecules from our NMRShiftDB data, we computed the number of modes by counting maxima in a kernel density estimate of the angle distribution. The plot above shows the frequency distribution of number of modes.}
\end{center}
\vskip -0.2in
\end{figure}

In this work, we make the assumption that most individual rotatable bond torsion angle distributions are multi-modal, typically with up to four distinct modes. In Figure \ref{fig:figure14}, we show that for a large subset of molecules from our NMRShiftDB data, most distributions have fewer than four distinct modes.

\section{Molecular Graph Featurization}\label{app:appF}

For each node in a molecular graph, we create a feature tensor to use in the graph neural network. These features are generated from basic molecular properties found in RDKit. They are padded with leading and/or trailing zeros so that each node has the same feature tensor shape whether it corresponds to a bond, atom or angle. In Tables \ref{tab:table2} and \ref{tab:table3} we describe the features used for bond and atom nodes, respectively. For angle nodes, we use a single feature. For the three atoms that correspond to an angle node, we compute the angle between them by using pairwise distances based on the average of upper and lower distance bounds provided by RDKit. 

\begin{table*}[t]
\caption{\label{tab:table2}\textit{Features for Bond Nodes.} Note: RDKit BondStereo types: \{ANY, CIS, E, NONE, TRANS, Z\}}
\vskip 0.15in
\begin{center}
\begin{small}
\begin{sc}
\scalebox{0.9}{
\begin{tabular}{llc}
\toprule
Feature & Description & Number of Elements \\
\midrule
Bond Type & One hot encoded from \{0, 1, 1.5, 2, 3\} & 5 \\
Conjugated & Is bond conjugated & 1 \\
In Ring & Is bond part of a ring & 1 \\
Stereo & Bond stereo type, one hot encoded from RDKit BondStereo types & 6 \\
Same Ring & Are endpoints of bond are in any ring together & 1 \\
Rotatable Bond & Is bond rotatable & 1 \\ 
Total & & 15 \\
\bottomrule
\end{tabular}
}
\end{sc}
\end{small}
\end{center}
\vskip -0.1in
\end{table*}

\begin{table*}[t]
\caption{\label{tab:table3}\textit{Features for Atom Nodes.} Note: RDKit ChiralTypes: \{UNSPECIFIED, Tetrahedral CW, Tetrahedral CCW, OTHER, Tetrahedral, Allene, Square Planar, Trigonal Bipyramidal, Octahedral\}. MMFF Atom Types selected: [1, 2, 3, 4, 5, 6, 7, 8, 9, 10, 11, 12, 15, 16, 17, 18, 20, 21, 22, 23, 24, 25, 26, 27, 28, 29, 30, 31, 32, 33, 37, 38, 39, 40, 42, 43, 44, 46, 48, 59, 62, 63, 64, 65, 66, 70, 71, 72, 74, 75, 78].}
\vskip 0.15in
\begin{center}
\begin{small}
\begin{sc}
\scalebox{0.9}{
\begin{tabular}{llc}
\toprule
Feature & Description & Number of Elements \\
\midrule
Atomic Number & One hot encoded from \{H, C, O, N, F, P, S, Cl\} & 8 \\
Valence & One hot encoded from 1-6 & 6 \\
Aromaticity & Whether atom is in aromatic structure, determined by RDKit & 1 \\
Hybridization & One hot encoded from \{$s, sp, sp^2, sp^3, sp^3d, sp^3d^2$, UNSPECIFIED\} & 7 \\
Partial Charge & Gasteiger Charge from RDKit (set to zero if not finite) & 1 \\
Formal Charge & Presence of net charge, one hot encoded from \{-1, 0, 1\} & 3 \\
Covalent Radius & RDKit covalent radius & 1 \\
van der Waals Radius & RDKit van der Waals radius & 1 \\
Default Valence & Valence of atom on periodic table, one hot encoded from 1-6 & 6 \\
Rings & Whether the atom is in a ring of size N for N from 3-8 & 6 \\
Chirality & One hot encoded from RDKit ChiralTypes & 9 \\
MMFF Atom Types & Atom type from RDKit's MMFFMolProperties, one hot encoded & 51 \\
Degree & One hot encoded from 1-6 & 6 \\
Number of Hydrogens & Total number of hydrogens on atom, one hot encoded from 0-3 & 4 \\
Radical Electrons & Number of radical electrons, one hot encoded from 0-2 & 3  \\
Total & & 113 \\
\bottomrule
\end{tabular}
}
\end{sc}
\end{small}
\end{center}
\vskip -0.1in
\end{table*}

Therefore, in total, each node has a feature tensor with 15 + 113 + 1 = 129 elements (\# bond + \# atom + \# angle). -10 is used as the sentinel value in place of features for other node types. 

\section{Computing and Setting Chiralities using RDKit}\label{app:appG}
To determine whether an atom has R or S chirality, we compute the oriented volume formed by the atom and its three neighbors \cite{GeoMol}. If the oriented volume is 1, then we say the atom has R chirality, and if the oriented volume is -1, we say the atom has S chirality. The oriented volume is given by:

\begin{align}
    \begin{split}
        OV(\textbf{p}_1, \textbf{p}_2, \textbf{p}_3, \textbf{p}_4) = \textit{sign}\left(\left|\begin{matrix}
        1 & 1 & 1 & 1\\
        x_1 & x_2 & x_3 & x_4 \\
        y_1 & y_2 & y_3 & y_4 \\
        z_1 & z_2 & z_3 & z_4 \\
        \end{matrix}
        \right|\right)
    \end{split}
\end{align}

We assign atoms to these four vectors in a consistent fashion by using the same CIP priority rules that we use for selecting the four atoms that define a rotatable bond torsion angle. The neighbors of the atom that have first, second, and third priority are assigned to $\textbf{p}_1$, $\textbf{p}_2$, and $\textbf{p}_3$, respectively. The atom itself is assigned to $\textbf{p}_4$.  

If we need to flip the chirality from R to S or vice versa, we do the following. If A is the atom, we 1) compute the plane formed by A's three neighbors 2) compute the projection of A onto the plane, 3) compute the plane formed by A, one of its neighbors, and A's projection onto the neighbor plane, and 4) reflect all atoms across this plane. 

\section{Additional Evaluations}\label{app:appH}
\begin{figure}[!hb]
\vskip 0.2in
\begin{center}
\subfigure[]{
  \includegraphics[scale = 1.]{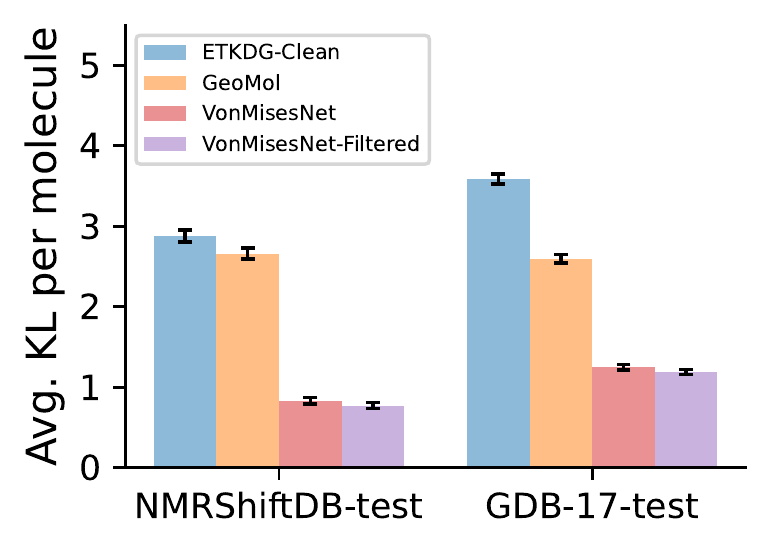}}%
  
\subfigure[]{
  \includegraphics[scale = 1]{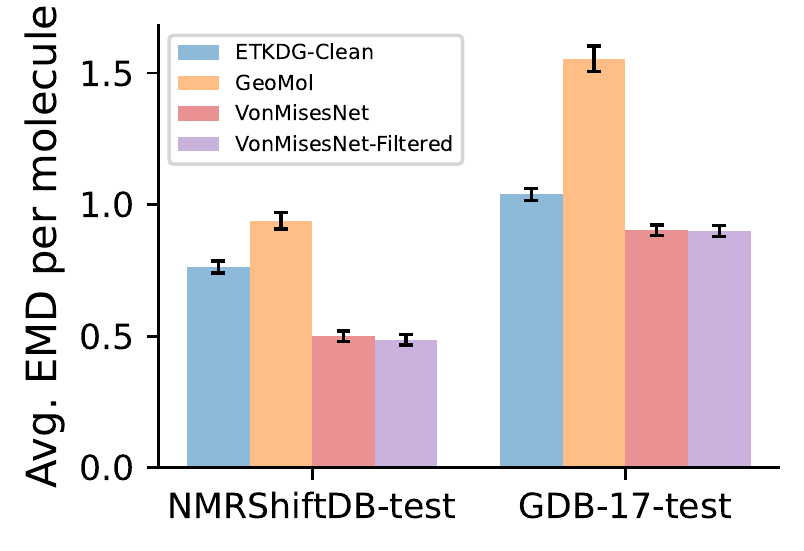}}%
\caption{\textit{Rotatable bond torsion angle distributions evaluation without Torsional Diffusion constraints.}\label{fig:figure15}\textbf{(a)} Average KL divergence of rotatable bond torsion angle distributions, per molecule, relative to PT-HMC ground truth in 997 random molecules from NMRShiftDB-test and 997 random molecules from GDB-17-test. Standard error bars are shown in black. The KL divergence is measured with 32 bins. \textbf{(b)} Average EMD of rotatable bond torsion angle distributions, per molecule, relative to PT-HMC ground truth for the same sets of molecules.}
\end{center}
\vskip -0.2in
\end{figure}

\begin{figure}[!ht]
\vskip 0.2in
\begin{center}
\subfigure[]{
\includegraphics[scale = 0.55]{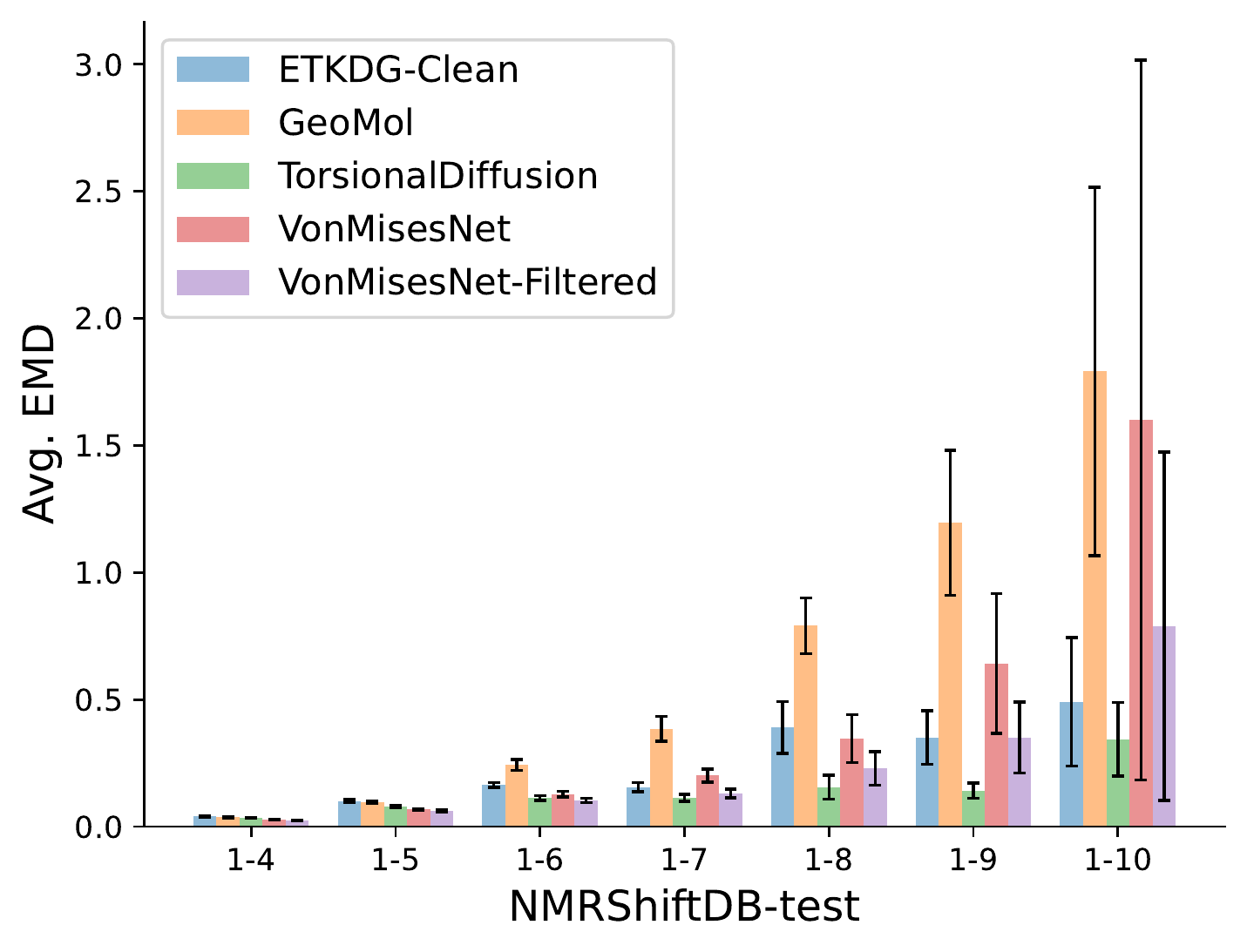}
}
\subfigure[]{
\includegraphics[scale = 0.55]{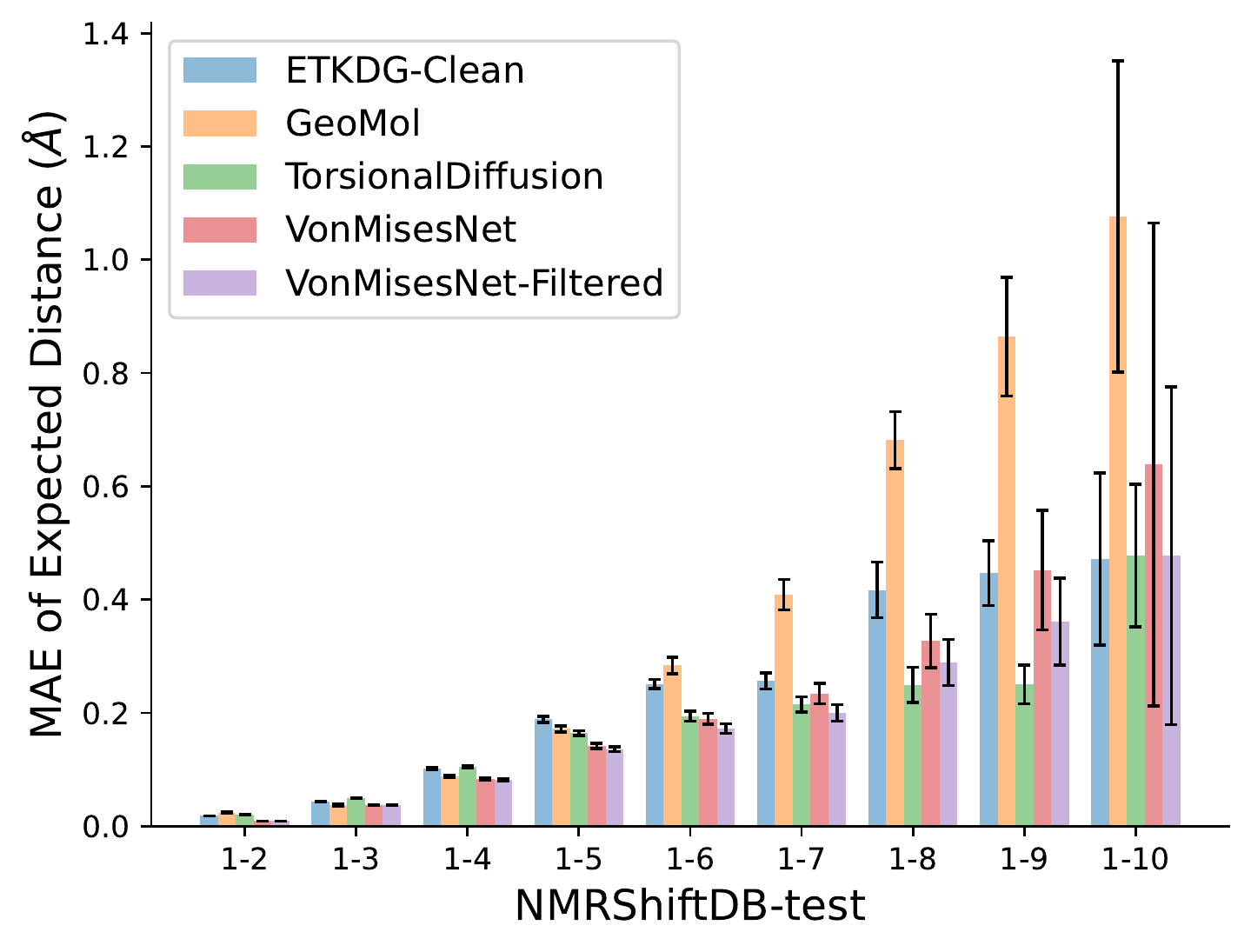}
}

\subfigure[]{
\includegraphics[scale = 0.55]{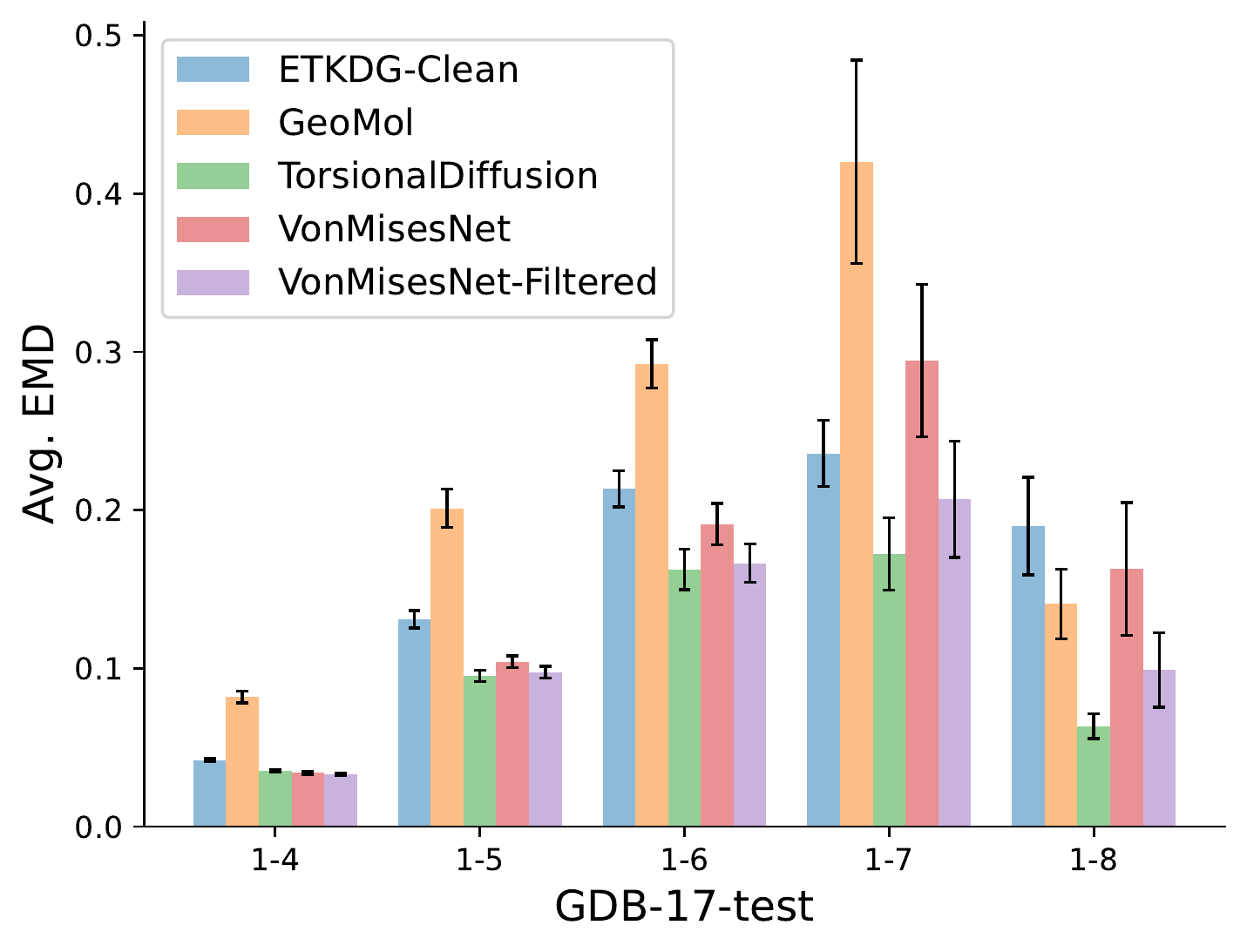}
}
\subfigure[]{
\includegraphics[scale = 0.55]{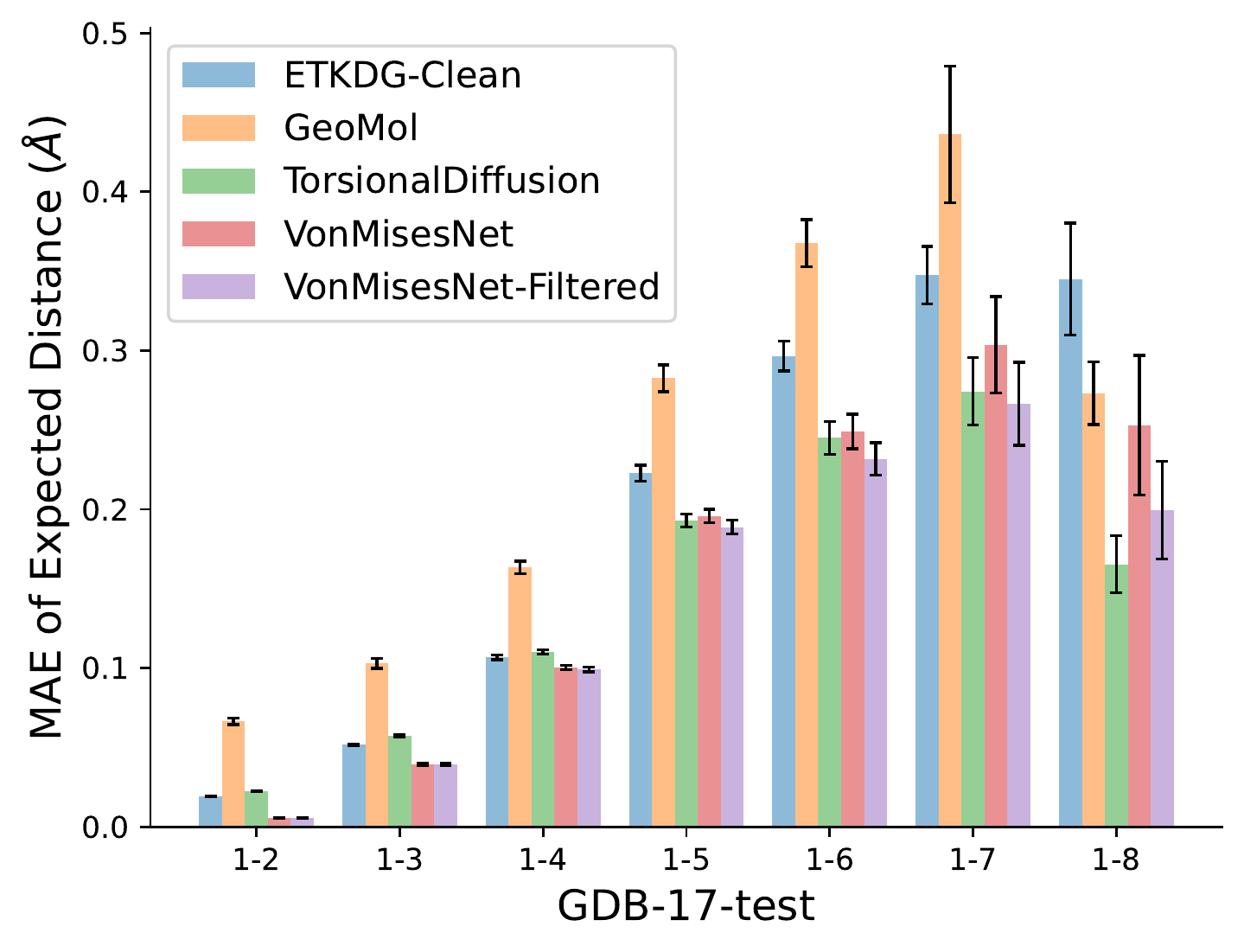}
}
\caption{\textit{Pairwise distance distributions evaluation, up to 1-10, rotatable.}\label{fig:figure16} We evaluate pairwise distance distributions relative to PT-HMC ground truth for 538 molecules from NMRShiftDB-test and 610 molecules from GDB-17-test. We evaluate distances for which every intermediate bond along the shortest path is rotatable. In \textbf{(a)} and \textbf{(c)} we compare the average EMD, per molecule, and in \textbf{(b)} and \textbf{(d)} we compare the MAE of the expected distance, per molecule. For the expected distance evaluations, we additionally include 1-2 distances that are not part of a ring and 1-3 distances for which at least one of the bonds is rotatable.}
\end{center}
\vskip -0.2in
\end{figure}

\begin{figure}[!ht]
\vskip 0.2in
\begin{center}
\subfigure[]{
\includegraphics[scale = 0.55]{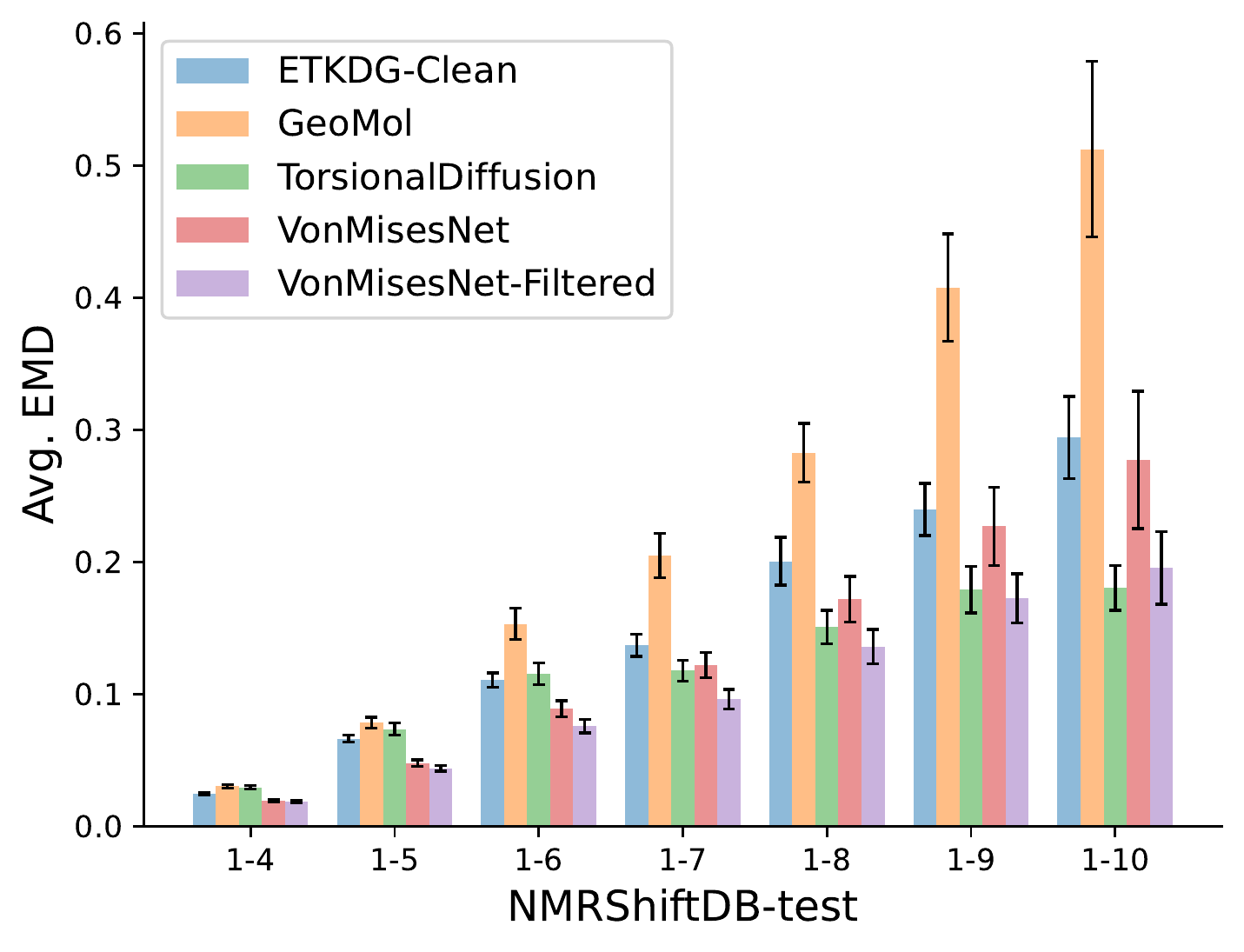}
}
\subfigure[]{
\includegraphics[scale = 0.55]{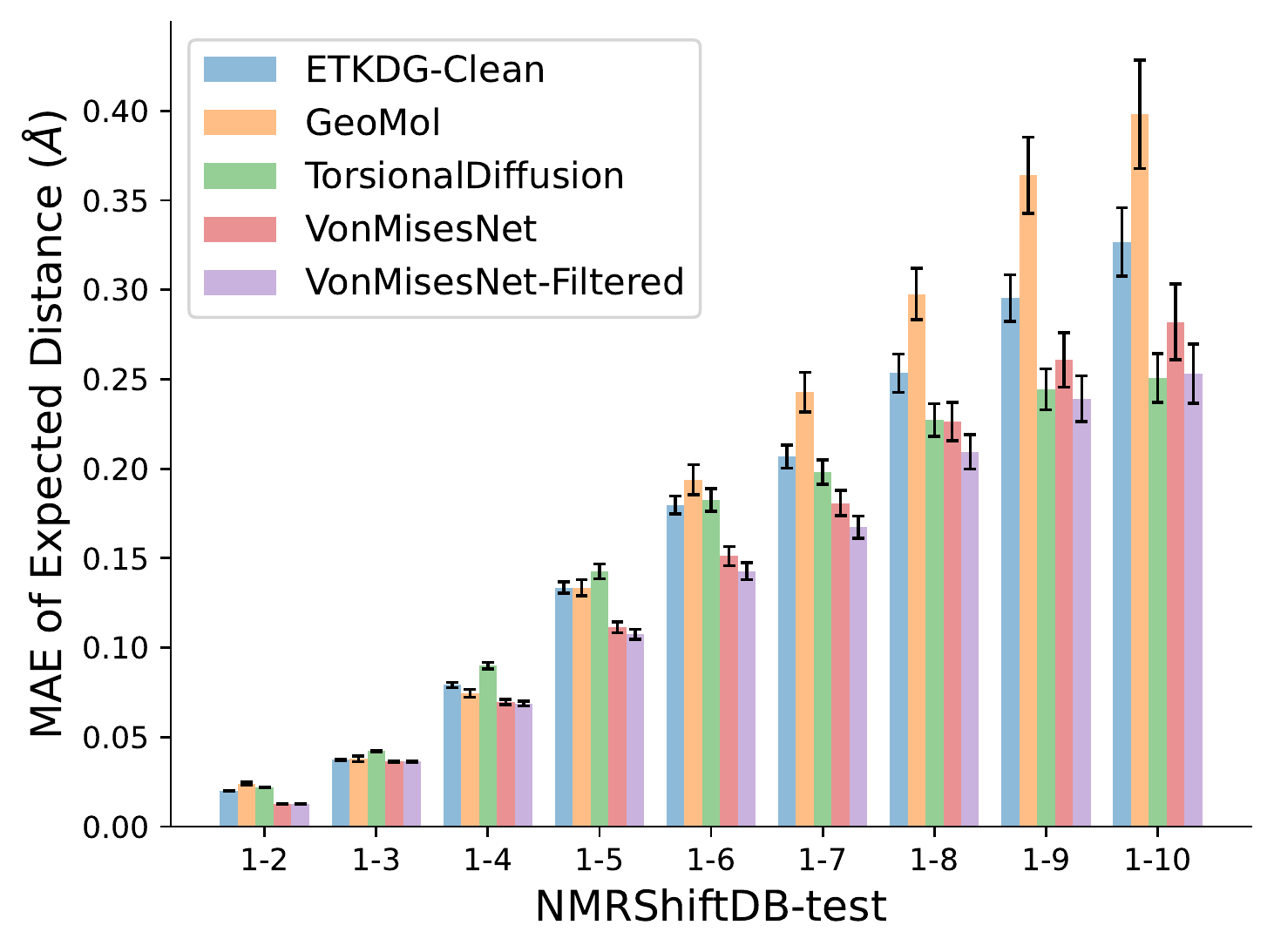}
}

\subfigure[]{
\includegraphics[scale = 0.55]{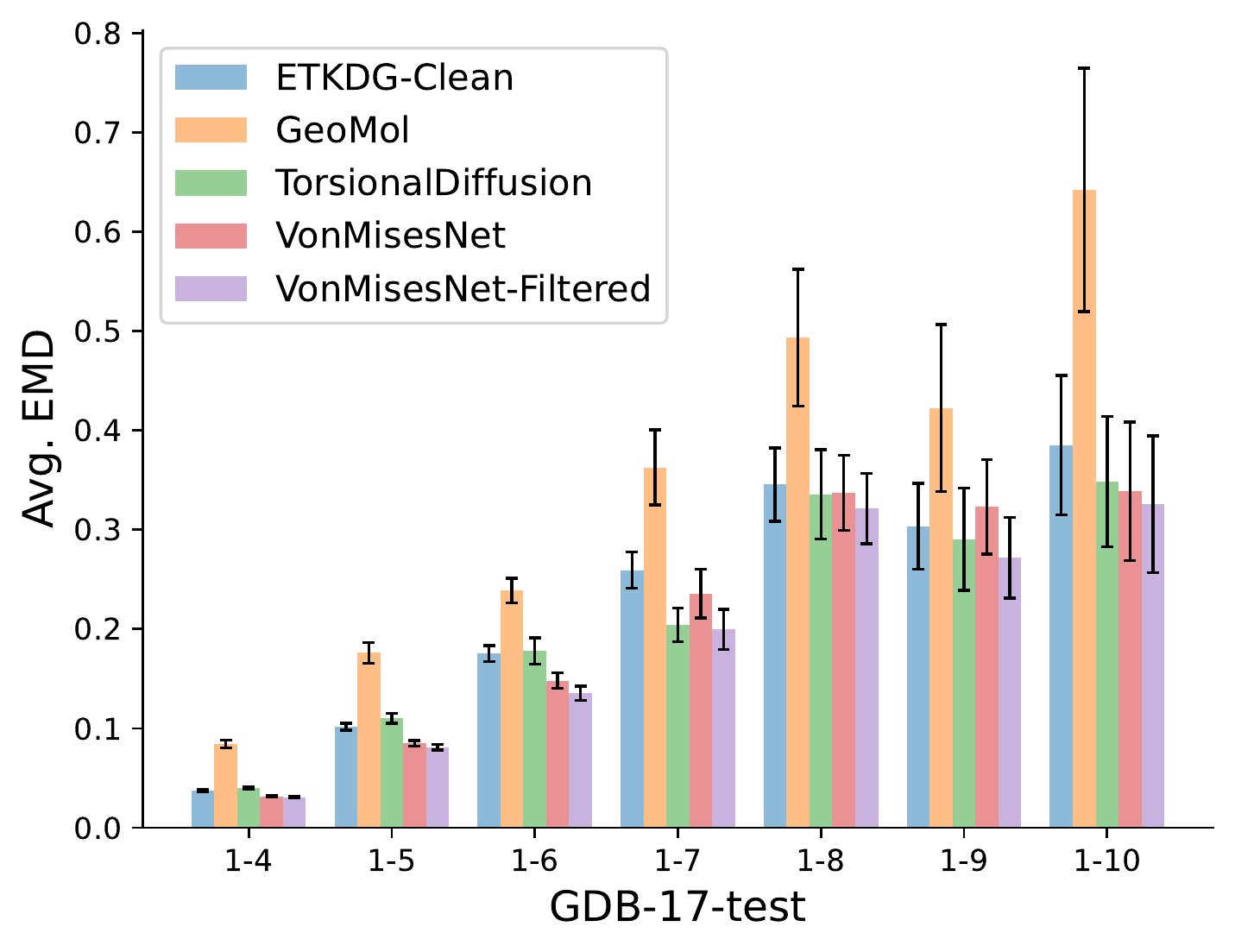}
}
\subfigure[]{
\includegraphics[scale = 0.55]{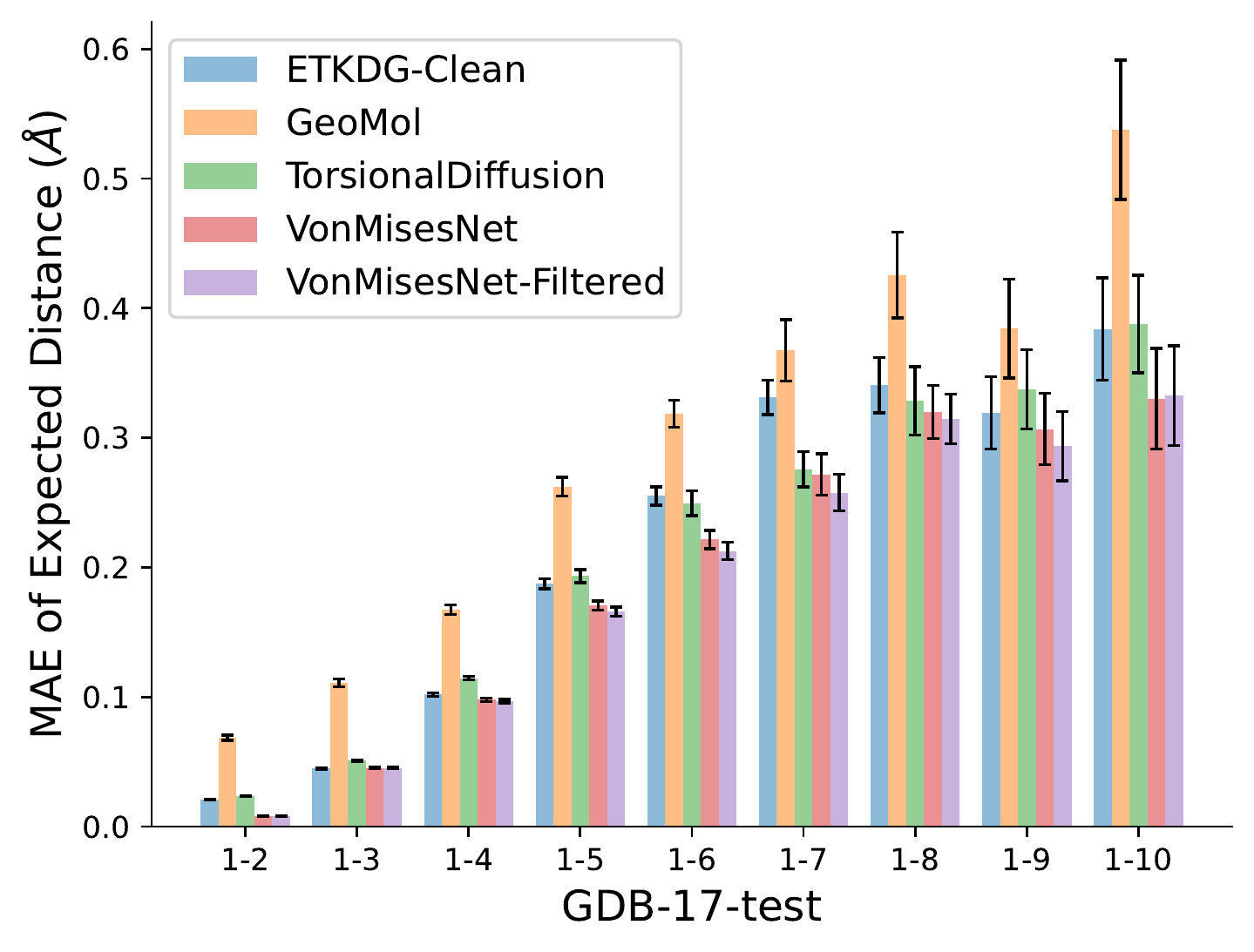}
}
\caption{\textit{Pairwise distance distributions evaluation, up to 1-10, excluding non-aromatic.}\label{fig:figure17} We evaluate pairwise distance distributions relative to PT-HMC ground truth for 538 molecules from NMRShiftDB-test and 610 molecules from GDB-17-test. We evaluate distances for which every intermediate bond along the shortest path is not part of a non-aromatic ring. In \textbf{(a)} and \textbf{(c)} we compare the average EMD, per molecule, and in \textbf{(b)} and \textbf{(d)} we compare the MAE of the expected distance, per molecule. For the expected distance evaluations, we additionally include 1-2 distances that are not part of a non-aromatic ring and 1-3 distances for which neither of the bonds is in a non-aromatic ring.}
\end{center}
\vskip -0.2in
\end{figure}

\begin{figure}[!ht]
\vskip 0.2in
\begin{center}
\subfigure[]{
\includegraphics[scale = 0.55]{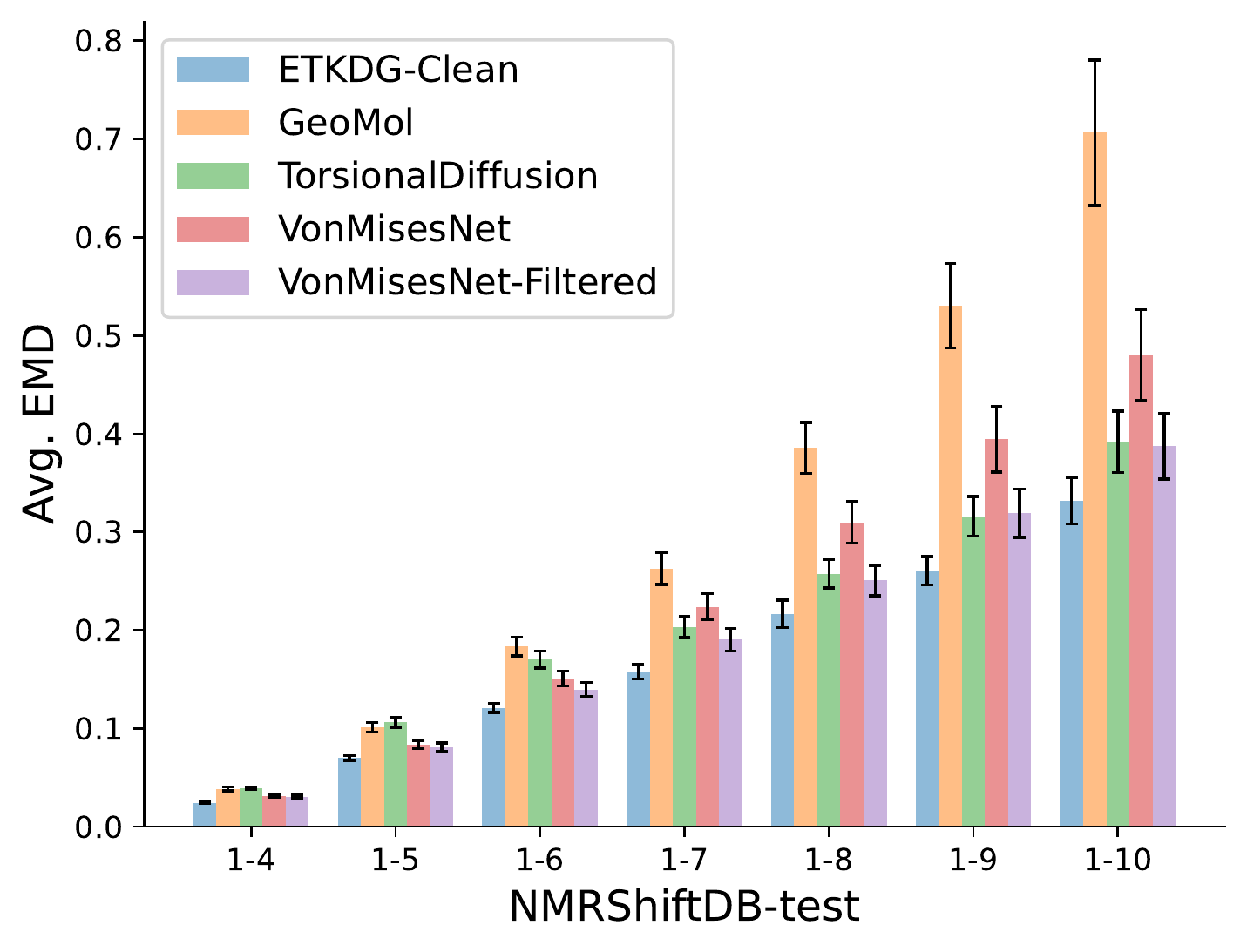}
}
\subfigure[]{
\includegraphics[scale = 0.55]{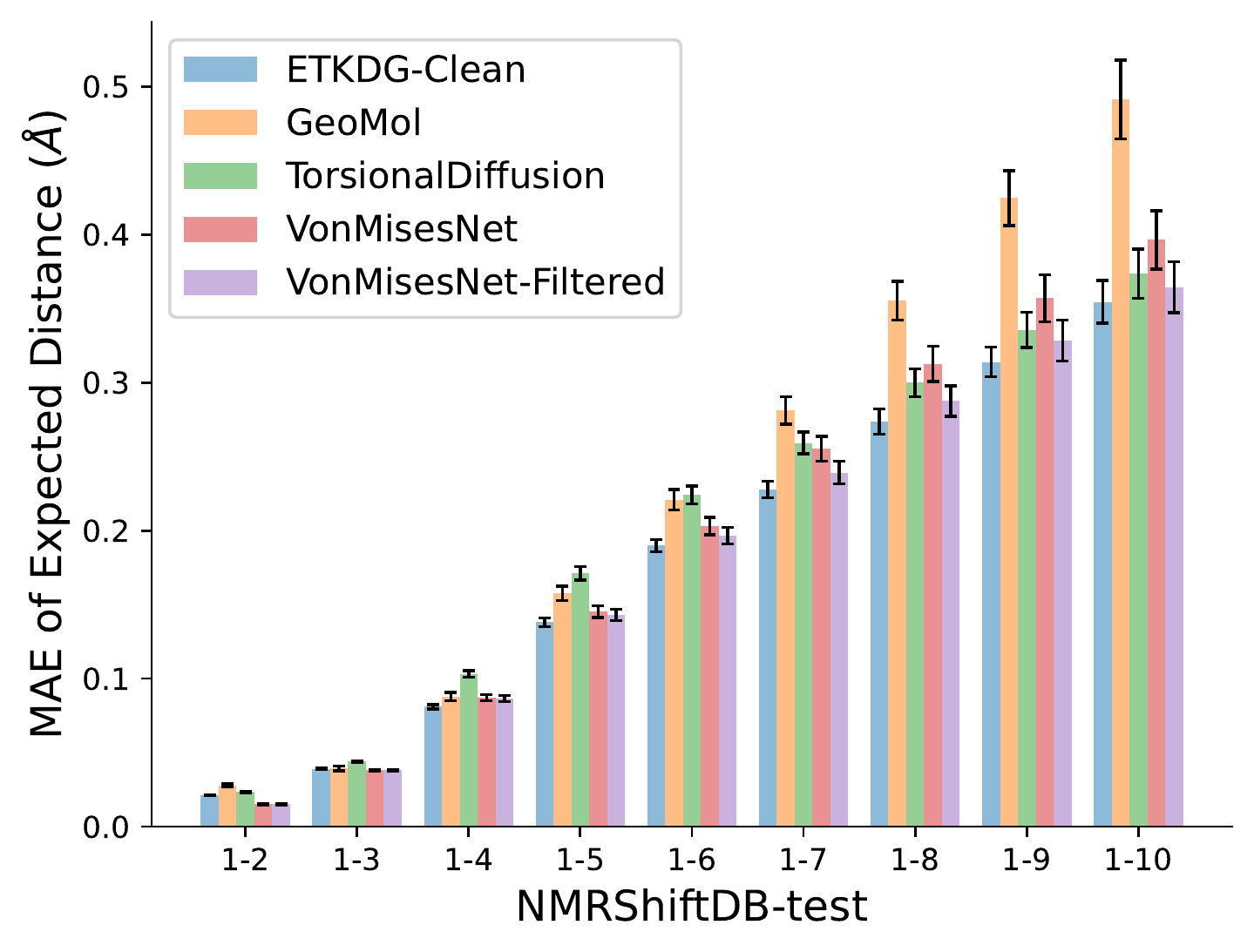}
}

\subfigure[]{
\includegraphics[scale = 0.55]{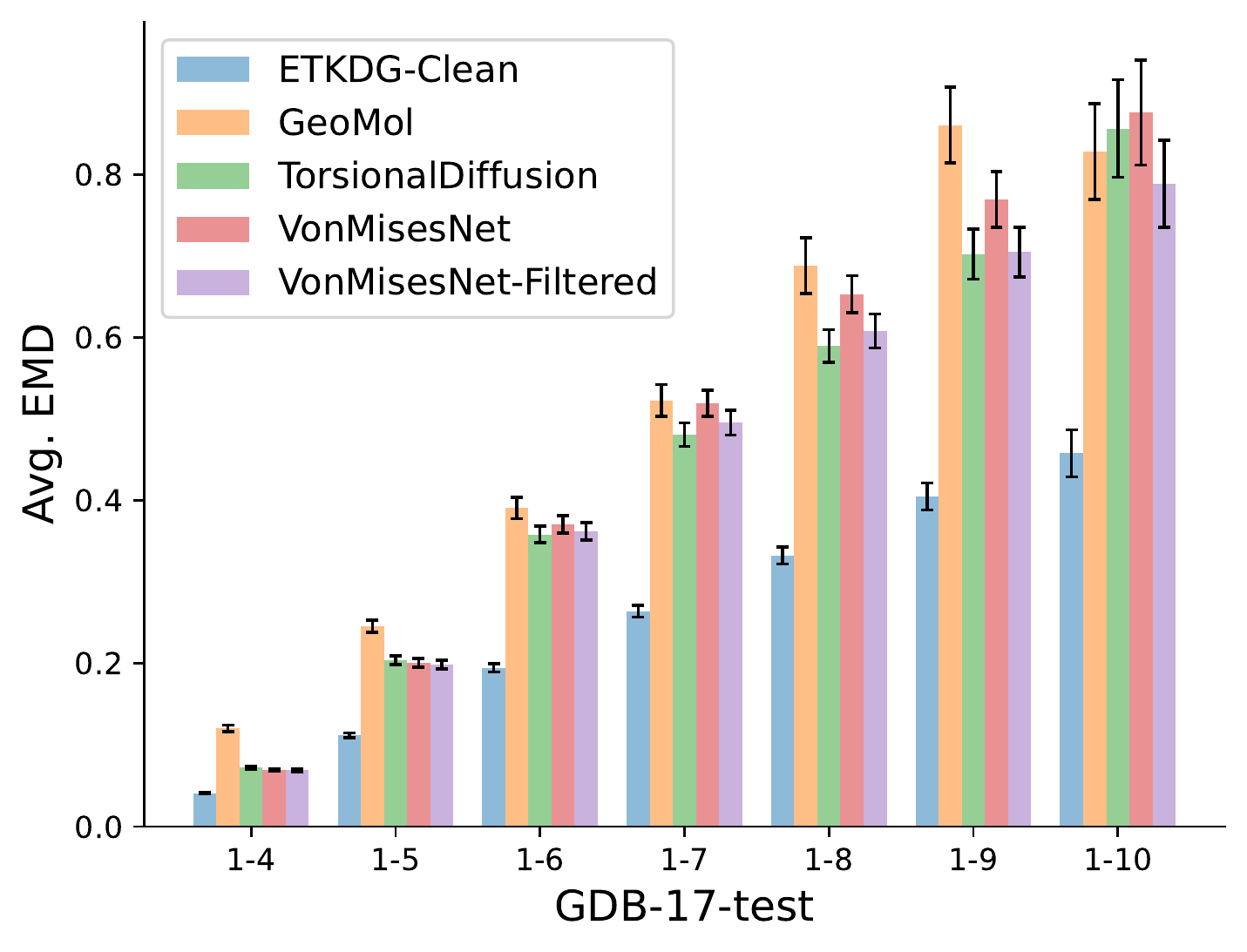}
}
\subfigure[]{
\includegraphics[scale = 0.55]{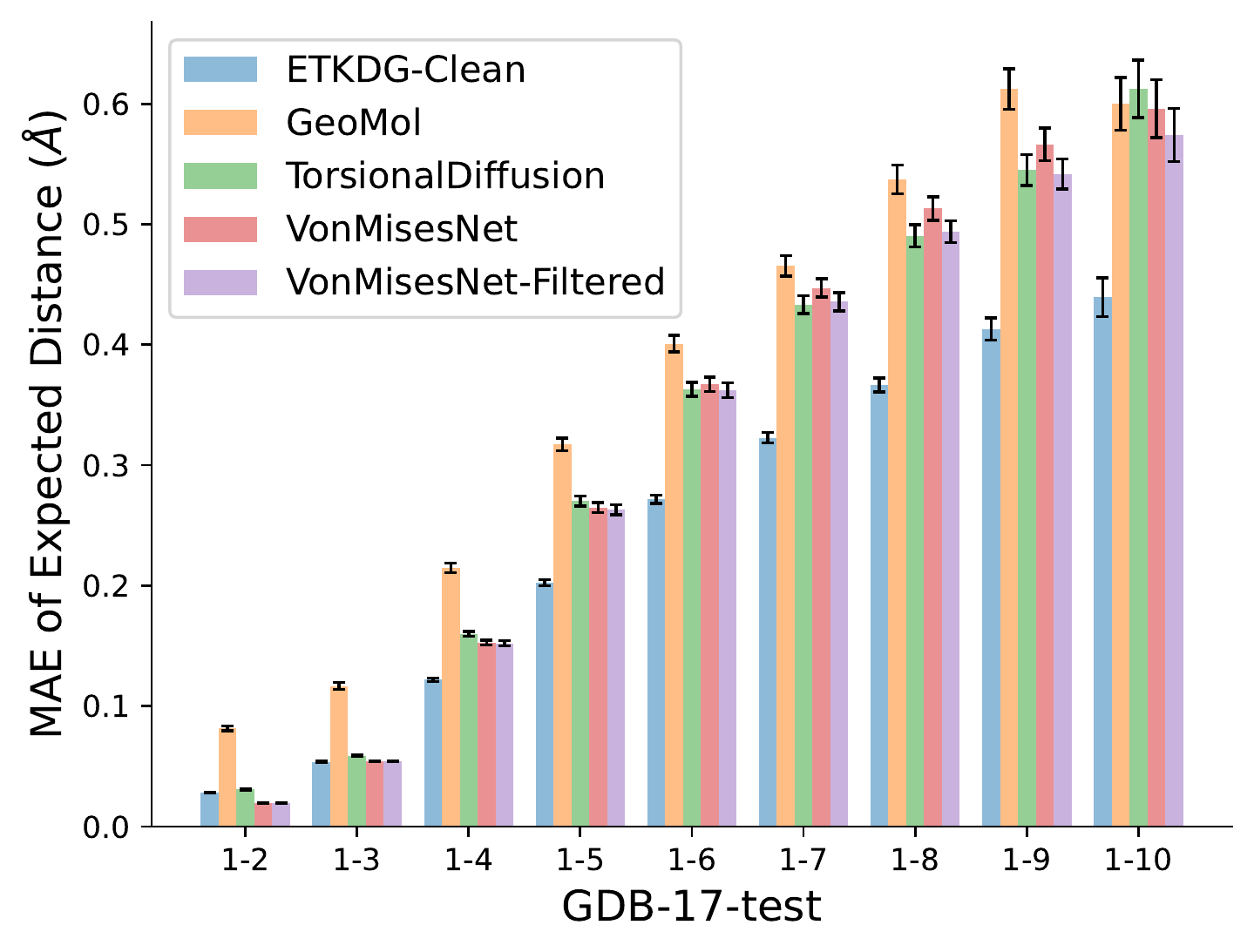}
}
\caption{\textit{Pairwise distance distributions evaluation, up to 1-10, no restrictions.}\label{fig:figure18} We evaluate pairwise distance distributions relative to PT-HMC ground truth for 538 molecules from NMRShiftDB-test and 610 molecules from GDB-17-test. We evaluate distances without restrictions. In \textbf{(a)} and \textbf{(c)} we compare the average EMD, per molecule, and in \textbf{(b)} and \textbf{(d)} we compare the MAE of the expected distance, per molecule. For the expected distance evaluations, we additionally include 1-2 distances and 1-3 distances without restrictions.}
\end{center}
\vskip -0.2in
\end{figure}

\begin{figure}[!ht]
\vskip 0.2in
\begin{center}
\subfigure[]{
\includegraphics[scale = 0.55]{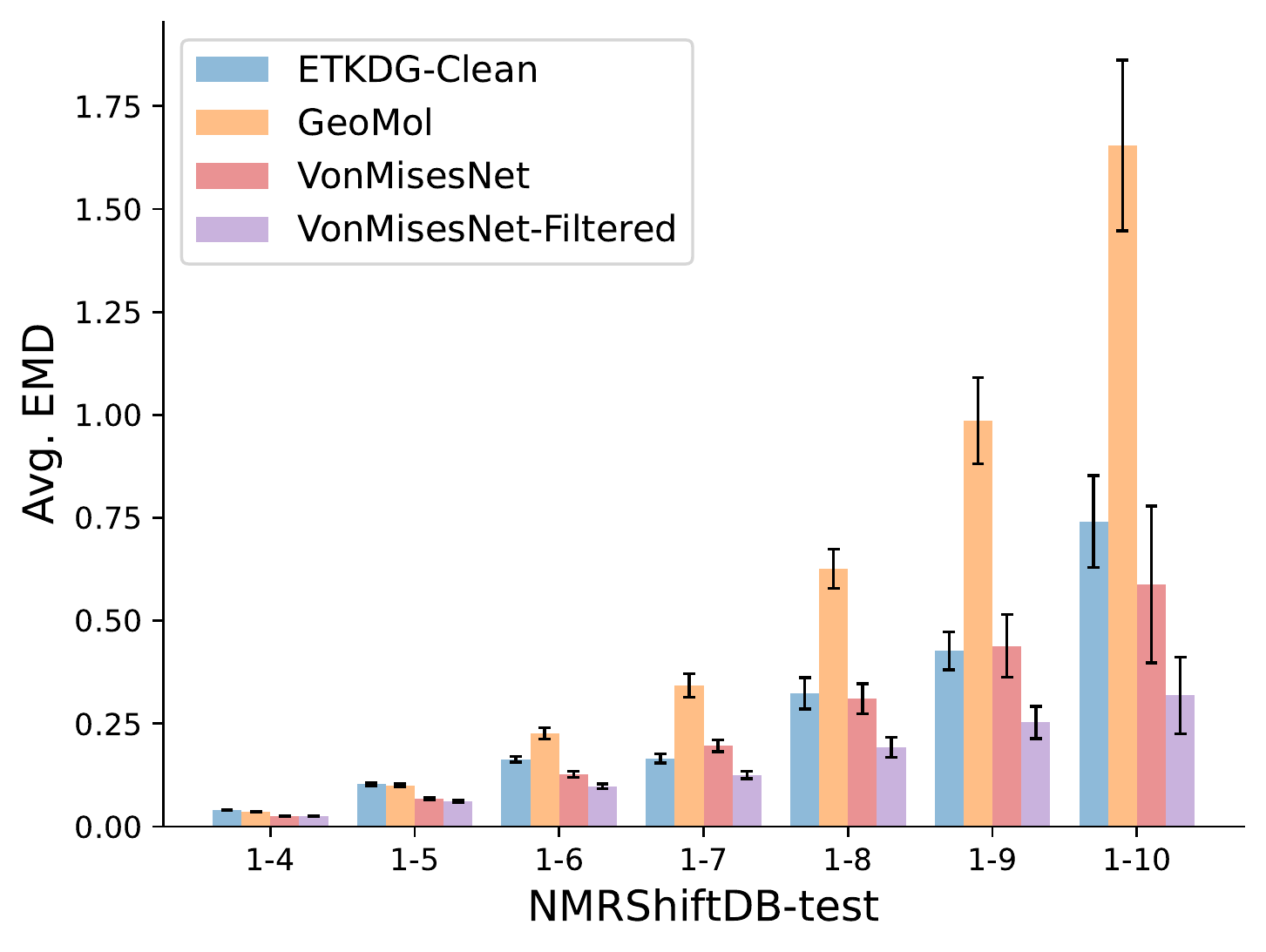}
}
\subfigure[]{
\includegraphics[scale = 0.55]{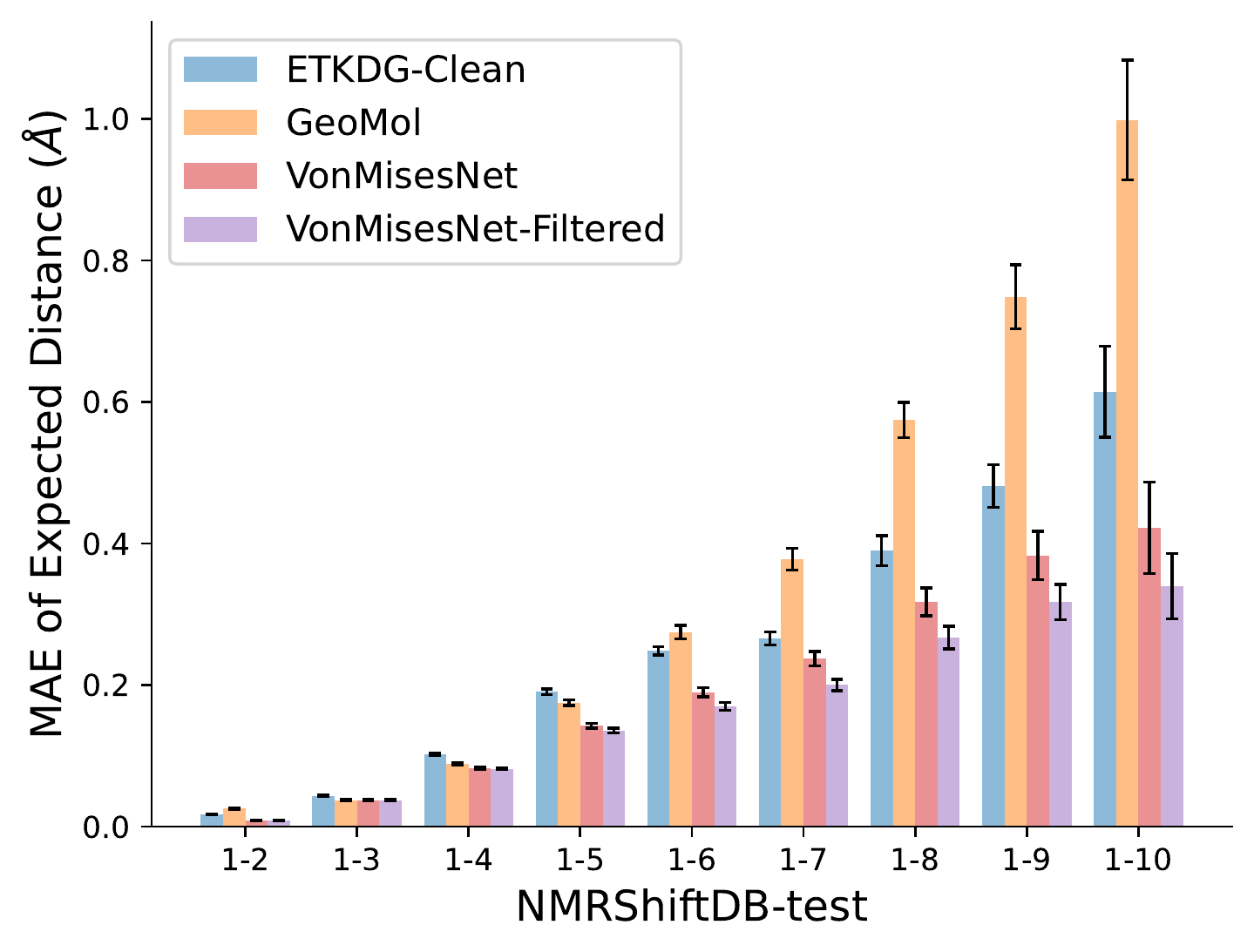}
}

\subfigure[]{
\includegraphics[scale = 0.55]{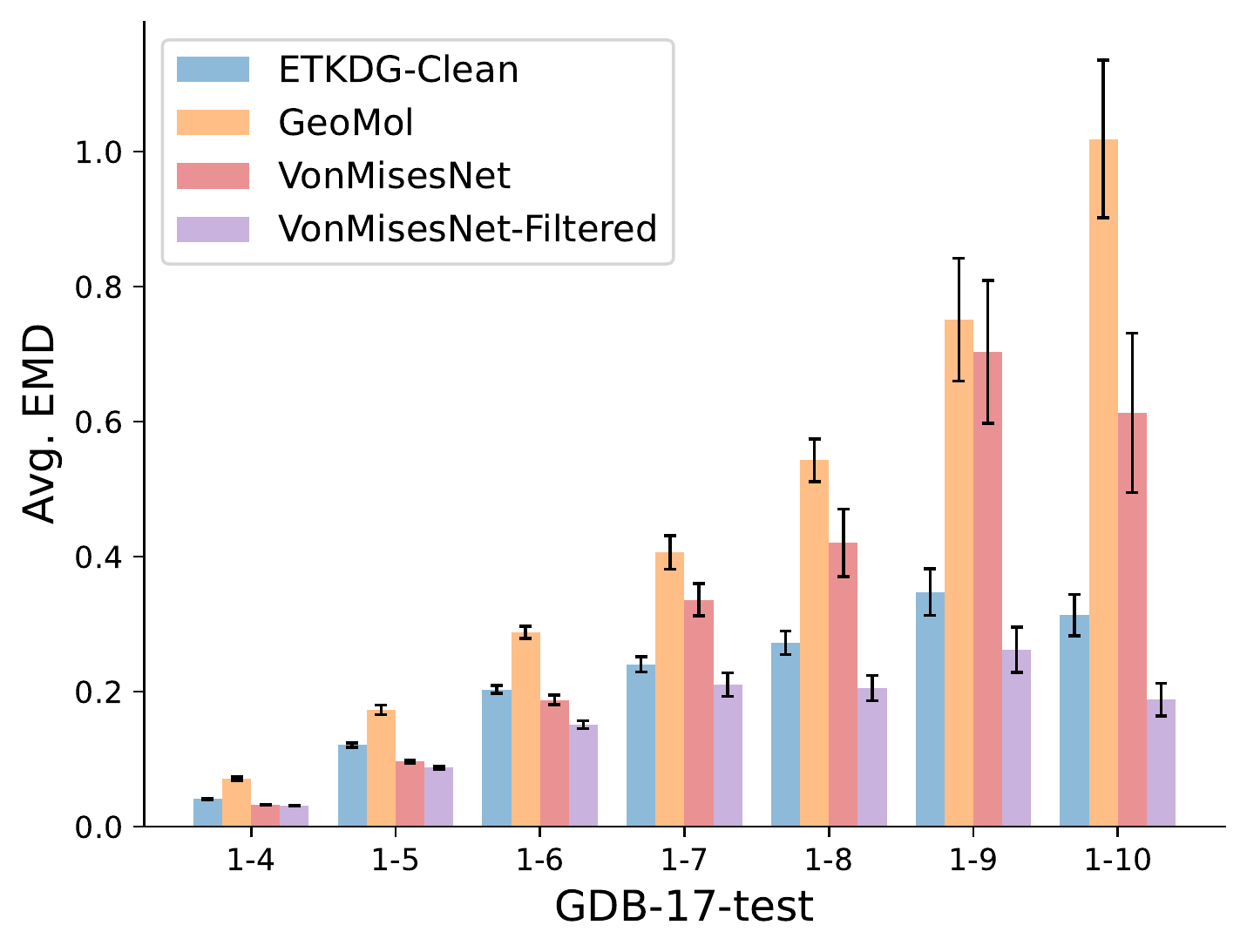}
}
\subfigure[]{
\includegraphics[scale = 0.55]{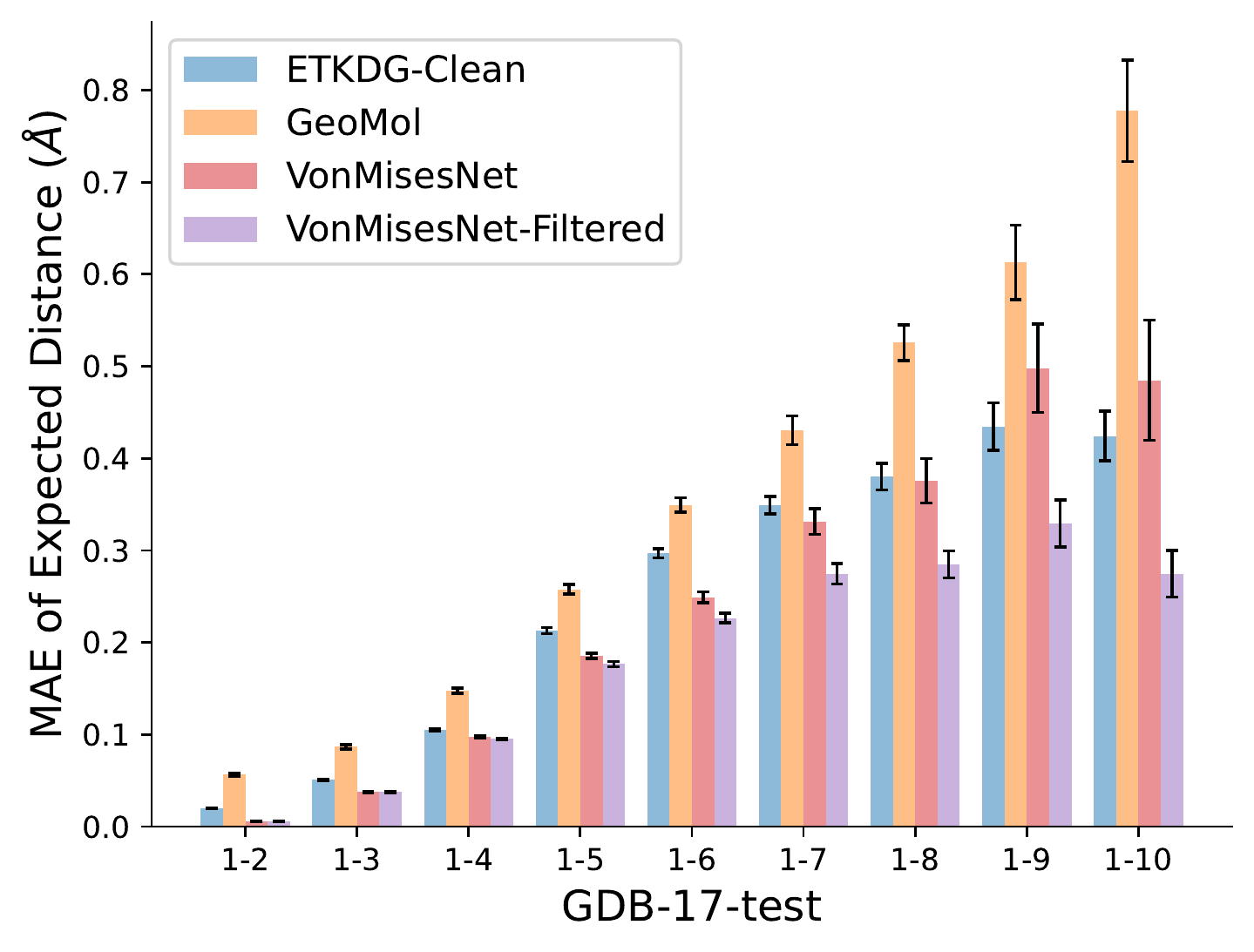}
}
\caption{\textit{Pairwise distance distributions evaluation, up to 1-10, excluding Torsional Diffusion constraints, rotatable.}\label{fig:figure19} We evaluate pairwise distance distributions relative to PT-HMC ground truth for 997 random molecules from NMRShiftDB-test and 997 random molecules from GDB-17-test. We evaluate distances for which every intermediate bond along the shortest path is rotatable. In \textbf{(a)} and \textbf{(c)} we compare the average EMD, per molecule, and in \textbf{(b)} and \textbf{(d)} we compare the MAE of the expected distance, per molecule. For the expected distance evaluations, we additionally include 1-2 distances that are not part of a ring and 1-3 distances for which at least one of the bonds is rotatable.}
\end{center}
\vskip -0.2in
\end{figure}

\begin{figure}[!ht]
\vskip 0.2in
\begin{center}
\subfigure[]{
\includegraphics[scale = 0.55]{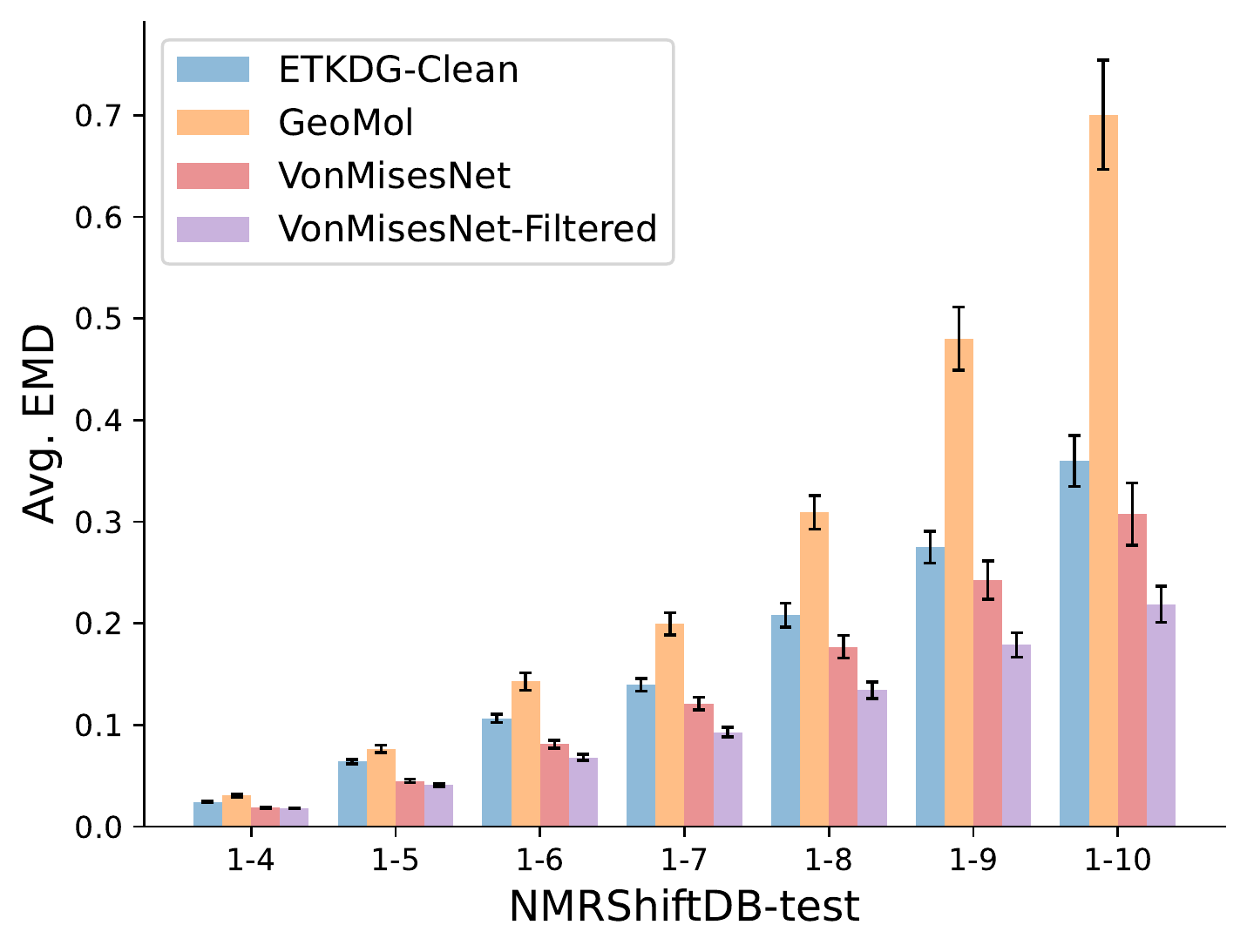}
}
\subfigure[]{
\includegraphics[scale = 0.55]{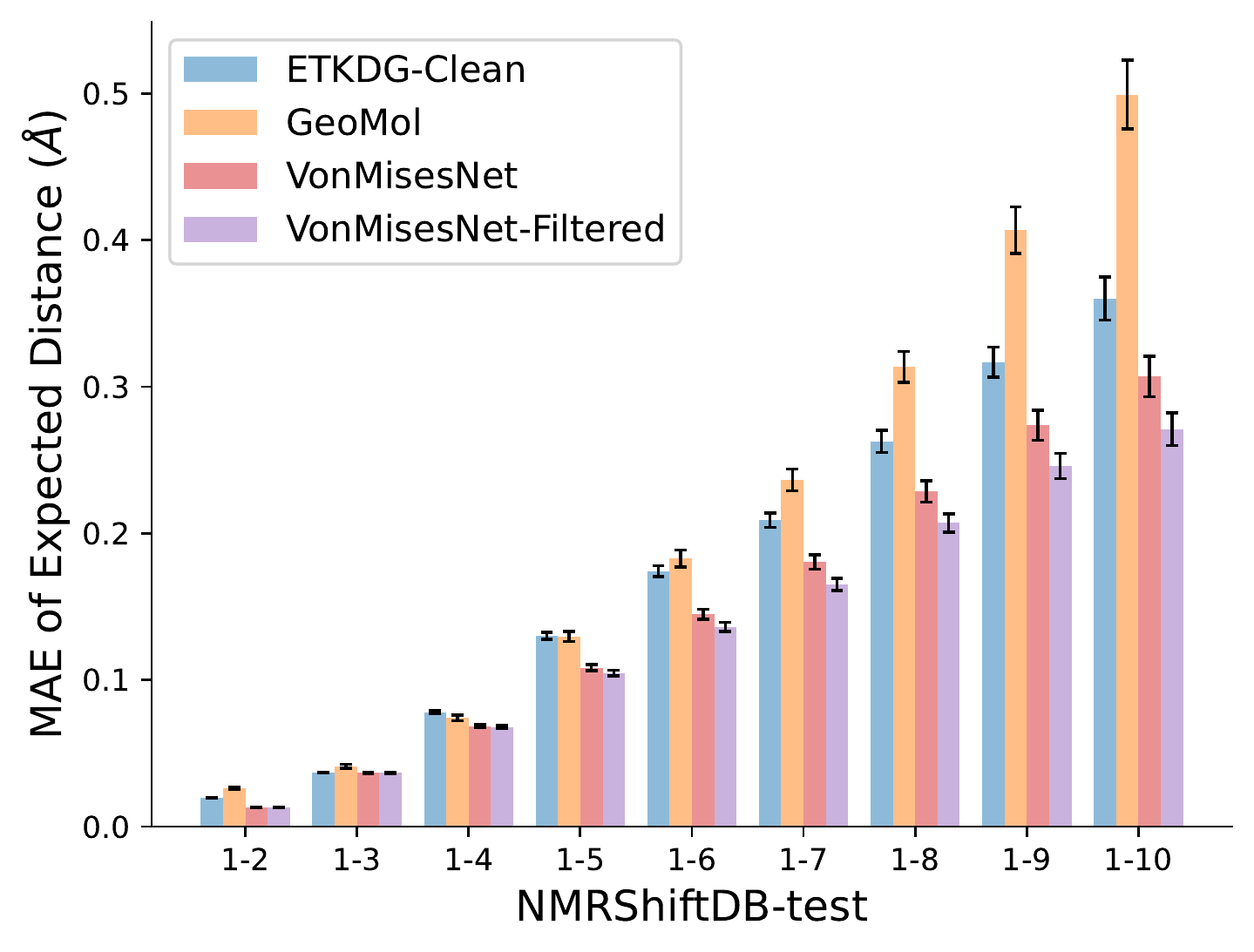}
}

\subfigure[]{
\includegraphics[scale = 0.55]{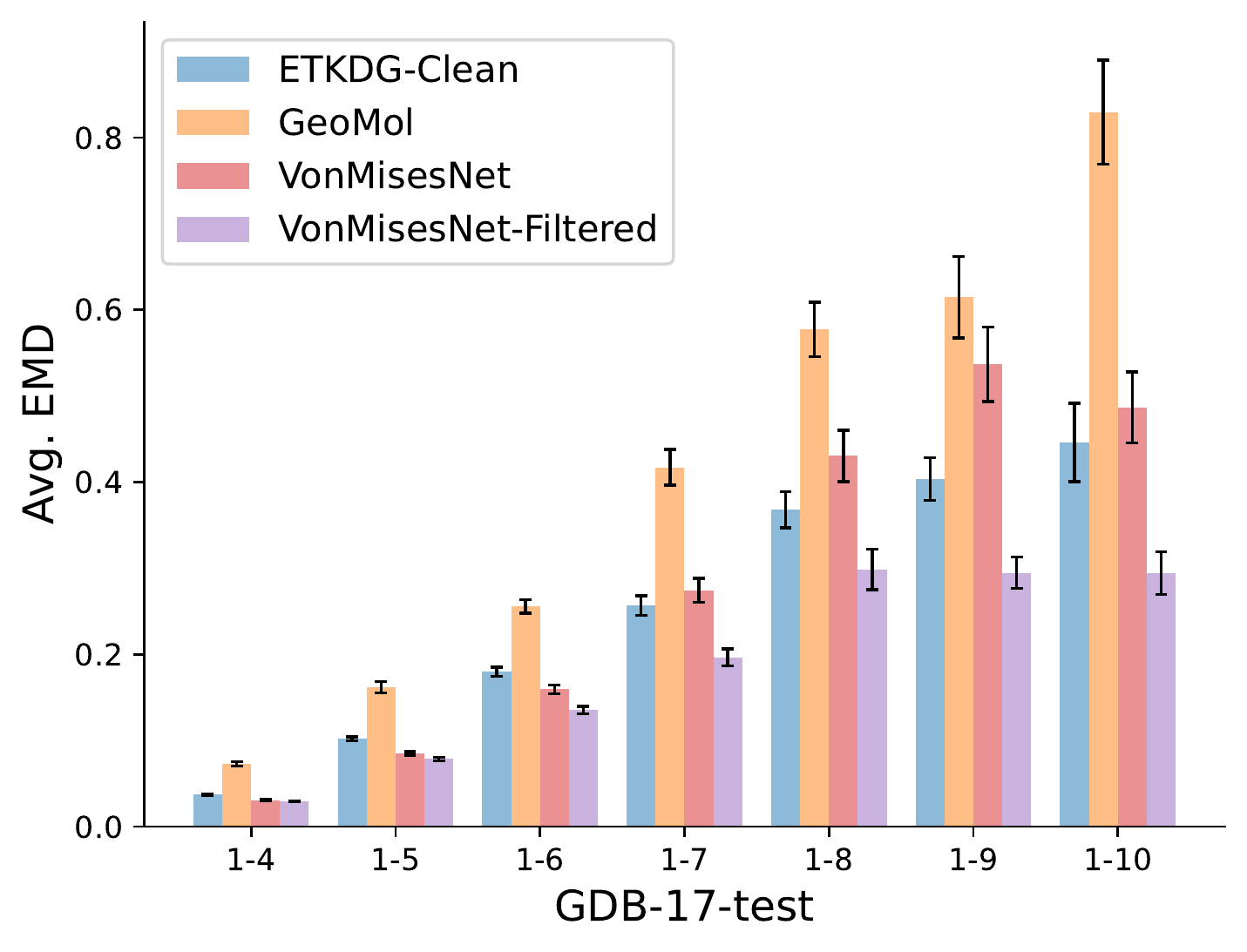}
}
\subfigure[]{
\includegraphics[scale = 0.55]{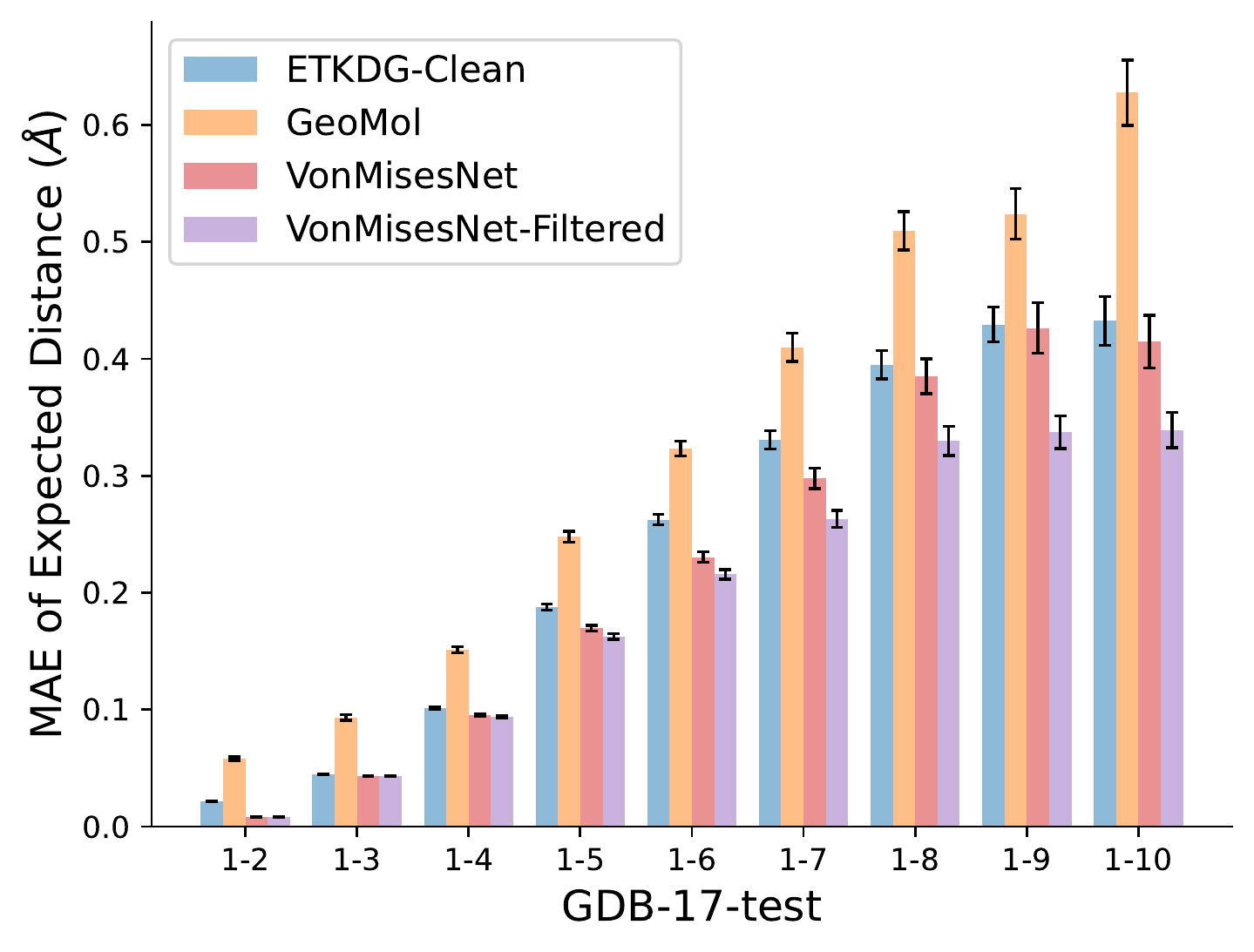}
}
\caption{\textit{Pairwise distance distributions evaluation, up to 1-10, excluding Torsional Diffusion constraints, excluding non-aromatic.}\label{fig:figure20} We evaluate pairwise distance distributions relative to PT-HMC ground truth for 997 random molecules from NMRShiftDB-test and 997 random molecules from GDB-17-test. We evaluate distances for which every intermediate bond along the shortest path is not part of a non-aromatic ring. In \textbf{(a)} and \textbf{(c)} we compare the average EMD, per molecule, and in \textbf{(b)} and \textbf{(d)} we compare the MAE of the expected distance, per molecule. For the expected distance evaluations, we additionally include 1-2 distances that are not part of a non-aromatic ring and 1-3 distances for which neither of the bonds is in a non-aromatic ring.}
\end{center}
\vskip -0.2in
\end{figure}

\begin{figure}[!ht]
\vskip 0.2in
\begin{center}
\subfigure[]{
\includegraphics[scale = 0.55]{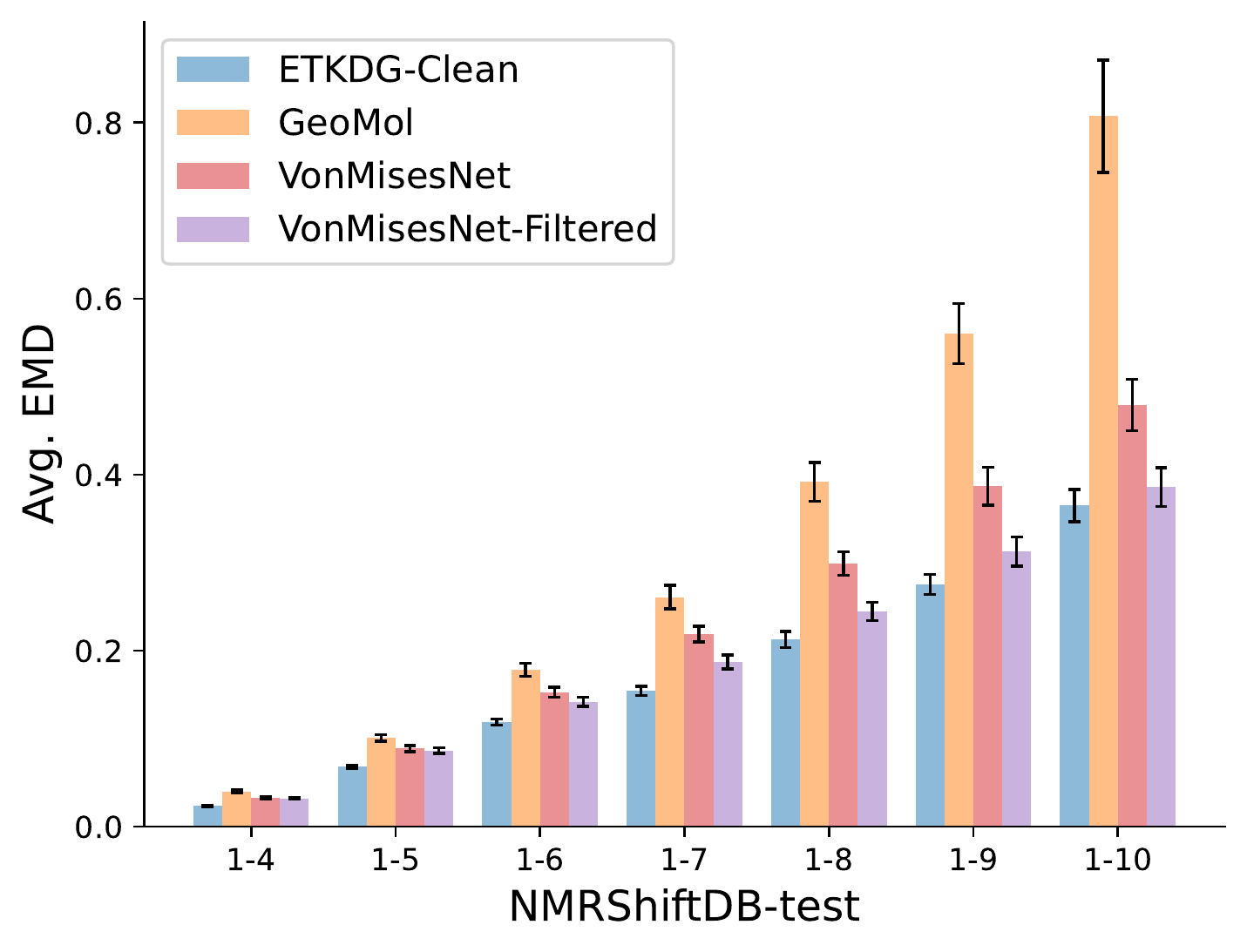}
}
\subfigure[]{
\includegraphics[scale = 0.55]{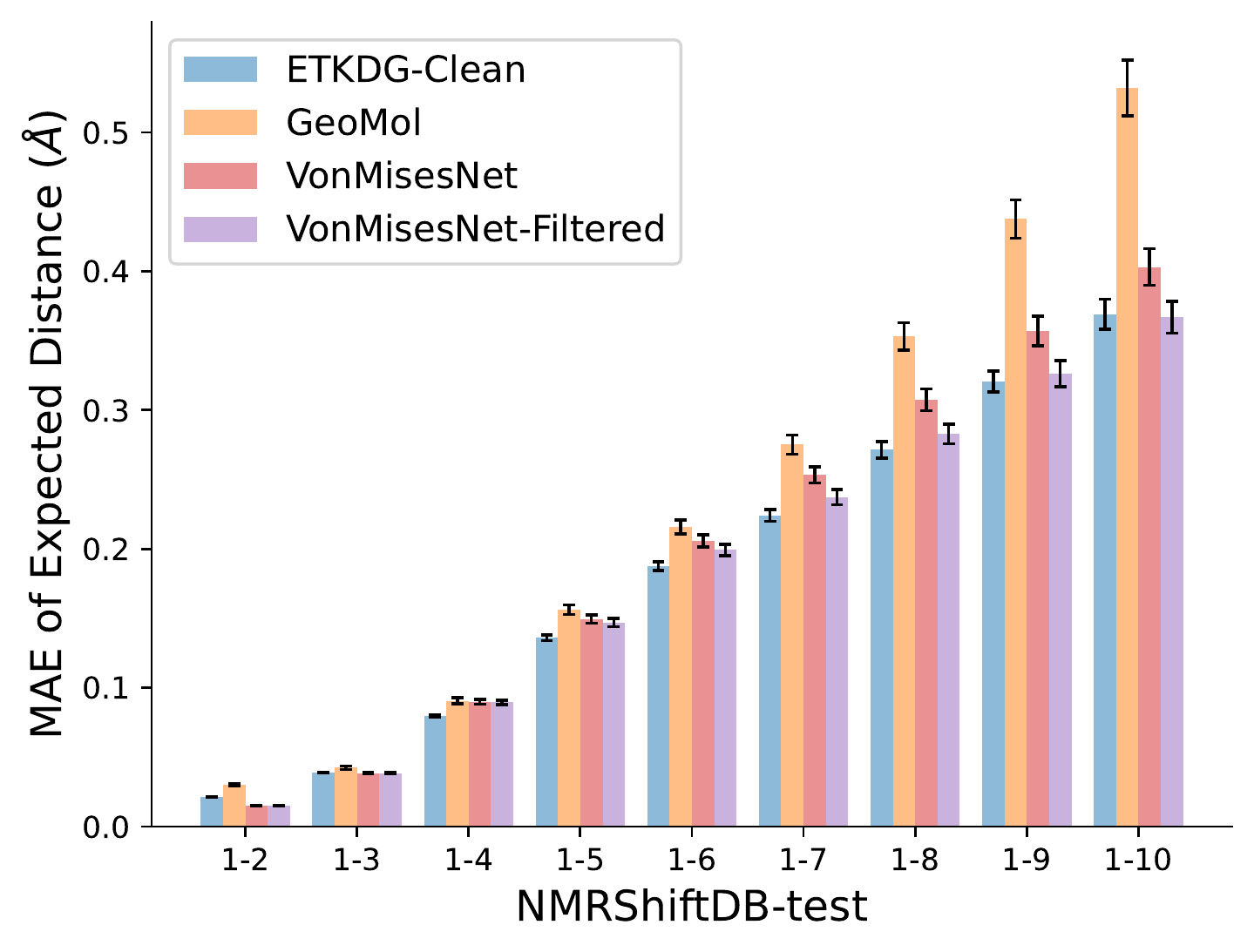}
}

\subfigure[]{
\includegraphics[scale = 0.55]{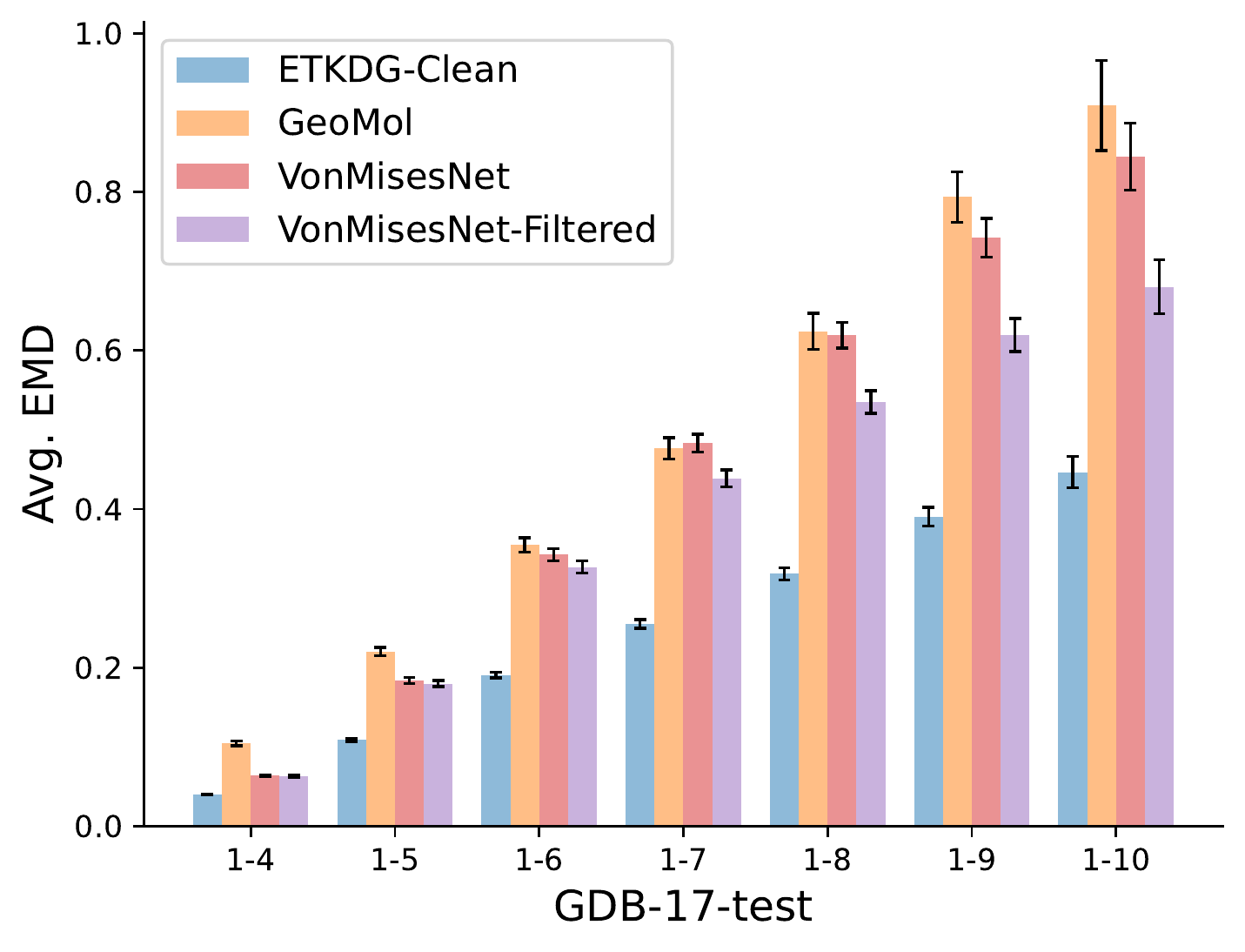}
}
\subfigure[]{
\includegraphics[scale = 0.55]{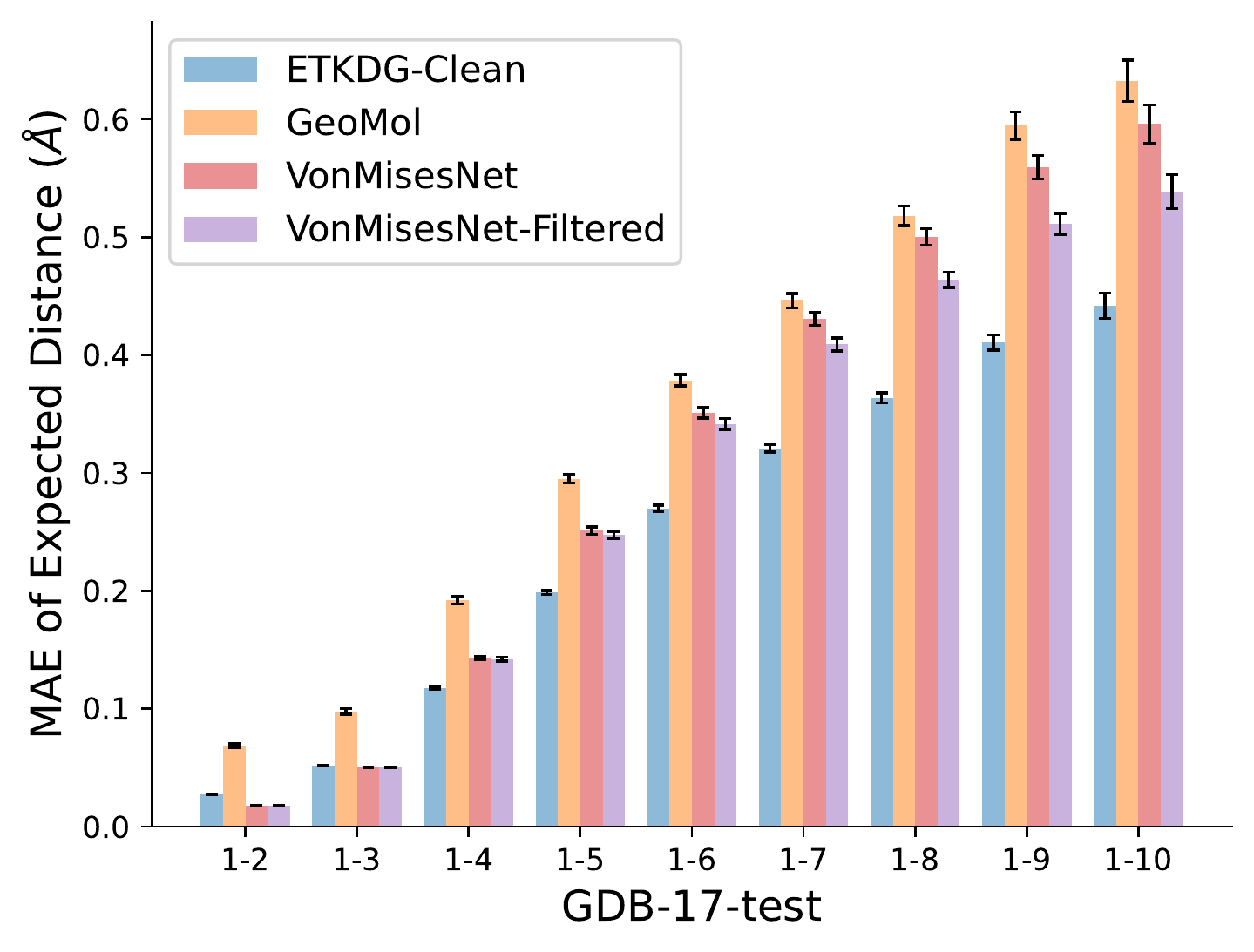}
}
\caption{\textit{Pairwise distance distributions evaluation, up to 1-10, excluding Torsional Diffusion constraints, no restrictions.}\label{fig:figure21} We evaluate pairwise distance distributions relative to PT-HMC ground truth for 997 random molecules from NMRShiftDB-test and 997 random molecules from GDB-17-test. We evaluate distances without restrictions. In \textbf{(a)} and \textbf{(c)} we compare the average EMD, per molecule, and in \textbf{(b)} and \textbf{(d)} we compare the MAE of the expected distance, per molecule. For the expected distance evaluations, we additionally include 1-2 distances and 1-3 distances without restrictions.}
\end{center}
\vskip -0.2in
\end{figure}

In this section we show additional evaluations. When evaluating without the Torsional Diffusion constraints, we use 997 out of the 1000 random molecules from NMRShiftDB-test and 997 out of the 1000 random molecules from GDB-17-test after removing molecules for which ETKDG embedding failed. In Figure \ref{fig:figure15}, we evaluate rotatable bond torsion angle distributions without the Torsional Diffusion constraint. As in section \ref{sec:sec33}, VonMisesNet and VonMisesNet-Filtered outperform the other methods. In Figures \ref{fig:figure16}, \ref{fig:figure17}, and \ref{fig:figure18}, we evaluate 1-$n$ pairwise distance distributions up to $n=10$. As in section \ref{sec:sec34}, VonMisesNet, VonMisesNet-Filtered, and Torsional Diffusion generally outperform the other methods when restricted to rotatable bonds, shown in Figure \ref{fig:figure16}. The performance of VonMisesNet-Filtered remains strong for $n \ge 6$, while the performance of VonMisesNet degrades. The performance of Torsional Diffusion tends to improve relative to VonMisesNet as a function of $n$. VonMisesNet and VonMisesNet-Filtered outperform the other methods on most metrics when all torsions are allowed except those belonging to non-aromatic rings, shown in Figure \ref{fig:figure17}, and ETKDG-Clean performs best with no restrictions, shown in \ref{fig:figure18}. In Figures \ref{fig:figure19}, \ref{fig:figure20}, and \ref{fig:figure21}, we evaluate on the random molecules from NMRShiftDB-test and GDB-17-test without the Torsional Diffusion constraints. Similar trends hold.  

\begin{figure}[!ht]
\vskip 0.2in
\begin{center}
\subfigure[]{
\includegraphics[scale = 0.27]{chirality_example_molecular_graph.pdf}
}

\subfigure[]{
\includegraphics[scale = 0.52]{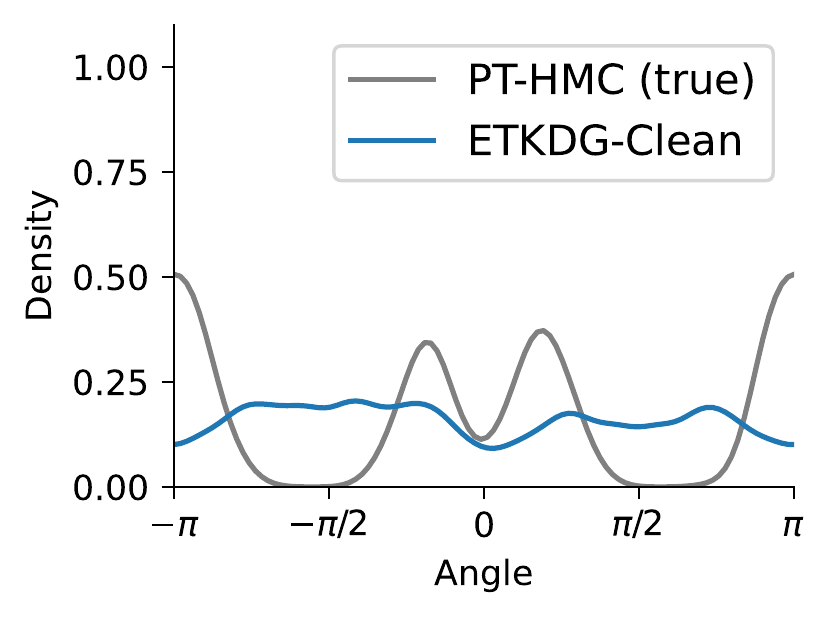}
}

\subfigure[]{
\includegraphics[scale = 0.52]{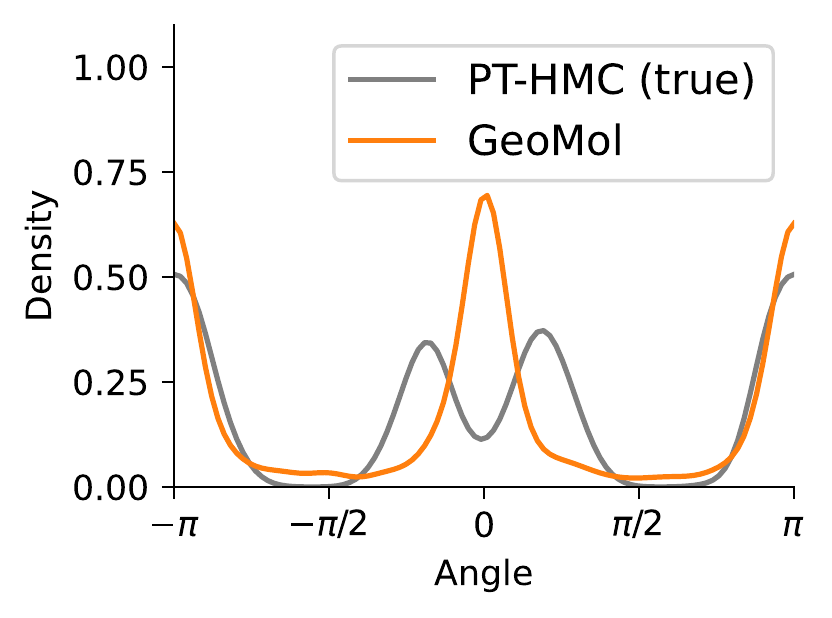}
}

\subfigure[]{
\includegraphics[scale = 0.52]{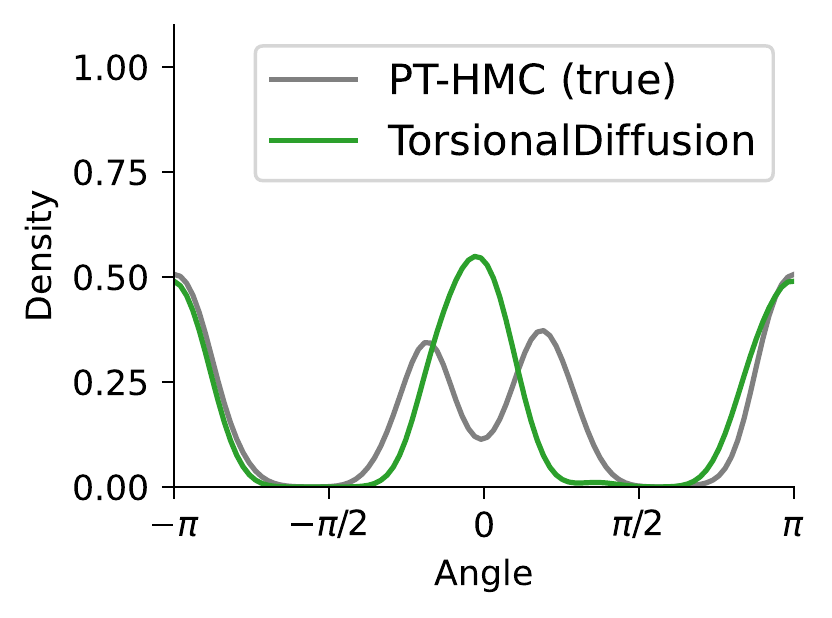}
}

\subfigure[]{
\includegraphics[scale = 0.52]{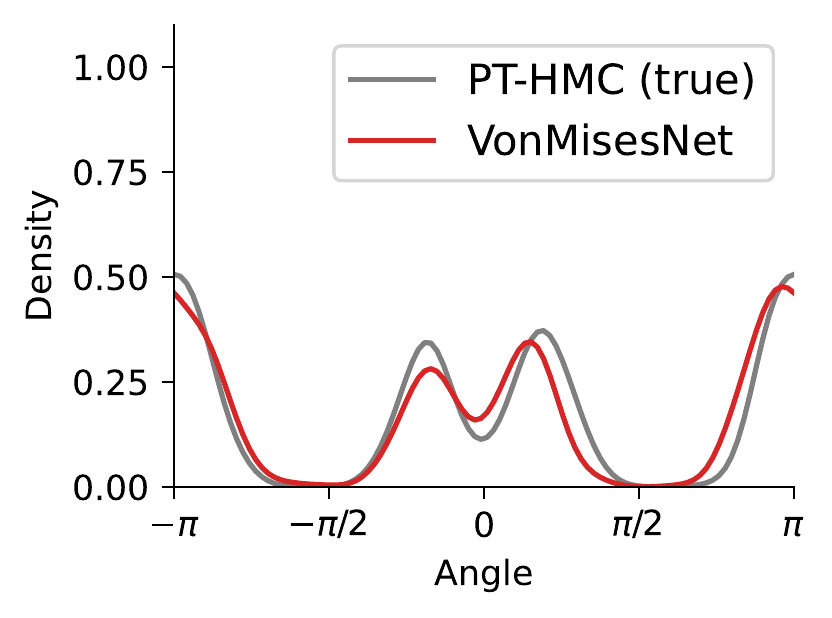}
}
\caption{\textit{Predicting chirality inversion example, extended.}\label{fig:figure22} \textbf{(a)} A molecule from NMRShiftDB-test, where the highlighted rotatable bond is between a carbon atom and a chirality inversion nitrogen atom. Kernel density estimates of the rotatable bond torsion angle from PT-HMC and from ETKDG-Clean, GeoMol, Torsional Diffusion, and VonMisesNet are shown in \textbf{(b)}, \textbf{(c)}, \textbf{(d)}, and \textbf{(e)}, respectively.}
\end{center}
\vskip -0.2in
\end{figure}

\section{Enforcing Consistent Atom Indexing}\label{app:appI}
When ETKDG generates multiple conformations for a single molecule, it can arbitrarily swap atom indices from one conformation to the next. This can happen specifically for atoms connected to a central atom that has four neighbors but does not have an existing chiral tag, or for neighbors of endpoint atoms of double bonds where the bond does not have a \textit{cis}/\textit{trans} stereochemistry tag and both endpoint atoms have more than one neighbor. Our metrics for computing pairwise distances require consistent atom indices, because we measure distances between specific indices across all conformations of a molecule. So, given an input molecule with an existing 3D geometry, we use the following procedure to prevent atom index swapping for any further conformations generated by ETKDG. For each atom that has four neighbors and no chiral tag, we compute the chirality based on the oriented volume and then assign an RDKit chirality tag accordingly. This will force future conformations generated by ETKDG to preserve the chirality and therefore preserve the current atom indexing. For each double bond where the bond does not have a \textit{cis}/\textit{trans} stereochemistry tag and both endpoint atoms have more than one neighbor, we set the double bond tag to \textit{cis} in a consistent manner. This will force future conformations generated by ETKDG to preserve the stereochemistry of the double bond and therefore preserve the current atom indexing. When we generate conformations using this procedure, we refer to the method as ETKDG-Clean.

When re-training GeoMol with our datasets, we noticed that failures occurred in the form of massive spikes in the train and validation loss for some molecules. We found that this occurred when the atom indices in RDKit were ordered in such a way that did not correspond to an ordering that results from creating the molecule directly from a SMILES string representation. This suggests that GeoMol requires atom indices to have the ordering based on a SMILES parsing. Therefore, for all input molecules to GeoMol for either training or inference, we first permute the atom indices so that they correspond to a SMILES ordering. Post-inference, we undo the permutation so that the atom indexing is consistent when comparing to other conformation generators. As a precaution, we used the same reordering procedure when running inference with Torsional Diffusion.  

\section{VonMisesNet Architecture}\label{app:appJ}
VonMisesNet is a graph neural network that processes a multi-partite graph representation of molecules. The neural network architecture is as follows. We process the node features in the input graph with a linear layer followed by Layer Norm. Subsequently, a series of graph convolution layers perform message passing between the graph nodes, where each message passing step is followed by a linear layer and ReLU activation, one for each step. Message passing is performed using matrix multiplication between the graph adjacency matrix and the node feature vectors, so that the messages are the current features and the aggregation function is a simple mean over neighbors. This results in a set of hidden vectors, one for each node in the graph. The ordering of the nodes is computed in a canonical way based on the ordering of atoms and bonds given by RDKit. For each prediction task, we process a relevant subset of the nodes using a separate readout function, i.e., feedforward neural network (FFN). Each of these FFNs applies Layer Norm to the inputs, applies a linear layer with size 128 followed by the ReLU activation, and then applies a final output linear layer. There is an FFN for the bond angle predictions that processes all of the angle nodes, an FFN for the bond length predictions that processes all of the bond nodes, and an FFN for the chirality probability predictions that processes all of the chirality inversion atom nodes. The output layer for each of these FFNs has size 1, and we also take the square of the outputs in order to enforce positive values. There are three FFNs that predict the von Mises mixture mean, concentration, and weight parameters, respectively, which process all of the rotatable bond nodes that do not have a chirality inversion endpoint atom. There are also two more sets of three FFNs that predict the von Mises mixture parameters, one for conditioning on R chirality and the other for S chirality, which process all of the rotatable bond nodes that have a chirality inversion endpoint atom. The output layer for each of these FFNs has size four. We scale the concentration outputs to be between one (mininum concentration) and 20 (maximum concentration) by applying BatchNorm1d and then a sigmoid activation to the concentration outputs, and then multiplying by the maximum and adding the minimum. We apply softmax to the weights outputs so that they sum to one. We use 20 graph convolution layers, a hidden size of 256, a batch size of 32, and the Adam optimizer with a learning rate of $0.0001$. We use gradient clipping for all of the model parameters with a cutoff value of 1.0. We scale the mean squared error losses for bond lengths, bond angles, and chirality probabilities by a factor of 32. Training on NMRShiftDB-train took 7.7 hours and training on GDB-17-train took 16.7 hours on a single NVIDIA GeForce RTX 2080 Ti GPU. 
\end{document}